\documentclass[11pt]{article}

\usepackage[nottoc,notlot,notlof]{tocbibind}  

\usepackage[normalem]{ulem}
\usepackage{graphicx} 
\usepackage{cite}     
\usepackage{amsthm}
\usepackage{float}
\restylefloat{table}
\usepackage{amsmath}
\usepackage{amssymb}
\usepackage{amsfonts}
\usepackage[usenames]{color}

\newcommand{\dd}{{\rm{d}}} 
\newcommand{\Dif}{{\rm D}}                                 
\newcommand{\rovno}{\!\!\!& = &\!\!\!} 
\newcommand{\eqdef}{\!\!\!& \equiv &\!\!\!} 
\newcommand{\ssqrt}{{\textstyle\frac1{\sqrt{2}}}} 

\newcommand{\X}{{\rho}}
\newcommand{\Kdt}{{\hbox{\tiny Kundt}}}

\newcommand{\boldu}{\mbox{\boldmath$u$}} 
\newcommand{\bolde}{\mbox{\boldmath$e$}} 
\newcommand{\boldk}{\mbox{\boldmath$k$}} 
\newcommand{\boldl}{\mbox{\boldmath$l$}} 
\newcommand{\boldm}{\mbox{\boldmath$m$}} 

\def \BE {\begin{equation}}
\def \EE {\end{equation}}
\def \BEA { \begin{eqnarray}}
\def \EEA {\end{eqnarray}}
\def \bea { \begin{eqnarray}}
\def \eea {\end{eqnarray}}
\def \be {\begin{equation}}
\def \ee {\end{equation}}

\def \pul {\textstyle{\frac{1}{2}}}
\def \ctvrt {\textstyle{\frac{1}{4}}}
\def \tre {\textstyle{\frac{1}{3}}}
\def \sest {\textstyle{\frac{1}{6}}}
\def \H {\mathcal{H}}
\def \T {\mathcal{T}}
\def \B {\mathcal{B}}
\def \k {m}
\def \oo {\omega}

\begin{document}

\title{Black holes and other spherical solutions in quadratic gravity\\ with a cosmological constant}

\author{V.~Pravda$^\diamond$, A.~Pravdov\' a$^\diamond$, J.~Podolsk\'y$^\star$, R.~\v{S}varc$^\star$
	\\
	\vspace{0.05cm} \\
	\vspace{0.05cm} \\
	{\small $^\diamond$ Institute of Mathematics, Academy of Sciences of the Czech Republic}, \\
	{\small \v Zitn\' a 25, 115 67 Prague 1, Czech Republic} \\[3mm]
	{\small $^\star$ Institute of Theoretical Physics, Faculty of Mathematics and Physics,} \\
	{\small Charles University, V~Hole\v{s}ovi\v{c}k\'ach~2, 180~00 Prague 8, Czech Republic
	} \\[3mm]
{\small  E-mail: \texttt{pravda@math.cas.cz, pravdova@math.cas.cz,}}\\
	{\small \texttt{podolsky@mbox.troja.mff.cuni.cz, robert.svarc@mff.cuni.cz }}\\}

\maketitle

\begin{abstract}
We study static spherically symmetric solutions to the vacuum field equations of quadratic gravity in the presence of a cosmological constant $\Lambda$. Motivated by the trace no-hair theorem, we assume the Ricci scalar to be constant throughout a spacetime. Furthermore, we employ the conformal-to-Kundt metric ansatz that is valid for all static spherically symmetric spacetimes and leads to a considerable simplification of the field equations. We arrive at a set of two ordinary differential equations and study its solutions using the Frobenius-like approach of (infinite) power series expansions. While the indicial equations considerably restrict the set of possible leading powers, careful analysis of higher-order terms is necessary to establish the existence of the corresponding classes of solutions. We thus obtain various non-Einstein generalizations of the Schwarzschild, (anti-)de~Sitter [or (A)dS for short], Nariai, and Pleba\'{n}ski--Hacyan spacetimes. Interestingly, some classes of solutions allow for an arbitrary value of~$\Lambda$, while other classes admit only discrete values of $\Lambda$. For most of these classes, we give recurrent formulas for all series coefficients. We determine which classes contain  the Schwarzschild--(A)dS black hole as a special case and briefly discuss the physical interpretation of the spacetimes. In the discussion of physical properties, we naturally focus on the generalization of the Schwarzschild--(A)dS black hole, namely the Schwarzschild--Bach--(A)dS black hole, which possesses one additional Bach parameter. We also study its basic thermodynamical properties  and observable effects on test particles caused by the presence of the Bach tensor.
This work is a considerable extension of our letter  [Phys. Rev. Lett., {\bf 121}  231104, 2018].
\end{abstract}
\vfil\noindent
PACS class:  04.20.Jb, 04.50.--h, 04.70.Bw




\newpage

\newpage
\tableofcontents
\newpage

\section{Introduction}

Despite great successes of Einstein's theory of gravity in giving predictions of various new physical phenomena, such as black holes and gravitational waves, one should keep in mind that so far this theory has been mostly tested in the weak-field regime and that
tests of strong gravity have started to appear only recently \cite{Abbott,Abuter,Akiyama}. Furthermore, even in the weak-field regime, there are attempts to incorporate the effects of dark matter and dark energy to the gravitational side of the equations by modifying the gravitational Lagrangian \cite{Capozziello}. Perhaps more importantly, there are also strong theoretical reasons to consider higher-order corrections to the Einstein--Hilbert action to address the non-renormalizability of Einstein's gravity (see e.g. \cite{Stelle:77,Salvio,Smilga,Stelle:1978}).

Thus it is of considerable interest to study vacuum solutions, and, in particular, black hole solutions, appearing in theories of gravity with higher-order corrections.
In this paper, we will focus on static spherically symmetric solutions of quadratic gravity (QG), for which quadratic terms in the curvature are added to the Einstein--Hilbert action,
\be
S = \int \dd^4 x\, \sqrt{-g}\, \Big(
\gamma \,(R-2\Lambda) +\beta\,R^2  - \alpha\, C_{abcd}\, C^{abcd}
\Big)\,,
\label{action}
\ee
with ${\gamma=1/G}$ ($G$ is the Newtonian constant), the cosmological constant $\Lambda $, and additional constant parameters $\alpha$, $\beta$ of the theory.

It is straightforward to show that all Einstein spaces $R_{ab}=\Lambda g_{ab}$  solve the corresponding field equations. Thus, in particular, the Schwarzschild--(A)dS black hole is a vacuum solution to QG. It is perhaps natural to expect that appropriate non-Einstein generalizations of Einstein spacetimes solving vacuum QG field equations might exist.\footnote{In this paper, we arrive at such generalizations of the Schwarzschild, (A)dS, Nariai, and Pleba\'{n}ski--Hacyan spacetimes.}
Indeed, it has been recently demonstrated (in the case of vanishing~$\Lambda$)  using in part numerical methods that another spherically symmetric black hole ``over and above'' the Schwarzschild solution exists in QG \cite{LuPerkinsPopeStelle:2015,LuPerkinsPopeStelle:2015b}. The uniqueness of the Schwarzschild black hole in this theory is thus lost. Nevertheless, there exist some no-hair type theorems forcing black holes in QG to share certain properties with the Schwarzschild black hole. In particular, the trace no-hair theorem of \cite{Nelson:10,LuPerkinsPopeStelle:2015b} implies that static, asymptotically flat solutions of QG with a horizon have ${R=0}$ throughout the spacetime. More generally, for static spacetimes with   $R$ sufficiently quickly approaching a constant, ${R=4 \Lambda}$ throughout the spacetime \cite{LuPerkinsPopeStelle:2015b}. These results lead to a considerable simplification of the field equations which, assuming ${R=4 \Lambda}$,
take the form
\BE
R_{ab}-\Lambda \, g_{ab}=4k\, B_{ab}\,,\qquad \hbox{with} \qquad
k \equiv \frac{\alpha}{\gamma+8\beta\Lambda} \,,
\label{fieldeqsEWmod}
\EE
where $B_{ab}$ is the Bach tensor defined in Eq.~\eqref{defBach}. Thus, all non-Einsteinian terms appearing in the fourth-order field equations of QG are combined into the Bach tensor.

This observation can be employed to simplify the field equations of QG for static spherically symmetric spacetimes. In particular, it has been pointed out that these spacetimes  are conformal to Kundt spacetimes (in fact this applies to a larger class of the Robinson--Trautman spacetimes \cite{PravdaPravdovaPodolskySvarc:2017}). Furthermore,
the Bach tensor is well-behaved under a conformal transformation, see (\ref{Bachproperties}). Therefore, it is convenient to express the Bach tensor in an appropriate Kundt background determined only by one free function, and then simply rescale it to obtain its expression in a spherically symmetric spacetime.
We have employed this approach leading to a considerable simplification of the field equations in \cite{PodolskySvarcPravdaPravdova:2018,SvarcPodolskyPravdaPravdova:2018,PodolskySvarcPravdaPravdova:2020}.
In the present work, we study the case with a nonvanishing cosmological constant $\Lambda$ in detail\footnote{Thus extending considerably the letter \cite{SvarcPodolskyPravdaPravdova:2018}.} using the Frobenius-like approach of infinite power series expansions.  Indicial equations significantly restrict  the set of possible leading powers.  However, careful analysis of the corresponding higher-order terms shows that some of the classes compatible with indicial equations are considerably restricted  or empty. Taking all higher orders into account, we arrive at eight classes of solutions in the Kundt coordinates\footnote{Out of these eight classes, seven (cf. Tables~\ref{tbl:011} and \ref{tbl:02}) can be transformed to the standard spherically symmetric coordinates and lead to twelve (cf. Table~\ref{tab:3} ) distinct classes of solutions in these coordinates with a continuous value of $\Lambda$.} allowing for an arbitrary value of $\Lambda$, and several additional classes of solutions allowing only for a single non-zero value or a  discrete set of values of $\Lambda$. For most classes, we derive recurrent formulas for all series coefficients and briefly discuss their physical interpretation. Interestingly, several classes contain various black hole solutions. Three of them (namely the classes $[-1,3]^\infty$,   $[0,1]$, and  $[0,0]$, corresponding to power series expansions in the vicinity of the origin,  the horizon, and an arbitrary finite point, respectively) describe (possibly distinct) generalizations of the Schwarzschild--(A)dS black hole with a nonvanishing Bach tensor and an \emph{arbitrary} cosmological constant $\Lambda$. We call them the Schwarzschild--Bach--(A)dS black holes. Furthermore, two other classes contain three QG generalizations of the Schwarzschild--(A)dS black hole with a nonvanishing Bach tensor and \emph{discrete} values of $\Lambda$, called  the higher-order discrete  Schwarzschild--Bach--(A)dS black holes (class $[-1,0]$), the extreme higher-order discrete  Schwarzschild--Bach--dS black holes  (a subcase of class $[0,2]$), and the extreme Bachian--dS black hole (subcase of class $[0,2]$). All the above solutions admit the Schwarzschild--(A)dS limit. In contrast, several additional classes, e.g. Bachian singularity $[1,0]$ and Bachian vacuum $[-1,2]^\infty$, do \emph{not} contain the Schwarzschild--(A)dS as a special case.
Note also that some of the solutions found in this paper, such as the Nariai spacetime, its Bachian generalization $[0,2]^\infty$ and Pleba\'{n}ski--Hacyan solutions $[0,0]^\infty$, $[0,1]^\infty$, are Kundt spacetimes which cannot be transformed into the standard static spherically symmetric coordinates.

Physical properties of the Schwarzschild--Bach--(A)dS black hole, which is a generalization of the classic Schwarzschild--(A)dS black hole possessing one additional (Bach) parameter, are discussed in more detail. In particular, we study basic thermodynamical properties of this black hole and observable effects on test particles caused by the presence of the Bach tensor.

Our paper is organized as follows. Section~\ref{sec_TB} focuses on the preliminary material, in particular the discussion of the field equations of QG, the conformal-to-Kundt ansatz for static spherical spacetimes, and curvature invariants for these spacetimes. Field equations of QG in the conformal-to-Kundt coordinates are presented in Section~\ref{derivingFE}. In addition, the classes compatible with indicial equations of the Frobenius-like analysis are summarized therein. In Section \ref{expansiont_0}, these classes are derived for the expansion in powers of  ${\Delta\equiv r-r_0}$, around a fixed point $r_0$, and in  Section~\ref{description}, they are analysed in detail. Similarly, Sections~\ref{expansiont_INF} and~\ref{description_INF} are devoted to the derivation and discussion of the classes obtained by the expansion in powers of $r^{-1}$ as ${r \rightarrow \infty}$, respectively. In Section~\ref{summary}, all solutions found  in this paper that can be transformed into the standard spherically symmetric coordinates are classified in the notation used in the literature. In concluding Sec. \ref{sec_sum2}, we give lists of all solutions sorted by (both Kundt and physical) regions in which the solutions are expanded.
Finally,   we review the derivation of the  field equations in Appendix~\ref{analysingFE}.


\section{Preliminaries: quadratic gravity, conformal-to-Kundt metric ansatz, and invariants}
\label{sec_TB}

\subsection{Quadratic gravity}
\label{DObackgrounds}

Quadratic gravity is a generalization of Einstein's theory whose action  contains additional terms which are quadratic in curvature. Since in four dimensions  the Gauss--Bonnet term does not contribute to the field equations,\footnote{However, let us remark that recently the specific regularization method taking the Gauss--Bonnet term into account even in four dimensions has been proposed in \cite{GlavanLin:2020}. This has been immediately followed by various explicit examples of such an approach as well as many doubts about its mathematical and physical relevance. However, here we stay on a classic level considering the Gauss--Bonnet term being irrelevant in a four-dimensional theory.} the action of QG in vacuum can be expressed in full generality by~\eqref{action}. The corresponding {vacuum field equations} are
\be
\gamma \left(R_{ab} - {\pul} R\, g_{ab}+\Lambda\,g_{ab}\right)-4 \alpha\,B_{ab}
 +2\beta\left(R_{ab}-\tfrac{1}{4}R\, g_{ab}+ g_{ab}\, \Box - \nabla_b \nabla_a\right) R = 0 \,, \label{fieldeqsEW}
\ee
where $B_{ab}$ is the {Bach tensor}
\BE
B_{ab} \equiv \big( \nabla^c \nabla^d + {\pul} R^{cd} \big) C_{acbd} \,, \label{defBach}
\EE
or equivalently
\BE
B_{ab}={\pul}\Box R_{ab} -{\sest}\big( \nabla_a \nabla_b +{\pul} g_{ab}\Box \big)R
-{\tre}RR_{ab}+R_{acbd}\,R^{cd}+{\ctvrt}\big({\tre}R^2-R_{cd}R^{cd}\big)g_{ab} \,.\label{BachRicci}
\EE
It is traceless, symmetric, conserved, and well behaved under a conformal transformation $g_{ab}=\Omega^2 \tilde g_{ab}$:
\begin{equation}
g^{ab}B_{ab}=0 \,, \qquad B_{ab}=B_{ba} \,, \qquad
\nabla^b B_{ab}=0
\,, \qquad B_{ab}=\Omega^{-2}\tilde B_{ab}\,.
\label{Bachproperties}
\end{equation}

Furthermore, from \eqref{BachRicci} it can be seen that the Bach tensor vanishes for Einstein spacetimes, i.e., for spacetimes obeying $R_{ab}=\tfrac{1}{4}R\, g_{ab}$, where $R$ is constant.
Consequently, in four dimensions, all vacuum solutions to the
Einstein theory (including solutions with  a cosmological constant) solve also the vacuum
equations of QG  \eqref{fieldeqsEW}.

The trace no-hair theorem of \cite{LuPerkinsPopeStelle:2015b} implies that
 for static spacetimes with   $R$ sufficiently quickly approaching a constant,
\BE
 R = 4 \Lambda  \,
 \label{traceEW}
\EE
throughout the spacetime. Then  the vacuum QG field equations  (\ref{fieldeqsEW}) simplify to \eqref{fieldeqsEWmod}.
For ${k=0}$, the field equations  
\eqref{fieldeqsEWmod} reduce to vacuum  Einstein's field equations of general relativity.
Another special case $k\rightarrow\infty$, i.e., ${\gamma=-8\beta\Lambda}$, was discussed in
\cite{PravdaPravdovaPodolskySvarc:2017}.


\subsection{Static spherically symmetric metrics}
\label{BH metric}

For our study of static spherically symmetric solutions to QG, instead of employing the standard  metric
\begin{equation}
\dd s^2 = -h(\bar r)\,\dd t^2+\frac{\dd \bar r^2}{f(\bar r)}+\bar r^2(\dd \theta^2+\sin^2\theta\,\dd \phi^2) \,,
\label{Einstein-WeylBH}
\end{equation}
we use the conformal-to-Kundt ansatz
\cite{PodolskySvarcPravdaPravdova:2018, SvarcPodolskyPravdaPravdova:2018, PodolskySvarcPravdaPravdova:2020}
\be
\dd s^2 \equiv \Omega^2(r) \,\dd s^2_{{\rm Kundt}}= \Omega^2(r)
\Big[\,\dd \theta^2+\sin^2\theta\,\dd \phi^2 -2\,\dd u\,\dd r+\H(r)\,\dd u^2 \,\Big]\,,
\label{BHmetric}
\ee
related to \eqref{Einstein-WeylBH} by the transformation
\begin{equation}
\bar{r} = \Omega(r)\,, \qquad t = u - \int\! \frac{\dd r}{\H(r)} \,,
\label{to static}
\end{equation}
with \be
h = -\Omega^2\, \H \,, \qquad
f = -\left(\frac{\Omega'}{\Omega}\right)^2 \H \,.\label{rcehf}
\ee
Prime denotes the derivative with respect to $r$, and  the argument $r$ of both $\Omega$
and $\H$ must be expressed in terms of $\bar{r}$ using the inverse of the relation ${\bar{r} = \Omega(r)}$.

In \eqref{BHmetric}, the seed metric ${\dd s^2_{\hbox{\tiny Kundt}}}$ is of Petrov type~D (consequently, also the full metric \eqref{BHmetric} is of type~D) and it is  a direct product  \cite{GriffithsPodolsky:2009, PravdaPravdovaPodolskySvarc:2017} of two 2-spaces. It belongs  to the {Kundt class}, admitting
nonexpanding, shear-free, and twist-free null congruence, see \cite{Stephanietal:2003, GriffithsPodolsky:2009}.
  The first spacelike part, spanned by ${\theta, \phi}$, is a round 2-sphere of Gaussian curvature ${K=1}$,  while the second part, spanned by ${u, r}$, is a two-dimensional Lorentzian spacetime. Using the stereographic representation of a 2-sphere given by  ${x+\hbox{i}\, y = 2\tan(\theta/2)\exp(\hbox{i}\phi)}$, the {Kundt seed} metric takes the form
\be
\dd s^2_{\hbox{\tiny Kundt}}=\frac{\dd x^2+\dd y^2}{\big(1+\frac{1}{4}(x^2+y^2)\big)^2}
-2\,\dd u\,\dd r+{\H}(r)\,\dd u^2 \,.
\label{Kundt seed xy}
\ee

The metric \eqref{BHmetric} admits a \emph{gauge freedom} given by a constant rescaling and a shift of~$r$,
\be
r \to \lambda\,r+\upsilon\,, \qquad u \to \lambda^{-1}\,u \,.
\label{scalingfreedom}
\ee

Note that for the classic Schwarzschild--(A)dS metric
\begin{equation}
f(\bar{r}) = h(\bar{r})=1-\frac{2m}{\bar{r}} -\frac{\Lambda}{3}{\bar r}^ 2\,,
\label{SchwarzschildAdSBH}
\end{equation}
the relations \eqref{to static}, \eqref{rcehf} imply
\begin{equation}
\bar{r}=\Omega(r)=-\frac{1}{r}\,,\qquad
\H(r) = \frac{\Lambda}{3} -r^2-2m\, r^3 \,.
\label{SchwAdS}
\end{equation}
The horizon is located at the zeros of the metric function
$\H $, where  $h({\bar r}) $ and $f({\bar r})$ also vanish
by \eqref{rcehf}.

Similarly, in a general case,  a horizon can be defined as the {Killing horizon} associated with the vector field ${\partial_u}$,
which coincides with ${\partial_t}$, i.e., located at
$r=r_h$ satisfying
\begin{equation}
\H \big|_{r=r_h}=0\,, \label{horizon}
\end{equation}
since $\Omega$ is everywhere nonvanishing.
Using \eqref{rcehf},
this condition corresponds to
${h({\bar r_h})=0=f({\bar r_h})}$.

Note that  a {time-scaling freedom} of the metric \eqref{Einstein-WeylBH}
\be
t\to  t/\sigma \,,
\label{scaling-t}
\ee
where ${\sigma \ne 0}$ is any constant,
can be used to
adjust a value of $h$ at a chosen radius ${\bar r}$ since  ${h\to h\,\sigma^2}$.

\subsection{Curvature invariants and geometric classification}
\label{invariants}

In \cite{PodolskySvarcPravdaPravdova:2020},  we have observed that for a geometrical and physical interpretation of solutions to the QG field equations,  the scalar curvature invariants constructed from the Ricci, Bach, and Weyl tensors play an  important role.

 For the static spherically symmetric metric \eqref{BHmetric}, we get
\begin{align}
R_{ab}\, R^{ab} &=  4\Lambda^2+16k^2\, B_{ab} B^{ab} \,, \label{invR}\\
B_{ab}\, B^{ab} &=  \tfrac{1}{72}\,\Omega^{-8}\,\big[(\B_1)^2 + 2(\B_1+\B_2)^2\big] \,,\label{invB}\\
C_{abcd}\, C^{abcd} &=  \tfrac{1}{3}\,\Omega^{-4}\,\big({\H}'' +2\big)^2 \,, \label{invC}
\end{align}
where
the functions $\B_1(r)$ and $\B_2(r)$ denote {two independent components of the Bach tensor},
\bea
&& \B_1 \equiv {\H}{\H}''''\,, \label{B1}\\
&& \B_2 \equiv {\H}'{\H}'''-\tfrac{1}{2}{{\H}''}^2 +2\,. \label{B2}
\eea
To derive \eqref{invR}--\eqref{B2}, we have used expressions
for the Ricci, Bach, and Weyl tensors given in Appendix~A and Appendix~B of \cite{PodolskySvarcPravdaPravdova:2020}
and the invariance of the Weyl tensor under conformal transformations, ${C_{abcd}\, C^{abcd}=\Omega^{-4}\, C_{abcd}^\Kdt\, C^{abcd}_\Kdt}$.

Note that the {Bach component} ${\B_1=\H \H''''}$
{vanishes on the horizon} where ${\H=0}$, see \eqref{horizon}.
Similarly as in \cite{PodolskySvarcPravdaPravdova:2020},
one can show that
\be
B_{ab}=0\quad \hbox{if, and only if,}\quad  B_{ab}\,B^{ab} =0\,,
\label{Bach=0iffINV=0}
\ee
and
\be
C_{abcd}\,C^{abcd}=0\quad \hbox{implies}\quad  B_{ab} =0\,.
\label{Weylinv=0thenBach=0}
\ee

There are two geometrically distinct classes of solutions to QG field equations,
depending on the Bach tensor
${B_{ab}}$, namely a simple case corresponding to ${B_{ab}=0}$,
and a case with ${B_{ab}\ne0}$,  not allowed in general relativity.
Later we will see that the Bach tensor influences various physical aspects
of the solutions, such as
the geodesic deviation equation for test particles and entropy of black holes.


\section{The field equations}
\label{derivingFE}

To  derive an explicit form of the field equations, we proceed as in  \cite{PodolskySvarcPravdaPravdova:2020}
using the conformal-to-Kundt metric ansatz \eqref{BHmetric},
with the Ricci and Bach tensors  for the Kundt seed metric $g_{ab}^\Kdt$
 and a spherically symmetric metric \eqref{BHmetric}
  given in  Appendices~A and~B of \cite{PodolskySvarcPravdaPravdova:2020},
 respectively.
Employing  also the Bianchi identities, the QG field equations
  \eqref{fieldeqsEWmod} reduce  to an autonomous system of two compact ordinary differential equations for the two metric functions $\Omega(r)$ and $\H(r)$, see Appendix~A here,
  \begin{align}
  \Omega\Omega''-2{\Omega'}^2 = &\ \tfrac{1}{3}k\, \B_1 \H^{-1} \,, \label{Eq1}\\
  \Omega\Omega'{\H}'+3\Omega'^2{\H}+\Omega^2
  -\Lambda\Omega^4= &\ \tfrac{1}{3}k \,\B_2  \,, \label{Eq2}
  \end{align}
where the functions $\B_1(r)$ and $\B_2(r)$ denote {two independent components of the Bach tensor}, \eqref{B1} and \eqref{B2},
respectively.

The trace \eqref{traceEW} of the field equations \eqref{fieldeqsEWmod}
takes the form
\begin{equation}
{\H}\Omega''+{\H}'\Omega'+{\textstyle \frac{1}{6}} ({\H}''+2)\Omega = \tfrac{2}{3}\Lambda\Omega^3 \,.
\label{trace}
\end{equation}
In fact,  \eqref{trace} can be obtained by subtracting \eqref{Eq1} multiplied by $\H'$ from the derivative of \eqref{Eq2} and dividing the result by $6\Omega'$.

Note that similarly as in  \cite{PodolskySvarcPravdaPravdova:2020},
vanishing of the Bach tensor implies the Schwarzschild--(A)dS solution
\eqref{SchwAdS} with the following scalar invariants
(\ref{invR})--(\ref{invC})
\be
R_{ab}\, R^{ab} = 4\Lambda^2\,,\qquad
B_{ab}\, B^{ab}=0\,,\qquad
C_{abcd}\, C^{abcd} = 48\,m^2\,r^6 \,.
\label{SchwarzInvariants}
\ee
For ${m\not=0}$, there is a curvature singularity at ${r\to\infty}$ corresponding to ${\bar{r}=\Omega(r)=0}$.\footnote{For brevity, in this paper the symbol ${r\to\infty}$ means ${|r|\to\infty}$, unless the sign of $r$ is explicitly specified.}

In the rest of this paper, we concentrate on \emph{solutions
with a nontrivial Bach tensor}.
In this case, the system \eqref{Eq1}, \eqref{Eq2} is coupled
in a complicated way and it seems hopeless to find explicit solutions in a closed form. Thus we focus on studying these solutions in terms of (infinite) power series. Since the system \eqref{Eq1}, \eqref{Eq2} is autonomous, there are only two natural possibilities --- the expansion in powers of the parameter ${\Delta \equiv r-r_0}$, which expresses the solution around any finite value $r_0$, and the expansion in powers of $r^{-1}$, which is applicable for sufficiently large values of $r$.

\subsection{Expansion in powers of~${\Delta \equiv r-r_0}$ }
\label{expansio_DElta}

We  search for solutions of  \eqref{Eq1}, \eqref{Eq2}
 in the form of an {expansion in powers\footnote{Note that other solutions containing for example $\log (r-r_0)$ terms may also exist. } of $r-r_0$  around any fixed value} ${r_0}$,
\begin{eqnarray}
\Omega(r) \rovno \Delta^n   \sum_{i=0}^\infty a_i \,\Delta^{i}\,, \label{rozvojomeg0}\\
\H(r)     \rovno \Delta^p \,\sum_{i=0}^\infty c_i \,\Delta^{i}\,, \label{rozvojcalH0}
\end{eqnarray}
where
\be
\Delta\equiv r-r_0\,, \label{DElta}
\ee
 $r_0$ is {a real constant}, and  ${i=0, 1, 2, \ldots}$ are integers, so that the metric functions are
expanded in integer steps of ${\Delta=r-r_0}$. On the other hand, the {dominant real powers}
$n$ and $p$ in the expansions (\ref{rozvojomeg0}) and (\ref{rozvojcalH0})
{need not be} integers. We only assume that ${a_0\not=0}$ and ${c_0\not=0}$, so that
the coefficients $n$ and $p$ are  the leading powers.

By inserting  \eqref{rozvojomeg0}--\eqref{DElta} into the field equations \eqref{Eq1} and \eqref{Eq2}, we will show in Section~\ref{expansiont_0} that only the following  \emph{eight classes of solutions} are compatible with the leading orders of the expansion in $\Delta$:
\be
 [n,p]=[-1,2]\,,\
[0,1]\,,\
[0,0]\,, \
[0,2]\,,\
[-1,0]\,,\
[1,0]\,,\
[0,>2]\,,\
[<0,2(n+1)<0]\,.
\label{4classes}
\ee

Section~\ref{description} contains more detailed analysis of these solutions, namely
\begin{itemize}
\item
	Sec. \ref{Schw_[n,p]=[-1,2]}:
 class $[-1,2]$, already discussed
in  \cite{PodolskySvarcPravdaPravdova:2020}, contains only
the Schwarzchild black hole for which the Bach tensor vanishes;
\item
Sec. \ref{SchwaBach_[n,p]=[0,1]}: class ${[0,1]}$ contains the Schwarzschild--Bach--(anti-)de Sitter  black hole
 (abbreviated as Schwarzschild--Bach--(A)dS, or even shorter as Schwa--Bach--(A)dS) with a nonvanishing Bach tensor;
\item
 Sec. \ref{SchwaBach_[n,p]=[0,0]}: class ${[0,0]}$ describes all other discussed solutions at generic points;
\item
 Sec. \ref{sec_[0,2]}: class  ${[0,2]}$, apart from the extreme Schwarzschild--dS solution with a generic $\Lambda$, includes also the extreme higher-order (discrete) Schwarzschild--Bach--dS black holes with restricted values of $\Lambda$, and the extreme Bachian black hole with $\Lambda=3/(8k)$;
\item
 Sec. \ref{IIIa}: class ${[-1,0]}$, apart from the Schwarzschild--(A)dS black hole for a generic $\Lambda$, contains also the higher-order (discrete) Schwarzschild--Bach--(A)dS black holes admitting only special values of $\Lambda$;
\item
 Sec. \ref{SchwaBach_[n,p]=[1,0]}: class ${[1,0]}$ describes a Bachian singularity at the origin;
 \item
 Sec. \ref{sec_[0,>2]}: class ${[0,>2]}$ is, in fact, empty;
 \item
 Sec. \ref{sec_[n,2n+2]}: classes ${[<0,2n+2]}$, requiring discrete values of $n$ and $\Lambda$, describe asymptotic regions of solutions with a strictly nonvanishing Bach tensor.
 \end{itemize}

\subsection{Expansion in powers of~$r^{-1}$ }
\label{expansion_INF}

Analogously, we may study and classify all possible solutions to the vacuum QG field equations for an asymptotic expansion as  ${r\rightarrow \infty}$. Instead of \eqref{rozvojomeg0}, \eqref{rozvojcalH0} with \eqref{DElta}, for very large $r$ we can assume that the metric functions $\Omega(r)$,  $\H (r)$  are expanded in \emph{negative powers} of $r$ as
\begin{eqnarray}
\Omega(r)      \rovno r^N   \sum_{i=0}^\infty A_i \,r^{-i}\,, \label{rozvojomegINF}\\
\H (r) \rovno r^P \,\sum_{i=0}^\infty C_i \,r^{-i}\,. \label{rozvojcalHINF}
\end{eqnarray}

	By inserting the series (\ref{rozvojomegINF}), (\ref{rozvojcalHINF}) into the field equations \eqref{Eq1}, \eqref{Eq2}, it can be shown that 	the following  five classes of solutions are compatible with the leading orders of the expansion in $r^{-1}$:
\be
[N,P]=[-1,3]^\infty\,,\
[-1,2]^\infty\,, \
[0,2]^\infty\,,\
[0,<2]^\infty\,,\
[>0,2N+2]^\infty\,,
\label{2classes}
\ee
see Section~\ref{expansiont_INF}. Subsequent Section~\ref{description_INF} contains more detailed analysis of the above solutions, namely
	\begin{itemize}
		\item Sec. \ref{Schw_[N,P]=[-1,3]}: class $[-1,3]^\infty$ describes the
		Schwarzschild--Bach--(A)dS black hole near the singularity at the origin;
		\item
		Sec. \ref{Schw_[N,P]=[-1,2]}:
		 class $[-1,2]^\infty$ describes Bachian--(A)dS vacuum near the origin --- a specific Bachian generalization of the (A)dS  space;
		\item Sec. \ref{sec_[0,2]infty}: class $[0,2]^\infty$ contains the exact Nariai spacetime with arbitrary $\Lambda$,  the spherically symmetric  higher-order discrete Nariai--Bach solutions near a finite point with a nonvanishing Bach tensor and discrete spectrum of~$\Lambda$, and another Bachian generalization of the Nariai spacetime belonging to the Kundt class;
		\item Sec. \ref{sec_[0,m2]infty}: classes $[0,<2]^\infty$ contain only exact Pleba\'{n}ski--Hacyan solutions in the Kundt class $[0,1]^\infty$ and $[0,0]^\infty$ with ${\Lambda=\frac{3}{8k}}$;
		\item Sec. \ref{sec_[N,2N+2]infty}: classes $[N>0,P=2N+2]^\infty$ contain solutions with regular Bachinan infinity, which require discrete values of $N$ and $\Lambda$ and describe asymptotic regions of solutions with a strictly  nonvanishing Bach tensor.
	\end{itemize} 	
\newpage

\section{Discussion of solutions using the expansion in powers of $\Delta$}
\label{expansiont_0}

The series \eqref{rozvojomeg0}, \eqref{rozvojcalH0}
together with  the first field equation (\ref{Eq1})  yield
\begin{align}
&\sum_{l=2n-2}^{\infty}\Delta^{l}\sum^{l-2n+2}_{i=0}a_i\, a_{l-i-2n+2}\,(l-i-n+2)(l-3i-3n+1) \nonumber \\
& \hspace{45.0mm}=\tfrac{1}{3}k \sum^{\infty}_{l=p-4}\Delta^{l}\,c_{l-p+4}\,(l+4)(l+3)(l+2)(l+1) \,,
\label{KeyEq1}
\end{align}
while the second field equation (\ref{Eq2})
gives
\begin{align}
&\sum_{l=2n+p-2}^{\infty}\Delta^{l}\sum^{l-2n-p+2}_{j=0}\sum^{j}_{i=0}a_i\,a_{j-i}\,c_{l-j-2n-p+2}\,(j-i+n)(l-j+3i+n+2) \nonumber \\
& \hspace{10.0mm} +\sum_{l=2n}^{\infty}\Delta^{l}\sum^{l-2n}_{i=0}a_i\,a_{l-i-2n}-\Lambda \sum_{l=4n}^{\infty}
\Delta^{l}\sum^{l-4n}_{m=0}\bigg(\sum^{m}_{i=0}a_i\,a_{m-i}\bigg)\bigg(\sum^{l-m-4n}_{j=0}a_j\,a_{l-m-j-4n}\bigg) \nonumber \\
& = \tfrac{1}{3}k \bigg[2+\sum^{\infty}_{l=2p-4}\Delta^{l}\sum^{l-2p+4}_{i=0}c_{i}\,c_{l-i-2p+4}\,(i+p)(l-i-p+4)
(l-i-p+3)(l-\tfrac{3}{2}i-\tfrac{3}{2}p+\tfrac{5}{2})\bigg]\,.
\label{KeyEq2}
\end{align}

It is also useful to consider  considerably simpler constraints  following from the trace equation
(\ref{trace})
\begin{align}
&\sum_{l=n+p-2}^{\infty}\Delta^{l}\sum^{l-n-p+2}_{i=0}c_i\,a_{l-i-n-p+2}\,\big[(l-i-p+2)(l+1)
+\tfrac{1}{6}(i+p)(i+p-1)\big] \nonumber \\
& \hspace{50mm} +\tfrac{1}{3}\sum^{\infty}_{l=n}\Delta^{l}\,a_{l-n} = \tfrac{2}{3}\Lambda
\sum^{\infty}_{l=3n}\Delta^{l}\sum^{l-3n}_{j=0}\sum^{j}_{i=0}a_i\,a_{j-i}\,a_{l-j-3n} \,.
\label{KeyEq3}
\end{align}

Coefficients of the same powers
of $\Delta^l$ in Eq. \eqref{KeyEq1} give expressions for
the coefficients $c_j$ in terms of~$a_j$. Since
 the lowest
orders on the left and right sides are ${l=2n-2}$ and ${l=p-4}$,
respectively, there are three cases to be considered
\begin{itemize}
	\item Case I: ${\ \ 2n-2<p-4}$\,, \ i.e.,  ${\ p>2n+2}$\,,
	\item Case II: ${\ 2n-2>p-4}$\,,  \ i.e.,  ${\ p<2n+2}$\,,
	\item Case III: ${2n-2=p-4}$\,,   \ i.e.,  ${\ p=2n+2}$\,.
\end{itemize}
Notice that Eq.~\eqref{KeyEq1} does not depend on the cosmological
constant~$\Lambda$ and thus the above cases do not differ from the  ${\Lambda=0}$
cases discussed systematically in \cite{PodolskySvarcPravdaPravdova:2020}.

In what follows, we   study various
solutions in these three cases.

\subsection{\textbf{Case I}}

In  Case I,
 the {lowest} order  in  (\ref{KeyEq1}) is on the {left hand} side ($\Delta^{2n-2}$)   and therefore, since $a_0\not=0$,
\begin{equation}
n(n+1)=0 \,,
\label{KeyEq1CaseI}
\end{equation}
leading to  two possible cases ${n=0}$ and ${n=-1}$. The lowest orders of Eq.  (\ref{KeyEq3}) are
\begin{equation}
\big[6n(n+p-1)+p(p-1)\big]c_0\,\Delta^{n+p-2}+\cdots+2\,\Delta^{n}+\cdots -4\Lambda a_0^2\Delta^{3n}+\cdots=0\,.
\label{KeyEq3CaseI}
\end{equation}

For ${n=0}$, these powers are ${\Delta^{p-2}}$,
 ${\Delta^{0}}$, and  ${\Delta^{0}}$, respectively, but in  Case~I, ${p-2>2n=0}$.
The lowest order ${(2-4\Lambda a_0^2)\Delta^{0}}$ thus leads to
\be
2\Lambda a_0^2=1\,,
\ee
and Eq. \eqref{KeyEq2} then gives
\be
\Lambda= \frac{3}{8k}\,.
\ee
Thus such class exists only for nonvanishing $\Lambda$ of this special value.

For ${n=-1}$,  Eq. \eqref{KeyEq3CaseI} reduces to
\begin{equation}
(p-3)(p-4)c_0\,\Delta^{p-3}+\cdots +2\,\Delta^{-1}+\cdots
-4\Lambda a_0^2\Delta^{-3}+\cdots=0
\,.
\label{KeyEq3CaseIn=-1}
\end{equation}
Since ${c_0\neq 0 \neq a_0}$, the only possibility is ${p=2}$,  $\Lambda=0$, with ${c_0=-1}$.
\vspace{5mm}

\noindent
\textbf{To summarize}: The only possible classes of solutions in Case~I are
\begin{align}
[n,p] &=[-1,2]\qquad  \hbox{with}\quad  \Lambda=0\,, \quad c_0=-1\,,\label{CaseI_summary1} \\
\ [n,p] &= [0,p>2]\quad \hbox{with}\quad \Lambda=\frac{3}{8k}\,,  \quad a_0^2=\frac{1}{2\Lambda}\,.
\label{CaseI_summary2}
\end{align}

\subsection{\textbf{Case II}}

In Case II, ${2n-2>p-4}$,  the {lowest} order in (\ref{KeyEq1}) is  $\Delta^{p-4}$, implying
\begin{equation}
p(p-1)(p-2)(p-3)=0 \,.
\label{KeyEq1CaseII}
\end{equation}
Therefore, as in \cite{PodolskySvarcPravdaPravdova:2020}, there are four possible cases ${p=0}$, ${p=1}$, ${p=2}$, and ${p=3}$. Equation \eqref{KeyEq3} has the following  lowest orders (see \eqref{KeyEq3CaseI})
\begin{align}
\hbox{for}\quad p=0:\quad &
\big[6n(n-1)\big]c_0\,\Delta^{n-2}+\cdots=-2\,\Delta^{n}+\cdots+4\Lambda a_0^2 \Delta^{3n}+\cdots&&
\hbox{necessarily}\quad n=0, 1\,,\\
\hbox{for}\quad p=1:\quad &
\big[6n^2\big]c_0\,\Delta^{n-1}+\cdots=-2\,\Delta^{n}+\cdots+4\Lambda a_0^2 \Delta^{3n}+\cdots&&
\hbox{necessarily}\quad n=0\,,\\
\hbox{for}\quad p=2:\quad &
\big[6n(n+1)+2\big]c_0\,\Delta^{n}+\cdots=-2\,\Delta^{n}+\cdots+4\Lambda a_0^2 \Delta^{3n}+\cdots&&
(3n^2+3n+1)c_0=-1\,,\label{contrp=2c0}
\\
\hbox{for}\quad p=3:\quad &
\big[6n(n+2)+6\big]c_0\,\Delta^{n+1}+\cdots=-2\,\Delta^{n}+\cdots+4\Lambda a_0^2 \Delta^{3n}+\cdots&&
\hbox{not compatible}\,.
\end{align}
The lowest orders  in  \eqref{KeyEq2} for ${p=2}$, implying ${n>0}$, are
\be
3a_0^2\,[n(3n+2)c_0+1]\,\Delta^{2n} +2k(c_0^2 -1)-3\Lambda a_0^4\Delta^{4n}
+ \cdots =0 \,, \label{eq2rozvoj0omeg}
\ee
and thus ${c_0=\pm 1}$, however, the constraint \eqref{contrp=2c0}, ${3n^2+3n+1=\pm 1}$, is in a contradiction with ${n>0}$ and therefore, the $p=2$ case is not allowed.
\vspace{5mm}

\noindent
\textbf{To summarize}: The only three possible  classes of solutions in Case~II are given by
\be
[n,p]=[0,1]\,,\qquad
[n,p]=[0,0]\,,\qquad
[n,p]=[1,0]\,.
\label{CaseII_summary}
\ee

\subsection{\textbf{Case III}}

In Case III,  ${2n-2=p-4}$, i.e.,  ${p=2n+2}$. Then the {lowest} order $\Delta^{p-4}$  in (\ref{KeyEq1})
implies
\be
p(p-2)\big[3a_0^ 2+4kc_0(p-1)(p-3)\big]=0\,.
\label{KeyEq1CaseIII}
\ee
Therefore, as in \cite{PodolskySvarcPravdaPravdova:2020}, there are three subcases
${p=0}$,  ${p=2}$, and ${3a_0^2=-4kc_0 (p-1)(p-3)}$ with ${p\not= 0,1,2,3}$, corresponding to
${n=-1}$,  ${n=0}$, and ${3a_0^2=-4kc_0(4n^2-1)}$ with ${n\not= -1,-1/2,0,1/2}$, respectively.
The leading orders of \eqref{KeyEq3} read
\bea
(11n^2+6n+1) c_0\,\Delta^{3n}  +\cdots \rovno -\, \Delta^n+\cdots +2\Lambda a_0^2\Delta^{3n}\cdots \,, \label{eqtr00omegIII}
\eea
 which implies {$n\leq 0$} and
\begin{align}
&\hbox{for}\quad n=-1\Leftrightarrow p=0:\hspace{8mm}
6c_0\,\Delta^{-3}+\cdots=-\,\Delta^{-1}+\cdots+2\Lambda a_0^2\Delta^{-3}+\cdots \qquad \Rightarrow c_0=\tfrac{\Lambda}{3}a_0^2\,,\label{contrp=2c0IIIa}\\
&\hbox{for}\quad n=0\Leftrightarrow p=2:\hspace{12mm} c_0+\cdots=-1+\cdots+2\Lambda a_0^2\cdots\qquad \Rightarrow c_0=2\Lambda a_0^2-1\,,\label{contrp=2c0IIIb}\\
&\hbox{for}\quad 3a_0^2=4kc_0(1-4n^2): \hspace{2mm} (11n^2+6n+1)c_0+\cdots=2\Lambda a_0^2+\cdots \,. \label{contrp=2c0IIIc}
\end{align}

In the case $[0,2]$, Eq. (\ref{KeyEq2}) implies
\be
{3a_0^2(1-\Lambda a_0^2)+2k(c_0^2-1)=0}\,,
\ee
that  gives either
\be
\Lambda a_0^2=1\ \Rightarrow\ c_0=1\,,
\ee
or
\be
\Lambda=\frac{3}{8k}\ \Rightarrow\ c_0=\frac{3}{4k}a_0^2-1\,.
\ee

In the case $3a_0^2=4kc_0(1-4n^2)$, with ${p\not= 0,1,2,3}$,
and thus ${n\not=-1,-1/2,0,1/2}$, respectively, Eq. \eqref{contrp=2c0IIIc}
with ${p\not=0,1}$ (${n\not=-1,-1/2}$, respectively, ${p,n<0}$), using ${a_0^2=\frac{4}{3}kc_0(1-4n^2)}$, gives
\be
\Lambda =\frac{3}{8k}\frac{11n^2+6n+1}{1-4n^2}\quad\Rightarrow\quad c_0=\frac{3}{4k}\frac{a_0^2}{1-4n^2}\,.\label{Lc0-n,2n+2}
\ee
Equation
(\ref{KeyEq2}) is then identically satisfied.

\vspace{5mm}

\noindent
\textbf{To summarize}: In Case III, there are three possible classes of solutions
\bea
[n,p] \rovno [-1,0]\hspace{5mm} \hbox{with} \quad
c_0=\frac{\Lambda}{3}a_0^2\,,\label{CaseIII_summary1}\\
\ [n,p] \rovno [0,2]\qquad \hbox{with}\quad
\hbox{either}\quad \Lambda a_0^2=1\,, \  c_0=1\,,
\quad \hbox{or}\quad
\Lambda=\frac{3}{8k}\,,\  \ c_0=\frac{3}{4k}a_0^2-1\,,
\label{CaseIII_summary2}\\
\ [n,p] \rovno [n<0,2n+2<0]\qquad \hbox{with}\quad
\Lambda =\frac{3}{8k}\frac{11n^2+6n+1}{1-4n^2}\,,\   c_0=\frac{3}{4k}\frac{a_0^2}{1-4n^2}\,.
\label{CaseIII_summary3}
\eea

\newpage

\section{Description and study of all possible solutions in powers of~$\Delta$}
\label{description}

In this section, we will investigate all the solutions contained in Cases~I, II, and~III, namely eight classes
\eqref{CaseI_summary1}, \eqref{CaseI_summary2} \eqref{CaseII_summary},  \eqref{CaseIII_summary1}, \eqref{CaseIII_summary2}, and \eqref{CaseIII_summary3}.


\subsection{Uniqueness of the Schwarzschild black hole in the class ${[n,p]=[-1,2]}$}
\label{Schw_[n,p]=[-1,2]}

The class ${[n,p]=[-1,2]}$ in Case~I, see \eqref{CaseI_summary1}, necessarily has ${\Lambda=0}$ and therefore it has been already studied in detail in \cite{PodolskySvarcPravdaPravdova:2020}. Therein, we have shown that the only solution in this class is the Schwarzschild solution given by
\begin{equation}
\Omega(r)=-\frac{1}{r}\,,\qquad
\H(r) = -r^2-2m\, r^3 \,,
\label{Schw}
\end{equation}
see \eqref{SchwAdS}.


\subsection{Schwarzschild--Bach--(A)dS black hole in the class ${[n,p]=[0,1]}$: near the horizon}
\label{SchwaBach_[n,p]=[0,1]}

In this section, we will present a detailed derivation of the metric of the class ${[0,1]}$  that represents a spherically symmetric non-Schwarzschild solution to QG, the Schwarzschild--Bach--(A)dS black hole
 with a nonvanishing Bach tensor and a cosmological constant, as we pointed out already in  \cite{SvarcPodolskyPravdaPravdova:2018}.   The first three terms in the expansion of  such a  solution read
\begin{align}
\Omega(r) & = -\frac{1}{r}+b\frac{(r-r_h)}{\X\,r_h^2}-\frac{b}{r_h}\left[ 2-\left(\frac{7}{3}\Lambda-\frac{1}{8k}\right)\frac{1}{r_h^2}+b\right] \Big(\,\frac{r-r_h}{\X\,r_h}\Big)^2+\cdots
\,, \label{IIbOmegaFULL}\\
\H (r) & = (r-r_h)\bigg\{\,\frac{r^2}{r_h}-\frac{\Lambda}{3\,r_h^3}\left(r^2+rr_h+r_h^2\right)
\nonumber \\
&\hspace{18.0mm}
+3b\,\X\,r_h\left[\Big(\,\frac{r-r_h}{\X\,r_h}\Big)+\frac{1}{3}\Big[4-\frac{1}{r_h^2}\Big(2\Lambda+\frac{1}{2k}\Big)+3b\Big] \Big(\,\frac{r-r_h}{\X\,r_h}\Big)^2+\cdots\right]
\,\bigg\}\,, \label{IIbH0FULL}
\end{align}
where
\begin{align}
\X &\equiv  1-\frac{\Lambda}{r_h^2}\,, \label{rho_[0,1]}
\end{align}
and
\be
r_0\equiv r_h \,\label{r0=rh}
\ee
 is the \emph{black hole horizon} since ${\H(r_h)=0}$.
For a vanishing \emph{``Bach parameter''}, ${b=0}$, the metric functions \eqref{IIbOmegaFULL}, \eqref{IIbH0FULL} immediately reduce to  the Schwarzschild--(A)dS metric functions \eqref{SchwAdS}
with a vanishing Bach tensor.

Let us derive the complete analytical form of this black hole  leading to \eqref{IIbOmegaFULL} and
 \eqref{IIbH0FULL}.
 For  ${[n,p]=[0,1]}$,  Eq.~\eqref{KeyEq1} reduces to
\be
\sum^{l+1}_{i=0}a_i\, a_{l+2-i}\,(l+2-i)(l+1-3i) =\tfrac{1}{3}k
\,c_{l+3}\,(l+4)(l+3)(l+2)(l+1) \,, \label{KeyEq1[n,p]=[01]}
\ee
where ${l\ge 0}$. Relabeling  ${l \to l-1 }$, Eq. \eqref{KeyEq1[n,p]=[01]} gives
\be
c_{l+2}=\frac{3}{k\,(l+3)(l+2)(l+1)l}\,\sum^{l}_{i=0}a_i \,
a_{l+1-i}(l+1-i)(l-3i)  \qquad \forall\ l\ge 1\,,
\label{nonSchwinitcondc}
\ee
and thus   all coefficients $c_{l+2}$, starting from $c_3$, can be expressed in terms of ${a_0,\ldots, a_{l+1}}$.

The lowest nontrivial order ${l=0}$ of
the ``trace equation'' \eqref{KeyEq3}  gives
\begin{align}
a_1=\frac{a_0}{3c_0}\left[2\Lambda a_0^2-(1+c_1)\right] ,
\label{nonSchwinitcond3}
\end{align}
and higher orders ${l=1, 2, \ldots}$ read
\begin{align}
(l+1)^2c_0a_{l+1}=\tfrac{2}{3}\Lambda \sum^{l}_{j=0}\sum^{j}_{i=0}a_ia_{j-i}a_{l-j}-\tfrac{1}{3}\,a_{l}
-\sum^{l+1}_{i=1}c_ia_{l-i+1}\left[(l-i+1)(l+1)+\tfrac{1}{6}i(i+1)\right]\, .
\end{align}
This  leads  (after relabeling  ${l \to l-1 }$) to
\be
a_{l}=\frac{1}{l^2c_0}\Big[\,\tfrac{2}{3}\Lambda \sum^{l-1}_{j=0}\sum^{j}_{i=0}a_i\,a_{j-i}\,a_{l-1-j}-\tfrac{1}{3}\,a_{l-1}
-\sum^{l}_{i=1}c_i\,a_{l-i}\left(l(l-i)+\tfrac{1}{6}i(i+1)\right)\Big] \ \ \
\forall \ l\geq 2\,,  \label{nonSchwinitconda}
\ee
which gives all the coefficients $a_{l}$ in terms of  ${a_0,\ldots,
	a_{l-1}}$ and ${c_0,\ldots, c_{l}}$.

In addition,  the lowest
nontrivial order ${l=0}$ of  \eqref{KeyEq2} gives the constraint
\be
 6kc_0c_2=3a_0[a_1c_0+a_0(1-\Lambda a_0^2)]+2k(c_1^2-1)\,,\label{6kc0}
\ee
which, using \eqref{nonSchwinitcond3},
becomes
\be
c_2 =\frac{1}{6kc_0}\left[2k(c_1^2-1)+a_0^2(2-c_1-\Lambda a_0^2)\right]\,.
\label{nonSchwinitcond2}
\ee

Therefore, this class of solutions has three free parameters
$a_0$, $c_0$, and $c_1$. Then ${a_1}$, ${c_2}$ are determined by   \eqref{nonSchwinitcond3},
\eqref{nonSchwinitcond2}, respectively, and all other coefficients $a_{l+1}$, $c_{l+2}$  for all ${l = 1, 2, \ldots }$ can be obtained using the recurrent relations \eqref{nonSchwinitconda}, \eqref{nonSchwinitcondc}, respectively.

The Bach and Weyl invariants (\ref{invB}), (\ref{invC})  at ${r=r_h\equiv r_0}$ read
\bea
B_{ab}\,B^{ab}(r_h) &=& \left(  \frac{1-c_1^2+3c_0c_2}{3 a_0^4} \right)^2
=\left(\frac{c_1-2+\Lambda a_0^2}{6k a_0^2}\right)^2
 = \frac{b^2}{4 k^2 a_0^4 } \,,\ \label{BachInvariant}\\
C_{abcd}\, C^{abcd}(r_h)& =& \frac{4}{3 a_0^4}(1 + c_1)^2 \,,\label{BInv2}
\eea
where we have introduced a key Bach parameter $b$ by
\be
b \equiv \frac{1}{3}(c_1-2+\Lambda a_0^2)\,. \label{b_definice}
\ee
The Bach tensor $B_{ab}$ is in general nonvanishing.
Using the recurrent relations \eqref{nonSchwinitconda} and \eqref{nonSchwinitcondc}, the first few coefficients read
\begin{align}
& a_{1} = -\frac{a_{0}}{c_{0}}\Big((1-\Lambda a_{0}^2)+b\Big) \,, \label{IIb_expansionaLambda} \\
& a_{2} = +\frac{a_{0}}{c_{0}^2}\Big((1-\Lambda a_{0}^2)^2+\Big[2-\Big(\frac{7}{3}\Lambda
-\frac{1}{8k}\Big)a_{0}^2\Big]b+b^2\Big) \,, \nonumber\\
& a_{3} = -\frac{a_{0}}{c_{0}^3}\Big((1-\Lambda a_{0}^2)^3+\frac{1}{9}\Big[25-\Big(\frac{179}{3}\Lambda-\frac{29}{8k}\Big)a_{0}^2
+\Big(\frac{298}{9}\Lambda^2-\frac{77}{24k}\Lambda+\frac{1}{16k^2}\Big)a_{0}^4\Big]b  \nonumber \\
& \hspace{37.0mm}  +\frac{1}{9}\Big[23-\Big(\frac{104}{3}\Lambda-\frac{35}{8k}\Big)a_{0}^2\Big]b^2
+\frac{7}{9}b^3 \Big) \,, \ldots\,,\nonumber\\
& c_1 = 2 - \Lambda a_0^2 + 3b \,, \label{IIb_expansioncLambda}\\
& c_2 = \frac{1}{3c_0}\Big((1-\Lambda a_0^2)(3-\Lambda a_0^2)+3\Big[4-\Big(2\Lambda+\frac{1}{2k}\Big)a_0^2\Big] b+9b^2\Big) \,, \nonumber\\
& c_3 = \frac{a_0^4}{2kc_0^2}\,b\Big(-\frac{1}{6}\Lambda+\frac{1}{16k}\Big) \,, \nonumber\\
& c_4 = \frac{a_0^2}{270kc_0^3}\,b\Big(9+\Big(12\Lambda-\frac{45}{4k}\Big)a_0^2
-\Big(14\Lambda^2-\frac{75}{8k}\Lambda+\frac{9}{32k^2}\Big)a_0^4
+3\Big[6+\Big(7\Lambda-\frac{39}{8k}\Big)a_0^2\Big]b+9b^2\Big) \,, \ldots\,,\nonumber
\end{align}
where $a_0$, $c_0$, and $b$ are free parameters.

\subsubsection{Identification of the Schwarzschild--(A)dS black hole}

To {identify the Schwarzschild--(A)dS black hole}, we must set ${B_{ab}=0}$, i.e., we choose ${b=0}$. Then the expansion coefficients
\eqref{IIb_expansionaLambda}, \eqref{IIb_expansioncLambda}
simplify to
\bea
&& a_i=a_0\,\bigg(\!-\frac{1-\Lambda a_0^2}{c_0}\bigg)^i
\quad \hbox{for all}\ i \ge 0 \,,\\
&&  c_1=2-\Lambda a_0^2\,, \quad c_2=\frac{1}{3c_0}(1-\Lambda a_0^2)(3-\Lambda a_0^2)\,, \quad c_i=0\quad \hbox{for all}\ i \ge 3\,,
\eea
where the first sequence is a {geometrical series} and  the second series is {truncated to
	a polynomial of the 3rd order}. Thus the metric functions  take the explicit closed form
\begin{eqnarray}
\Omega(r)      \rovno  a_0 \,\sum_{i=0}^\infty  \,\Big(-(1-\Lambda a_0^2)\frac{\Delta}{c_0}\Big)^i
=\frac{a_0\,c_0}{c_0+(1-\Lambda a_0^2)\Delta}=\frac{a_0\,c_0}{(1-\Lambda a_0^2)(r-r_h)+c_0}\,, \label{IIbOmega}\\
\H (r) \rovno  c_0(r-r_h)+c_1(r-r_h)^2+c_2(r-r_h)^3 \,. \label{IIbH0}
\end{eqnarray}
Using the gauge freedom \eqref{scalingfreedom},
we can always set
\begin{equation}
a_0 =-\frac{1}{r_h} \,, \qquad c_0 =r_h-\frac{\Lambda}{r_h} \,, \label{IIb_a0}
\end{equation}
so that the metric functions reduce to
\be
{\bar r}=\Omega(r) = -\frac{1}{r}\,, \qquad
\H (r) = \frac{\Lambda}{3} -r^2-\Big(\frac{\Lambda}{3} -r_h^2 \Big)\frac{r^3}{r_h^3} \,. \label{IIbH0Schw}
\ee
Thus for ${b=0}$ we recover the Schwarzschild--(A)dS solution \eqref{SchwAdS}, with the black hole horizon
located at ${r=r_h}$, so that ${2m=(\frac{\Lambda}{3} - r_h^2)/r_h^3}$.

\subsubsection{More general Schwarzschild--Bach--(A)dS black hole }

For ${b\not=0}$,
the solution  \eqref{nonSchwinitconda}, \eqref{nonSchwinitcondc},
that is \eqref{IIb_expansionaLambda}, \eqref{IIb_expansioncLambda}, represents
 a {generalization of the Schwarzschild--(A)dS black hole with a nontrivial Bach tensor}  with the Bach invariant
${B_{ab}\,B^{ab}}$  proportional to $b^2$ at the horizon \eqref{BachInvariant}.
It reduces to the Schwarzschild--(A)dS solution \eqref{IIbH0Schw} for ${b \to 0}$. After summing the ``background'' terms independent of $b$ as in \eqref{IIbOmega}, and using the same gauge  \eqref{IIb_a0}, one obtains the explicit form of the solution \eqref{IIbOmegaFULL}, \eqref{IIbH0FULL} with ${r=r_h}$ still being  the horizon.

As in \cite{PodolskySvarcPravdaPravdova:2020}, we rewrite this class of solutions in an alternative and more explicit form. Let us  introduce coefficients $\alpha_i, \gamma_i$ as those parts of $a_i, c_i$, respectively, that \emph{do not} involve the ``${b=0}$'' Schwarzschild--(A)dS background, i.e.,
\bea
 a_i \eqdef a_i(b=0)-\frac{b}{r_h}\,\frac{\alpha_i}{(-r_h\rho)^i}\,,\qquad\hbox{where}\quad
a_i(b=0)\equiv\frac{1}{(-r_h)^{1+i}}  \ \ \ i\geq 0 \,,\label{def_alphai}\\
 c_1 \eqdef 2-\frac{\Lambda}{r_h^2} + 3b\,\gamma_1 ,\quad
c_2 \equiv \frac{r_h}{3(r_h^2-\Lambda)}\left(1-\frac{\Lambda}{r_h^2}\right)\left(3-\frac{\Lambda}{r_h^2}\right)+3b\,\frac{r_h\gamma_2}{r_h^2-\Lambda} ,\
\nonumber\\
  c_i \eqdef  3b\,\frac{\gamma_i \,r_h^{i-1}}{(r_h^2-\Lambda)^{i-1}}
\ \ \ i \ge 3.
\label{def_gammai}
\eea
Then ${\Omega}$ and ${\H}$ take the explicit form
\begin{align}
\Omega(r) & = -\frac{1}{r}-\frac{b}{r_h}\sum_{i=1}^\infty\alpha_i\Big(\,\frac{r_h-r}{\X\,r_h}\Big)^i \,, \label{Omega_[0,1]}\\
\H (r) & = (r-r_h)\bigg[\,\frac{r^2}{r_h}-\frac{\Lambda}{3r_h^3}\left(r^2+rr_h+r_h^2\right)
+3b\,\X\,r_h\sum_{i=1}^\infty\gamma_i\Big(\,\frac{r-r_h}{\X\,r_h}\Big)^i\,\bigg] \,, \label{H_[0,1]}
\end{align}
where ${\X}$ is given by \eqref{rho_[0,1]} and
\begin{align}
\alpha_1 &\equiv 1\,,\quad \gamma_1=1\,,\quad
\gamma_2 = \frac{1}{3}\Big[4-\frac{1}{r_h^2}\Big(2\Lambda+\frac{1}{2k}\Big)+3b\Big] \,. \label{alphasgammainitial_[0,1]}
\end{align}
Using \eqref{nonSchwinitconda} and \eqref{nonSchwinitcondc}, the remaining coefficients $\alpha_l, \gamma_{l+1}$ for ${l \ge 2}$ are given by the recurrent relations
(defining ${\alpha_0=0}$)
\begin{align}
&\alpha_{l}= \, \frac{1}{l^2}\Big[-\frac{2\Lambda}{3r_h^2}\,\sum_{j=0}^{l-1}\sum_{i=0}^{j}\big[\alpha_{l-1-j}\X^j+\big(\X^{l-1-j}
+b\,\alpha_{l-1-j}\big)\big(\alpha_i \X^{j-i}+\alpha_{j-i}(\X^i+b\,\alpha_{i})\big)\big]-\frac{1}{3}\alpha_{l-2}(2+\X)\X(l-1)^2 \nonumber\\
& \hspace{14.0mm} +\alpha_{l-1}\big[\frac{1}{3}+(1+\X)\big(l(l-1)+\frac{1}{3}\big)\big]
-3\sum_{i=1}^{l}(-1)^i\,\gamma_i\,(\X^{l-i}+b\,\alpha_{l-i})\big(l(l-i)+\frac{1}{6}i(i+1)\big)\Big]\,, \nonumber\\
&\gamma_{l+1}= \, \frac{(-1)^{l}}{kr_h^2\,(l+2)(l+1)l(l-1)}\sum_{i=0}^{l-1}\big[\alpha_i \X^{l-i}
+\alpha_{l-i}\big(\X^i+b\,\alpha_i\big) \big](l-i)(l-1-3i) \, \quad \hbox{for}\quad l\geq2\,.
\label{alphasIIbgeneral}
\end{align}
Then the first few terms read
\begin{align}
& \alpha_2 =
2-\left(\frac{7}{3}\Lambda-\frac{1}{8k}\right)\frac{1}{r_h^2}+b
\,, \nonumber\\
& \alpha_3 =
\frac{1}{9}\left[25+\left(\frac{29}{8k}-\frac{179}{3}\Lambda\right)\frac{1}{r_h^2}+\left(\frac{1}{16k^2}-\frac{77}{24k}\Lambda+\frac{298}{9}\Lambda^2\right)\frac{1}{r_h^4}\right]\nonumber\\
&\qquad+\frac{1}{9}\left[23+\left(\frac{35}{8k}-\frac{104}{3}\Lambda\right)\frac{1}{r_h^2}\right]\,b+\frac{7}{9}\,b^2
\,, \quad \ldots\,,
\label{alphasIIb0}\\
& \gamma_3 =
 \frac{1}{96k^2r_h^4}\left(1-\frac{8k}{3}\Lambda\right)
 \,, \nonumber\\
& \gamma_4 =
\frac{1}{18kr_h^2}\left[\frac{1}{5}+\left(-\frac{1}{4k}+\frac{4}{15}\Lambda\right)\frac{1}{r_h^2}-\frac{1}{160k^2r_h^4}-\frac{1}{45r_h^4}\left(14\Lambda^2-\frac{75}{8k}\Lambda\right)
\right]\nonumber\\
&\qquad+\frac{1}{720kr_h^2}\left[16+\left(-\frac{13}{k}+\frac{56}{3}\Lambda\right)\frac{1}{r_h^2}\right]\,b
 +\frac{1}{90kr_h^2}\,b^2
 \,, \quad \ldots\,,
\label{gammasIIb0}
\end{align}
leading to \eqref{IIbOmegaFULL}, \eqref{IIbH0FULL}.

This  spherically symmetric Schwarzschild--Bach--(A)dS black-hole spacetime \eqref{Omega_[0,1]}, \eqref{H_[0,1]} in  QG admits  \emph{three independent parameters}: the cosmological constant $\Lambda$,
 the horizon position $r_h$ ($r_h$ is a root of $\H(r)$ given by \eqref{H_[0,1]}), and
	the dimensionless {Bach parameter} $b$,  chosen  in such a way that it
{determines the value of the Bach tensor \eqref{B1}, \eqref{B2} on the
	horizon} $r_h$, namely
\be
\B_1(r_h) = 0\,, \qquad
\B_2(r_h) = -\frac{3}{kr_h^2}\,b \,. \label{bonhorizon}
\ee
Then  the invariants \eqref{invB} and \eqref{invC} reduce to
\be
B_{ab}\,B^{ab}(r_h) = \frac{r_h^4}{4 k^2}\,b^2 \,,\qquad
C_{abcd}\, C^{abcd}(r_h) = 12\,\left[r_h^2\,(1+b)-\frac{\Lambda}{3}\right]^2  \,\label{BCInvariants_[0,1]}
\ee
on the horizon.

As ${\Delta\equiv r-r_h\to 0}$, the dominant terms of the two metric functions are
${\Omega=a_0=-\frac{1}{r_h}}$ and\newline ${\H=c_0\,\Delta=(r_h-\frac{\Lambda}{r_h})(r-r_h)}$ (cf. \eqref{SchwAdS}), in which case  the relations
\eqref{to static}, \eqref{rcehf}
give
\bea
&& {\bar r}={a_0}+\cdots\ \rightarrow\ \mbox{const.}
\,,\\
&& h \ \to\ \bar\Delta(\equiv (\bar r-a_0))
\ \rightarrow\ 0\,,\quad
f \ \to\ \bar\Delta(\equiv (\bar r-a_0))
\ \rightarrow\ 0\,,
\eea
confirming the existence of the horizon at $r=r_h$, i.e., ${\bar{r}_h=a_0=-\frac{1}{r_h}}$.

Finally let us check that the ${k=0}$ limit (corresponding to the Einstein limit of QG) of this solution  reduces to  the Schwarzschild--(A)dS solution. The condition ${k=0}$ implies ${a_1c_0+a_0(1-\Lambda a_0^2)=0}$, see the constraint  \eqref{6kc0}. This, substituted into \eqref{nonSchwinitcond3}, yields  ${c_1=2-\Lambda a_0^2}$, and thus ${b=0}$ due to our definition \eqref{b_definice}.

\vspace{5mm}

\noindent
\textbf{To conclude}: The class ${[n,p]=[0,1]}$,
expressed in terms of the series \eqref{Omega_[0,1]}, \eqref{H_[0,1]} {around the horizon}~$r_h$, represents the spherically symmetric Schwarzschild--Bach--(A)dS  black hole 
 generalizing  the Schwarzschild--(A)dS black hole.

\newpage

\subsection{Analysis of physical properties of the Schwarzschild--Bach--(A)dS black hole}
\label{analysisofScwaBacgAdS}

In this section, we study various aspects of the class of black holes found in Sec.~\ref{SchwaBach_[n,p]=[0,1]}.

\subsubsection{{Behaviour of the metric functions}  }
\label{discussion-and-figures}

This part is devoted to the analysis of the Schwarzschild--Bach--(A)dS metric
functions and their comparison with the classic Schwarzschild--(A)dS solution.

\begin{figure}[h!]
	\includegraphics[scale=0.44]{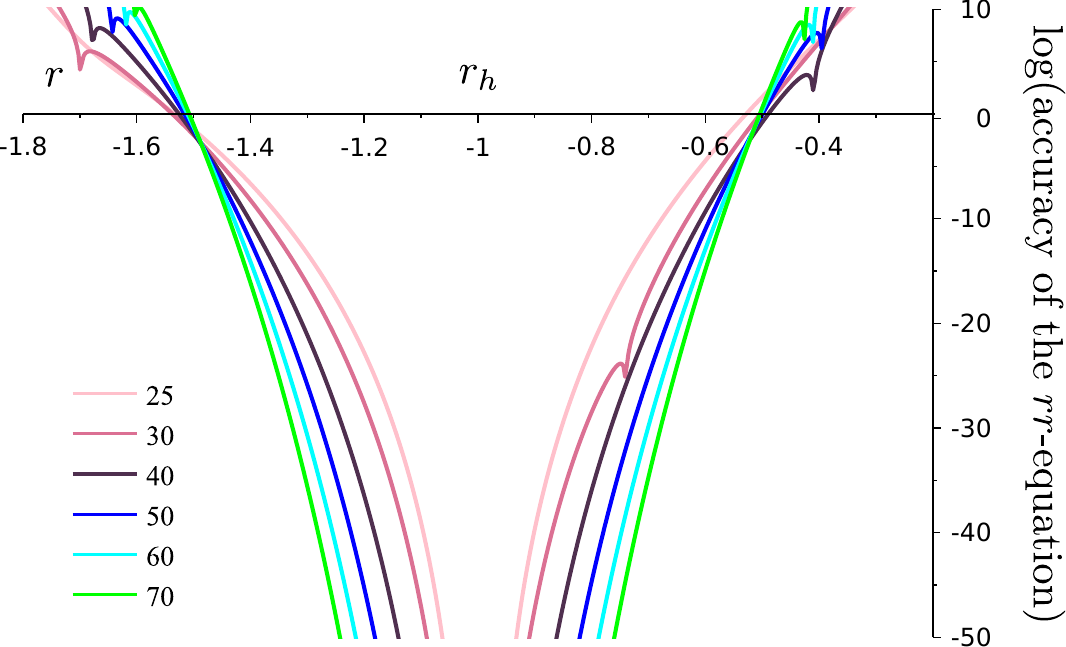}
    \hspace{10.0mm}
	\includegraphics[scale=0.44]{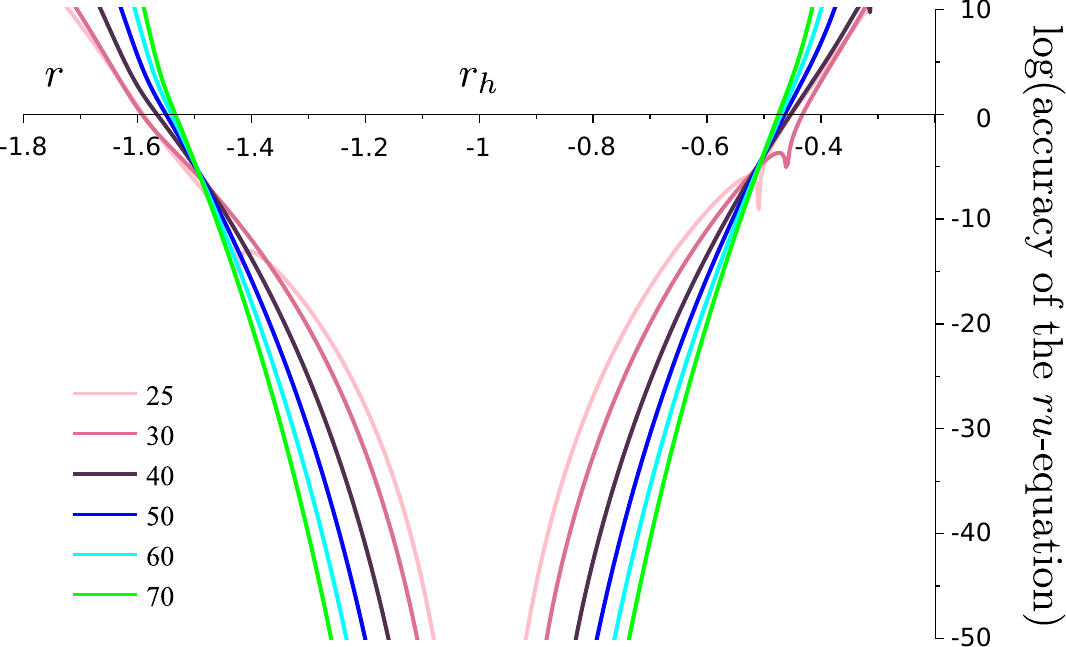}
	\caption{\label{fig:test}
		The difference between the left and right hand side of the field equation \eqref{Eq1} (left) and \eqref{Eq2} (right) for the metric functions ${\Omega}$ and ${\H}$ given by \eqref{Omega_[0,1]}, \eqref{H_[0,1]}. The plotted values approach zero in a certain range of the radial coordinate $r$ around the horizon ${r_h}$  with the growing order~$n$. This indicates that the equations are satisfied with a reasonable precision, depending on the finite order of the polynomial approximation in some convergence radius. Here we set ${r_h=-1,\, k=0.5}$, ${b=0.3}$, and     ${\Lambda=0.2}$.}
\end{figure}

First, let us verify that the metric functions ${\Omega(r)}$ and ${\H(r)}$, represented by the power series \eqref{Omega_[0,1]} and \eqref{H_[0,1]}, really solve the field equations \eqref{Eq1} and \eqref{Eq2} in some neighbourhood around the horizon ${r_h}$, with a reasonable precision depending on the assumed finite order {$n$}. To do so, the difference between the left and right hand sides of the field equations \eqref{Eq1} and \eqref{Eq2} is plotted in Fig.~\ref{fig:test}. It clearly approaches zero with growing order {$n$} of the polynomial approximation.

As a next natural step, we examine the \emph{convergence radius} of the power series \eqref{Omega_[0,1]} and \eqref{H_[0,1]}. These radii can be roughly deduced already from Fig.~\ref{fig:test}. However, to be more precise we employ the standard d'Alembert ratio test for two different sets of parameters. The ratio between two subsequent coefficients $\alpha_n/\alpha_{n-1}$ and ${-\gamma_n/\gamma_{n-1}}$
is visualized in Fig.~\ref{fig:1} (this figure and some of the following figures are taken from out letter \cite{SvarcPodolskyPravdaPravdova:2018}). The plots clearly indicate that such a ratio approaches a specific constant. We may thus conclude that the solution \emph{asymptotically approaches a geometric series}, for which the radius of convergence can be simply estimated.

\begin{figure}[h!]
	\begin{center}
		\includegraphics[scale=0.44]{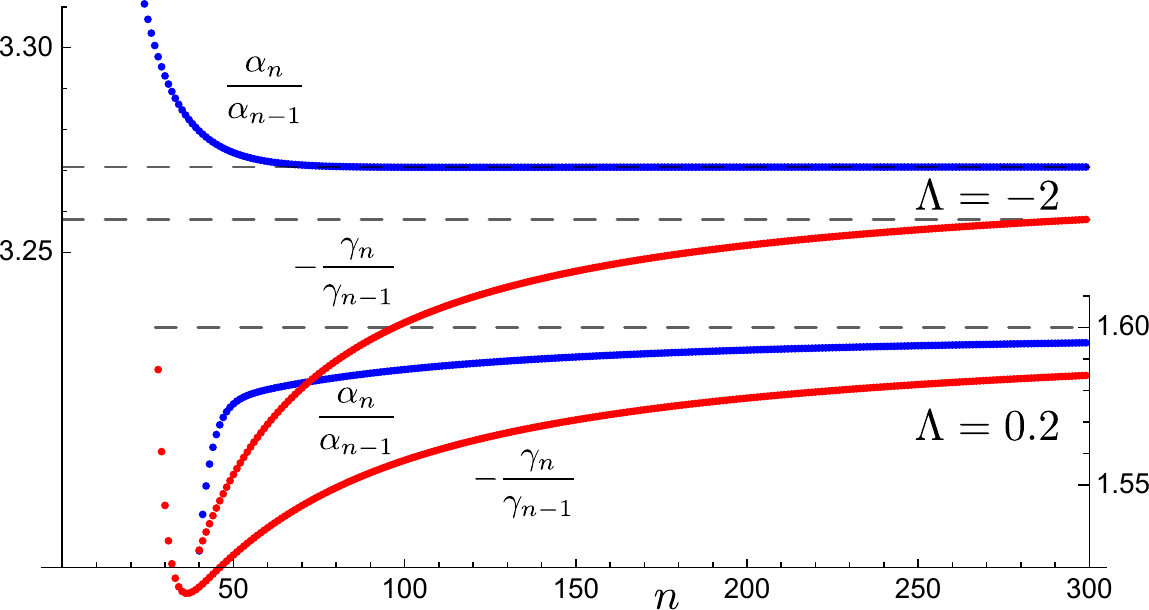}
	\end{center}
	\caption{\label{fig:1}
		The convergence radius for the power series \eqref{Omega_[0,1]} and \eqref{H_[0,1]} representing the metric functions ${\Omega}$ and ${\H}$ can be estimated using the d'Alembert ratio test demonstrating the asymptotic behaviour similar to the geometric series. Here we assume ${r_h=-1,\, k=0.5}$ with ${b=0.3,\, \Lambda=0.2}$ (bottom) and ${b=0.2,\, \Lambda=-2}$ (top).}
\end{figure}

\begin{figure}[h!]
	\includegraphics[scale=0.44]{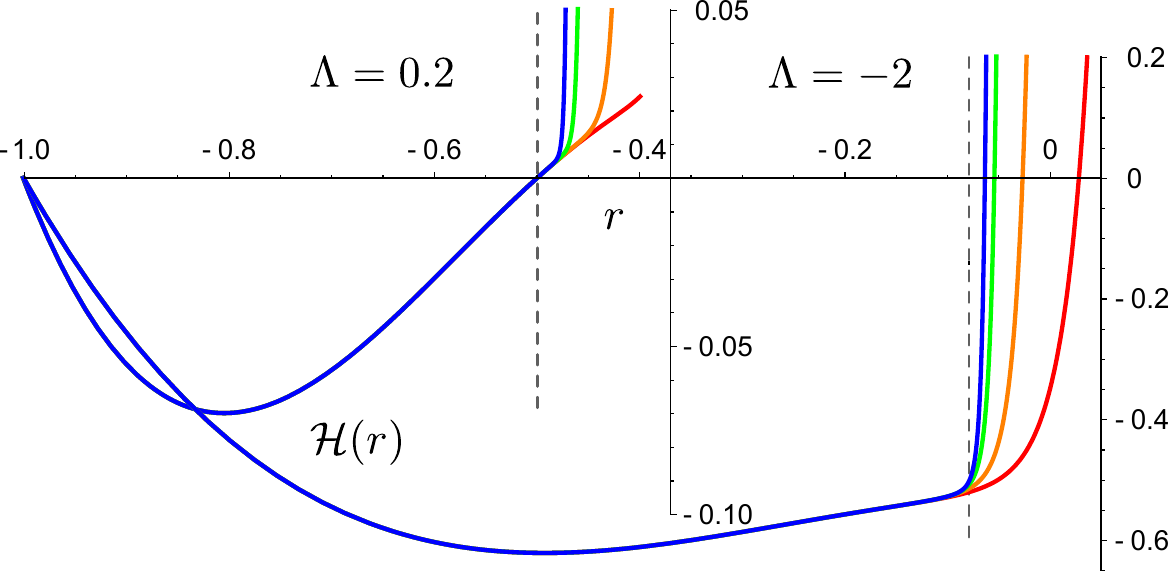}
    \hspace{10.0mm}
	\includegraphics[scale=0.44]{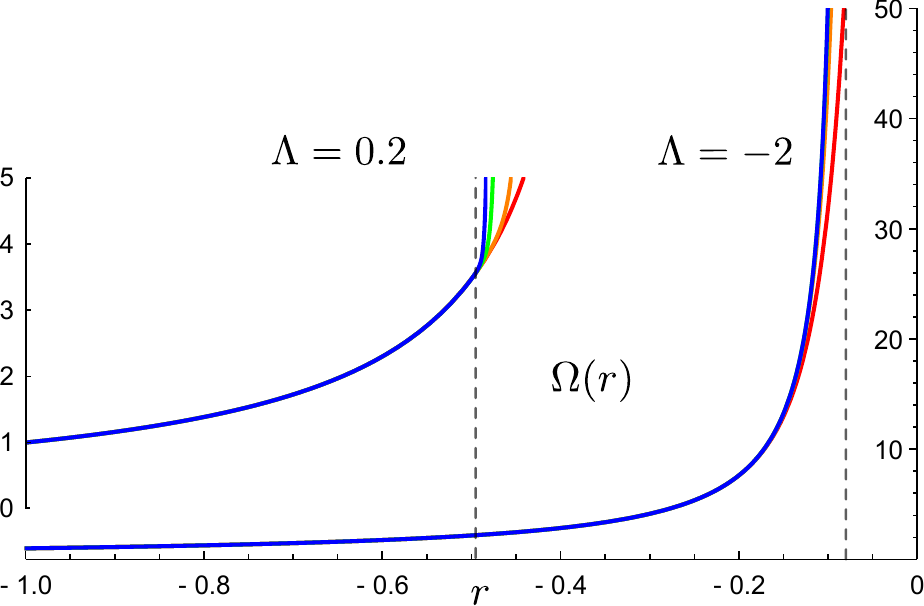}
	\caption{\label{fig_H_Omega}
		The functions ${\cal H}(r)$ given by \eqref{H_[0,1]} (left) and $\Omega(r)$ given by \eqref{Omega_[0,1]} (right) for two values of the cosmological constant ${\Lambda}$ (with the same parameters as in Fig.~\ref{fig:1}).
The black-hole horizon ${r_h=-1}$ is located at the centre of the radius of convergence. Here we plot the functions ${\cal H}(r)$ in the regions outside the  black-hole horizon up to the outer radii of convergence indicated by  dashed vertical lines. For ${\Lambda>0}$, the function ${\cal H}(r)$ seems to have another root corresponding to the cosmological horizon, while for ${\Lambda<0}$, it remains nonvanishing. First 50 (red), 100 (orange), 200 (green), 300 (blue) terms in the expansions are used. The results fully agree with the numerical solutions up to the dashed lines, where such simulations also fail. The function $\Omega(r)$ grows monotonously.}
\end{figure}

Now, let us plot the typical behaviour of the metric functions ${\Omega(r)}$ and ${\cal H}(r)$ near the black-hole horizon ${r_h}$, see Fig.~\ref{fig_H_Omega}. The qualitative behaviour of the function ${\cal H}$ depends on  the sign of the cosmological constant $\Lambda$.
 For any value of $\Lambda$, the black-hole horizon separates static (${r>r_h}$) and non-static (${r<r_h}$) regions of the spacetime. However, for positive cosmological constant ${\Lambda>0}$, an additional outer boundary of the static region appears, corresponding to the \emph{cosmological horizon} given by the second root of the function $\H$, similarly as for the Schwarzschild--de Sitter black hole.  For ${\Lambda<0}$, the outer static region seems to be unbounded, as in the Einstein theory.

\begin{figure}[h!]
	\begin{center}
		\includegraphics[scale=0.44]{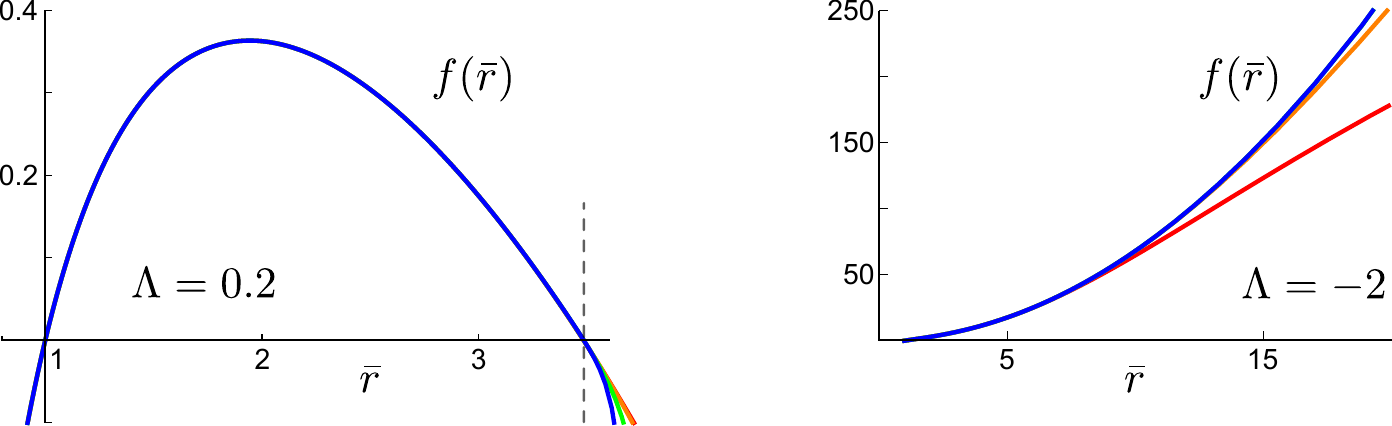}
	\end{center}
	\caption{\label{fig:3}
		The function ${f(\bar{r})}$ of the standard metric form \eqref{Einstein-WeylBH} related to the solution \eqref{Omega_[0,1]}, \eqref{H_[0,1]} via the transformation \eqref{rcehf}. The \emph{positive} ${\Lambda}$ case (left) indicates the presence of the cosmological horizon at the boundary of the convergence interval (the dashed line). For \emph{negative} ${\Lambda}$ (right), the series converges in the whole plotted range, corresponding to a static region everywhere above the black-hole horizon. Here we use the same values of parameters as in Figs.~\ref{fig:1}, \ref{fig_H_Omega}.}
\end{figure}

To give a complementary picture of the Schwarzschild--Bach--(A)dS metric behaviour, which can be more intuitively compared with the classic Schwarzschild--(A)dS case,  we employ the standard spherically symmetric line element \eqref{Einstein-WeylBH} and plot the metric function ${f(\bar{r})}$ in Fig.~\ref{fig:3}. Obviously, there is not any significant qualitative difference with respect to the Einstein theory. The second root for the positive value of $\Lambda$ again indicates the presence of the cosmological horizon, separating static and non-static regions.

\begin{figure}[h!]
	\begin{center}
		\includegraphics[scale=0.44]{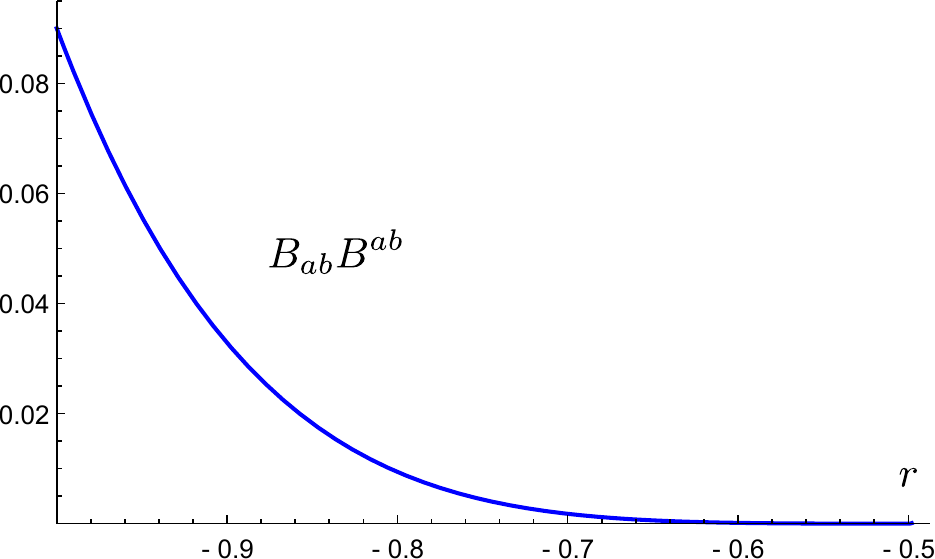}
				\hspace{15.0mm}
		\includegraphics[scale=0.44]{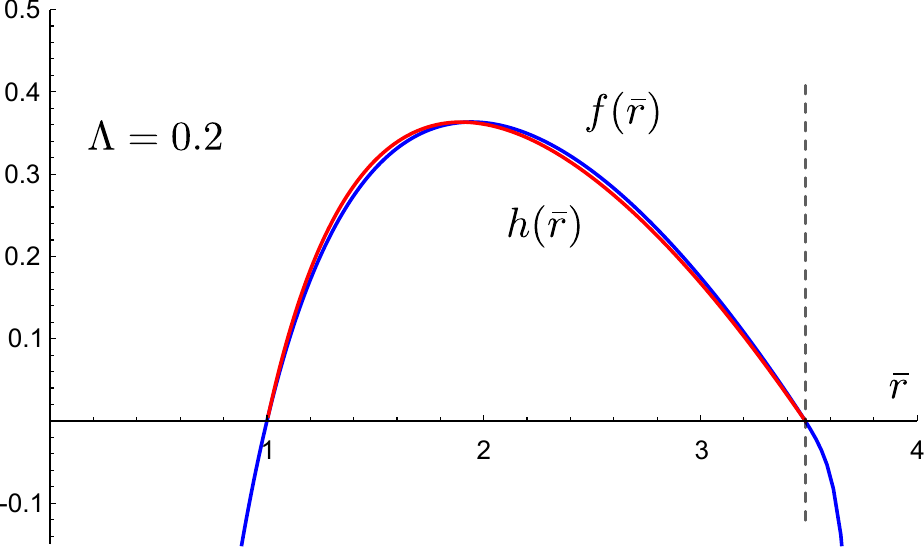}
	\end{center}
	\caption{\label{fig:41}
		The Bach invariant \eqref{invB} as a function of the Kundt coordinate $r$ (left), and the metric functions ${f(\bar{r})}$ and ${h(\bar{r})}$ of standard line element \eqref{Einstein-WeylBH} corresponding to the solution \eqref{Omega_[0,1]}, \eqref{H_[0,1]} via \eqref{rcehf} (right), with the same parameters as in Figs.~\ref{fig:1}, \ref{fig_H_Omega} for ${\Lambda=0.2}$.}
\end{figure}

To clarify the difference between the Schwarzschild--(A)dS and Schwarzschild--Bach--(A)dS solutions, in Fig.~\ref{fig:41} we plot the \emph{Bach invariant} \eqref{invB} as a function of the Kundt coordinate $r$. Recall that this invariant identically vanishes for the Schwarzschild--(A)dS black hole. In addition, we also plot the metric functions ${f(\bar{r})}$ and ${h(\bar{r})}$ of standard spherically symmetric line element \eqref{Einstein-WeylBH} corresponding to the solution \eqref{Omega_[0,1]}, \eqref{H_[0,1]} via \eqref{rcehf}. In contrast with the Schwarzschild--(A)dS black hole, these functions are \emph{not} identical.

\begin{figure}[h!]
	\begin{center}
		\includegraphics[scale=0.5]{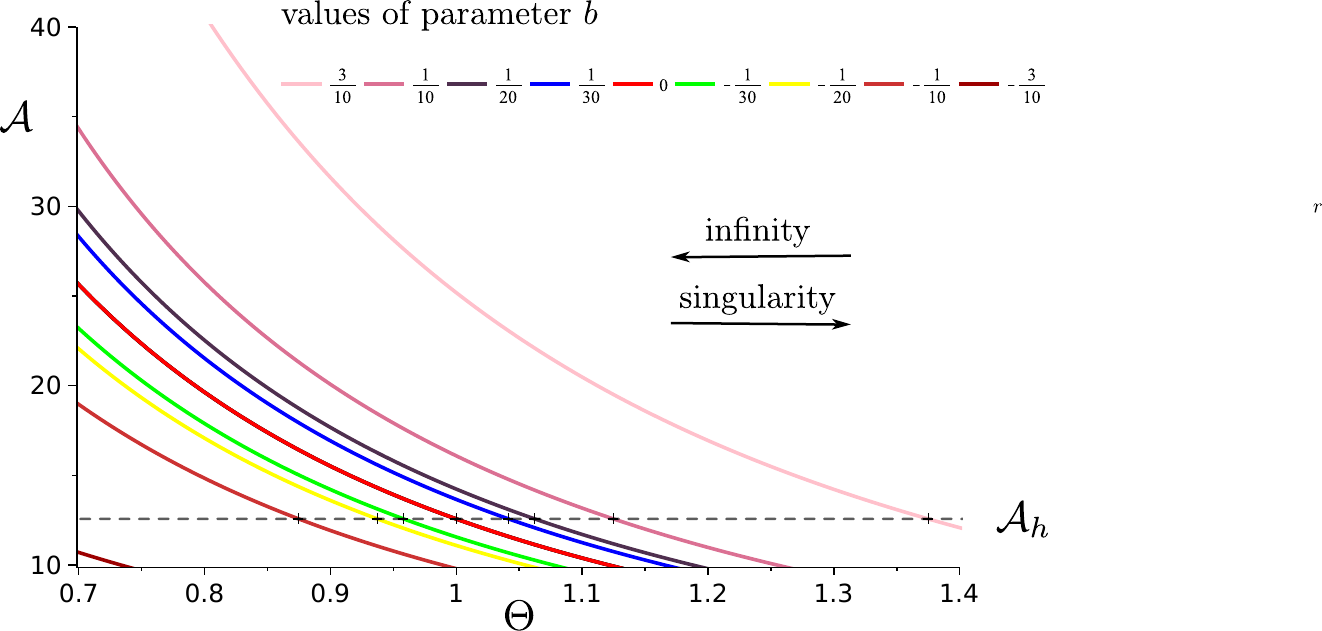}
	\end{center}
	\caption{\label{fig:area}
		Relation between the sphere area $\mathcal{A}$ (of constant~$r$) and the corresponding value of the expansion~$\Theta$ of the privileged null vector field ${\bold{k}=\partial_r}$, for different values of the Bach parameter $b$. The dashed line indicates the horizon area $\mathcal{A}_h$ which is $b$-independent. Here ${r_h=-1,\, k=0.5}$, ${\Lambda=0.2}$.}
\end{figure}

Finally, we can also compare the Schwarzschild--Bach--(A)dS solution with its classic analogy on an invariant level of \emph{spacetime geometry} by plotting the  dependence of the area of a sphere with $r$ fixed on the value of the expansion ${\Theta=\frac{1}{2}{k^a}_{;a}=\Omega^{-3}\,\Omega_{,r}}$ of the privileged null congruence $\bold{k}=\partial_r$ inducing a spacetime foliation, see Fig.~\ref{fig:area}. Here we may identify specific quantitative differences. Namely, any observer following the geodesics generated by $\bold{k}$ will measure a larger spherical area when ${b>0}$ (and smaller for ${b<0}$) for any fixed value of the congruence expansion. Put it in a opposite way, on  spheres of the same area, the expansion of $\bold{k}$ increases with $b$ growing, and vice versa.

\subsubsection{Thermodynamic properties: horizon area, temperature, entropy}

In this section, let us study  geometrical and thermodynamic
properties  of the Schwarzschild--Bach--(A)dS black hole,
extending the concise discussion in the letter
\cite{SvarcPodolskyPravdaPravdova:2018}.

The horizon  generated by the  null Killing vector ${\xi\equiv\sigma\partial_u=\sigma\partial_t}$ (considering the time-scaling freedom \eqref{scaling-t} represented by the parameter $\sigma$)  is given by vanishing of the norm of $\xi$, i.e., by ${\H(r)=0}$ at ${r=r_h}$ \eqref{horizon},   using
\eqref{H_[0,1]}.  Integrating the angular coordinates of the metric \eqref{BHmetric}, we obtain the horizon area
\be
{\cal A} = 4\pi\,\Omega^2(r_h)= \frac{4\pi}{r_h^2}= 4\pi\,{\bar r}_h^2 \,.
\label{horizon_area}
\ee

With the only nonvanishing derivatives of  $\xi$ being $\xi_{u;r}\! =\!-\xi_{r;u}\! =\!\frac{1}{2}\sigma(\Omega^2\H)'$,
${\xi^{\,r;u}\!=\!-\xi^{\,u;r}\!= \!\Omega^{-4}\xi_{u;r}}$, using \eqref{H_[0,1]}, the surface gravity
${\kappa^2\equiv-\frac{1}{2}\,\xi_{\mu;\nu}\,\xi^{\,\mu;\nu}}$
\cite{Wald:1984} takes the form
\be
\kappa/\sigma =-\frac{1}{2}(\H'+2\H\,\Omega'/\Omega)_{|r=r_h} = -\frac{1}{2}\,\H'(r_h) =
-\frac{1}{2}\rho\, r_h
=\frac{1}{2}\, {\bar r}_h^{\,-1}(1-\Lambda {\bar r}_h^2)
\,.
\label{surface_gravity}
\ee	
Therefore, the temperature  of
the black hole horizon ${T\equiv\kappa/(2\pi)}$  \cite{FanLu:2015} reads
\be
T/\sigma
= -\frac{1}{4\pi}\rho\,r_h
= \frac{1}{4\pi}\,{\bar r}_h^{\,-1}(1-\Lambda {\bar
	r}_h^2)  \,.
\label{temperature}
\ee
Note that  the Bach parameter $b$ does not enter the expressions \eqref{surface_gravity} and \eqref{temperature}, and thus they are same as for the  Schwarzschild--(A)dS solution. Both are zero for ${\bar r=1/\sqrt{\Lambda}}$ corresponding to the extreme Schwarzschild--dS solution with coinciding black-hole and cosmological horizons.

To determine the black-hole horizon
entropy, we employ the generalized
formula for higher-derivative theories derived by Wald \cite{Wald:1993,IyerWald:1994}
\be
S=\frac{2\pi}{\kappa}\,\oint \mathbf{Q}\,,
\label{WaldS}
\ee
with  the Noether-charge 2-form $\mathbf{Q}$ on the horizon (corresponding to the Lagrangian ${\cal L}$ of the theory) being
\bea
{\bf Q} \rovno \pul \varepsilon_{\mu\nu\alpha\beta}\,Q^{\mu\nu}\,\dd x^{\alpha}\wedge\dd x^{\beta} \,,\nonumber\\
Q^{\mu\nu} \rovno 2X^{\mu\nu\rho\sigma}\,\xi_{\rho;\sigma}
+4{X^{\mu\nu\rho\sigma}}_{;\rho}\,\xi_\sigma   \qquad \hbox{and}\qquad
X^{\mu\nu\rho\sigma}\equiv\frac{\partial {\cal L}}{\partial R_{\mu\nu\rho\sigma}}
\,. \label{NoetherCharge}
\eea
For quadratic gravity  \eqref{action}, we obtain
\bea
X^{\mu\nu\rho\sigma} \rovno
\frac{1}{16\pi}\bigg[
\Big(\gamma+\frac{2}{3}(2\alpha+3\beta) R \Big)g^{\nu[\sigma}g^{\rho]\mu}
-4\alpha\, g^{\kappa[\nu}g^{\mu][\rho}g^{\sigma]\lambda}R_{\kappa\lambda}
\bigg]\,. \label{Xabcd}
\eea
A lengthy calculation for the metric \eqref{BHmetric} then gives
the Noether-charge 2-form on the horizon
\be
\mathbf{Q} = -\frac{\Omega^2\, \H'}{16\pi}\left[\gamma+\frac{4}{3}\Lambda(\alpha+6\beta)+\frac{4}{3}k\alpha\,\frac{\B_1+\B_2}{\Omega^4}\right]\!\!\bigg|_{r=r_h}
\,\sin\theta\,\dd\theta\wedge\dd\phi \,. \label{Noether}
\ee
Finally,  using \eqref{horizon_area}, \eqref{surface_gravity}, and \eqref{bonhorizon}, the Schwarzschild--Bach--(A)dS black hole horizon entropy \eqref{WaldS} is given by a simple explicit formula  (cf. Eq.~(6.5) of~\cite{SvarcPodolskyPravdaPravdova:2018})
\be
S = \frac{1}{4}{\cal A}\left[\gamma+\frac{4}{3}\Lambda\,(\alpha+6\beta) -4\alpha\,\frac{b}{{\bar r}_h^{\,2}}\right]
\,. \label{entropy}
\ee
For ${\Lambda=0}$, it agrees with  \cite{LuPerkinsPopeStelle:2015} (with the identification
${k=\alpha}$ and ${b=\delta^*}$),
\cite{PodolskySvarcPravdaPravdova:2018}, and
\cite{PodolskySvarcPravdaPravdova:2020}. Therefore,
the ``non-Schwarzschild parameter''
$\delta^*$ of \cite{LuPerkinsPopeStelle:2015}
is the dimensionless Bach parameter $b$
determining the value of the Bach tensor on the horizon
$r_h$, see \eqref{bonhorizon}.
The standard entropy expression for the Schwarzschild black hole,
${S=\frac{1}{4G}\,{\cal A}}$,
is recovered either for ${b=0, \Lambda=0}$ (the Schwarzschild black hole in QG)
or for ${\alpha=0=\beta}$ (the Schwarzschild black hole in Einstein's
gravity). The result of \cite{LuPope:2011}, ${S =
	\tfrac{1}{4G}\,{\cal A}\,\big(1+\tfrac{4}{3}k\Lambda \big)}$, for the Schwarzschild--(A)dS black hole in the Einstein--Weyl gravity is recovered by setting ${b=0}$ and ${\beta=0}$ in \eqref{entropy}.
The	entropy \eqref{entropy} for the Schwarzschild--(A)dS black hole ($b=0$) vanishes in
critical gravity with {{${\beta=0}$}, ${\alpha=k\gamma}$}, ${\Lambda=-\frac{3}{4k}<0}$ \cite{LuPope:2011}.
Note also that  the deviations from {{${S = \tfrac{1}{4}{\cal A}\,\big[\gamma+\tfrac{4}{3}\Lambda (\alpha+6\beta)\big]}$}} are larger for smaller black holes because they have smaller~${\bar r}_h^{\,2}$.

\vspace{5mm}

\subsubsection{Specific motion of test particles caused by the Bach tensor}
\label{geodeviation}

This section generalizes Sec.~13 of~\cite{PodolskySvarcPravdaPravdova:2020} to the case of a nonvanishing cosmological constant $\Lambda$, and extends Sec.~V. and~VI. of~ \cite{SvarcPodolskyPravdaPravdova:2018}.
We will demonstrate that the effect  of the Bach tensor parts $\B_1$, $\B_2$ given by~\eqref{B1}, \eqref{B2}, entering the Bach invariant \eqref{invB}, can be directly observed through the relative motion of freely falling particles described by the equation of geodesic deviation.
\vspace{2mm}

\noindent
{\bf A: Interpreting solutions in quadratic gravity using geodesic deviation}

\noindent
Projecting the equation of geodesic deviation onto an {orthonormal frame} ${\{\bolde_{(0)}, \bolde_{(1)}, \bolde_{(2)}, \bolde_{(3)}}\}$ with the time-like vector being an observer's $4$-velocity ${\bolde_{(0)}=\boldu}$  and satisfying ${\bolde_{(a)}\cdot\bolde_{(b)}=\eta_{ab}}$, we obtain
\be
\ddot Z^{(\rm{i})}= R^{(\rm{i})}_{\quad(0)(0)(\rm{j})}\,Z^{(\rm{j})} \,,\qquad
\rm{i},\,\rm{j}=1,\,2,\,3\,, \label{InvGeoDev}
\ee
where
\begin{equation}
\label{PhysAccel}
\ddot Z^{(\rm{i})} \equiv e^{(\rm{i})}_a\,\frac{\Dif^2 Z^a}{\dd\, \tau^2}
=e^{(\rm{i})}_a\,{Z^a}_{;cd}\, u^c u^d  \,, \qquad \hbox{and} \qquad
{R_{(\rm{i})(0)(0)(\rm{j})}\equiv R_{abcd} \,e^a_{(\rm{i})}u^b u^c e^d_{(\rm{j})}} \,.
\end{equation}
The decomposition of the Riemann tensor  into the
traceless Weyl tensor, the Ricci tensor, and the scalar curvature $R$ gives
\begin{align}\label{DecompFrame}
R_{(\rm{i})(0)(0)(\rm{j})}=C_{(\rm{i})(0)(0)(\rm{j})}+\tfrac{1}{2}\big(R_{(\rm{i})(\rm{j})}
-\delta_{\rm{i}\rm{j}}\,R_{(0)(0)}\big)-\tfrac{1}{6}\,R\,\delta_{\rm{i}\rm{j}} \,.
\end{align}
Furthermore, employing the vacuum field equations \eqref{fieldeqsEWmod} with \eqref{traceEW} in \eqref{DecompFrame},  Eq.~(\ref{InvGeoDev}) takes the form
\begin{align}
\ddot{Z}^{(\rm{i})}=\frac{\Lambda}{3}\,Z^{(\rm{i})} +  C_{(\rm{i})(0)(0)(\rm{j})}\,Z^{(\rm{j})}
+2k\big(B_{(\rm{i})(\rm{j})}\,Z^{(\rm{j})}-B_{(0)(0)}\,Z^{(\rm{i})}\big)\,. \label{InvGeoDevExpl}
\end{align}
Applying the Newman--Penrose scalar decomposition (obtained in \cite{BicakPodolsky:1999,PodolskySvarc:2012})  with respect to a  (real) {null frame}  ${\{\boldk, \boldl, \boldm_{i} \}}$ with two future-oriented null vectors  $\boldk$ and $\boldl$  and  two spatial vectors  $\boldm_{i}$ orthogonal to them, normalized as
${\boldk\cdot\boldl=-1}$ and ${\boldm_{i}\cdot\boldm_{j}=\delta_{ij}}$, such that
\begin{align}
\boldk=\ssqrt(\boldu+\bolde_{(1)})\,, \qquad \boldl=\ssqrt(\boldu-\bolde_{(1)})\,,
\qquad \boldm_{i}=\bolde_{(i)} \quad \hbox{for} \quad i=2,\,3 \,, \label{NullFrame}
\end{align}
we arrive at  the equation of geodesic deviation \eqref{InvGeoDevExpl} in the QG theory in the form
\begin{align}
\ddot{Z}^{(1)} = &
\frac{\Lambda}{3}\,Z^{(1)} + \Psi_{2S}\,Z^{(1)}+ \tfrac{1}{\sqrt{2}}(\Psi_{1T^j}-\Psi_{3T^j})\,Z^{(j)} \nonumber\\
& \quad \qquad +2k\,\big[(B_{(1)(1)}-B_{(0)(0)})\,Z^{(1)}+B_{(1)(j)} \,Z^{(j)}\,\big]\,,\label{InvGeoDevFinal1}\\
\ddot{Z}^{(i)} = &\frac{\Lambda}{3}\,Z^{(i)} - \tfrac{1}{2}\Psi_{2S}\,Z^{(i)} + \tfrac{1}{\sqrt{2}}(\Psi_{1T^i}-\Psi_{3T^i})\,Z^{(1)} -\tfrac{1}{2}(\Psi_{0^{ij}}+\Psi_{4^{ij}})\,Z^{(j)} \nonumber\\
& \quad \qquad +2k\,\big[\,B_{(i)(1)} \,Z^{(1)}+B_{(i)(j)} \,Z^{(j)}-B_{(0)(0)}\,Z^{(i)}\,\big]\,, \label{InvGeoDevFinal2}
\end{align}
where  ${\Psi_{2T^{(ij)}}=\tfrac{1}{2}\Psi_{2S}\,\delta_{ij}}$, see \cite{PodolskySvarc:2012}.
The system \eqref{InvGeoDevFinal1}, \eqref{InvGeoDevFinal2}	has a clear physical interpretation.
Classical effects include the Newtonian tidal deformations, longitudinal motions, and  the transverse effects of gravitational waves (propagating in the directions $\bolde_{(1)}, -\bolde_{(1)}$)
caused by the scalar components $\Psi_{2S}$,  $\{ \Psi_{3T^i}, \Psi_{1T^i}\}$, and $\{\Psi_{4^{ij}}, \Psi_{0^{ij}}\}$, respectively, and isotropic radial motions of test particles caused by
the cosmological constant $\Lambda$. Apart from these classical effects, there are  additional specific effects caused by the nonvanishing Bach tensor, encoded in the frame components $B_{(a)(b)}$.

\vspace{2mm}

\noindent
{\bf B: Geodesic deviation in the Schwarzschild--Bach--(A)dS black hole spacetime}

\noindent
The general results obtained in the previous section can now be applied to the spherically symmetric black hole metric \eqref{BHmetric}. We choose an orthonormal frame associated with a radially falling observer, i.e.,  ${\dot{x}=0=\dot{y}}$, namely
\begin{align}
& \bolde_{(0)}\equiv \boldu= \dot{r}\,\partial_r +\dot{u}\,\partial_u \,, \nonumber \\
& \bolde_{(1)}= \tfrac{1}{2}\big[( {\Omega^2\dot{u}} )^{-1}-{\H}\dot{u}\big]\partial_r -\dot{u}\,\partial_u \,, \nonumber \\
& \bolde_{(i)}= \Omega^{-1}\big[1+\tfrac{1}{4}(x^2+y^2)\big]\partial_i \,, \label{OrtFrame}
\end{align}
where ${\dot{r}=\tfrac{1}{2}\big[({\Omega^2\,\dot{u}})^{-1}+{\H}\dot{u}\big]}$ due to  the normalisation of an observer's four-velocity ${\boldu\cdot\boldu=-1}$. Then the associated null interpretation frame  (\ref{NullFrame}) reads \begin{equation}
\boldk = \frac{1}{\sqrt{2}\,\dot{u}\,\Omega^2}\,\partial_r \,, \qquad
\boldl = \frac{\dot{u}\,{\H}}{\sqrt{2}}\,\mathbf{\partial}_r+\sqrt{2}\dot{u}\,\partial_u\,, \qquad
\boldm_i = \Omega^{-1}\big[1+\tfrac{1}{4}(x^2+y^2)\big]\partial_i \,. \label{NullIntFrame}
\end{equation}
Since the spherically symmetric black hole metric \eqref{BHmetric} is of algebraic type D, there is only one nonvanishing Weyl-tensor component with respect to \eqref{NullIntFrame}, namely
\begin{equation}
\Psi_{2S} \equiv  C_{abcd}\; k^a\, l^b\, l^c\, k^d
=\tfrac{1}{6}\,\Omega^{-2}({\H}''+2)\,. \label{Psi2Int}
\end{equation}
Using the Bach tensor projections with respect to the orthonormal frame (\ref{OrtFrame}),
\begin{align}
B_{(0)(0)}&= \frac{1}{24\,\Omega^6\dot{u}^2}\Big[-(1-\Omega^2{\H}\dot{u}^2)^2\,{\H}''''+2\Omega^2\dot{u}^2({\H}'{\H}'''-\tfrac{1}{2}{{\H}''}^2+2)\Big]\,, \\
B_{(1)(1)}&= \frac{1}{24\,\Omega^6\dot{u}^2}\Big[-(1+\Omega^2{\H}\dot{u}^2)^2\,{\H}''''-2\Omega^2\dot{u}^2({\H}'{\H}'''-\tfrac{1}{2}{{\H}''}^2+2)\Big]\,, \\
B_{(0)(1)}&= -\frac{1}{24\,\Omega^6\dot{u}^2}\,(1-\Omega^4{\H}^2\dot{u}^4)\,{\H}'''' \,,
\qquad B_{(0)(i)}= 0\,, \\
B_{(i)(j)}&= \frac{\delta_{ij}}{12\,\Omega^4}({\H}{\H}''''+{\H}'{\H}'''-\tfrac{1}{2}{{\H}''}^2+2)\,,
\qquad B_{(1)(i)}= 0 \,,
\end{align}
the equations of geodesics deviation \eqref{InvGeoDevFinal1}, \eqref{InvGeoDevFinal2} take the form
\begin{eqnarray}
\ddot{Z}^{(1)} \rovno
\frac{\Lambda}{3}\,Z^{(1)} + \, \frac{1}{6}\, \Omega^{-2}\big({\H}''+2\big)\,Z^{(1)}\,
-\,\frac{1}{3}\,k\,\Omega^{-4}\big({\H}{\H}''''+{\H}'{\H}'''-\tfrac{1}{2}{{\H}''}^2+2\big)Z^{(1)} \,, \label{InvGeoDevBH1}\\
\ddot{Z}^{(i)} \rovno \frac{\Lambda}{3}\,Z^{(i)} - \frac{1}{12}\,\Omega^{-2}\big({\H}''+2\big)\,Z^{(i)}
+\frac{1}{12}\,k\,\Omega^{-4}\big((\Omega^2\H\dot{u}^2)^{-1}+\Omega^2{\H}\dot{u}^2\big){\H}{\H}''''\,Z^{(i)} \,. \label{InvGeoDevBHi}
\end{eqnarray}
Therefore, apart from the classical effects, namely the isotropic influence of the cosmological constant $\Lambda$ and  a classical tidal deformation caused by the Weyl curvature \eqref{Psi2Int} proportional to ${\Omega^{-2}({\H}''+2)}$, i.e., the square root of the invariant \eqref{invC}, there are two additional effects of quadratic gravity  caused by the
presence of a nonvanishing Bach tensor. They appear in the longitudinal \eqref{InvGeoDevBH1} and transverse
\eqref{InvGeoDevBHi} components
of the acceleration and are proportional (up to a constant) to the square roots of the two parts of the invariant \eqref{invB}, to the amplitudes $\B_1$, $\B_2$, see  \eqref{B1}, \eqref{B2}.

When considering a test particle which is initially static (${\dot{r}=0}$), the geodesic deviation equations simplify to
\begin{align}
\ddot{Z}^{(1)} = &
\frac{\Lambda}{3}\,Z^{(1)} + \frac{1}{6}\, \Omega^{-2}\big({\H}''+2\big)\,Z^{(1)}
-\frac{1}{3}\,k\,\Omega^{-4}\big(\B_1+\B_2\big)Z^{(1)} \,, \label{InvGeoDevBH1r0}\\
\ddot{Z}^{(i)} = & \frac{\Lambda}{3}\,Z^{(i)}- \frac{1}{12}\,\Omega^{-2}\big({\H}''+2\big)\,Z^{(i)}
-\frac{1}{6}\,k\,\Omega^{-4}\,\B_1\,Z^{(i)} \,, \label{InvGeoDevBHir0}
\end{align}
thanks to the 4-velocity normalization ${\Omega^2\H\,\dot{u}^2=-1}$.
Note that the first component $\B_1$ of the Bach tensor
can be directly observed in the transverse components of the
acceleration \eqref{InvGeoDevBHir0} along ${\bolde_{(2)}, \bolde_{(3)}}$, i.e.,
${\partial_x, \partial_y}$ (equivalent to ${\partial_\theta, \partial_\phi}$),
while the second component $\B_2$ enters the radial component \eqref{InvGeoDevBH1r0} along
$\bolde_{(1)}= -\dot{u}\,(\partial_u+\H\,\partial_r)= -\H\,\Omega'\,\dot{u}\,\partial_{\bar r}$,
proportional to $\partial_{\bar r}$. Note also that on the horizon, there is only the radial effect given by ${\B_2(r_h)}$ since
${\B_1(r_h)=0}$ due to \eqref{B1} and \eqref{horizon}, see also \eqref{bonhorizon}.

As has been already argued in \cite{PodolskySvarcPravdaPravdova:2020}, the effects of  $\B_1, \B_2$ can be distinguished from the Newtonian tidal effect in the Schwarzschild solution. Thus, by observing a free fall of a set of test particles, one can distinguish the Schwarzschild--(A)dS black hole from the Schwarzschild--Bach--(A)dS black hole with a nonvanishing Bach tensor.

\vspace{2mm}


\subsection{Schwarzschild--Bach--(A)dS black hole in the class ${[n,p]=[0,0]}$: near a generic point}
\label{SchwaBach_[n,p]=[0,0]}

This class of solutions has the \emph{highest number of free parameters} (see Table~\ref{tab:3}) and for a special choice of these parameters, it represents the previously discussed Schwarzschild--Bach--(A)dS black hole near \emph{any} regular point ${r=r_0\not= r_h}$. The first few terms of the expansion read
	\begin{eqnarray}
	\Omega(r)      \rovno
			-\frac{1}{r} - \frac{b_1}{2r_0\big(r_0^2\nu +\tfrac{1}{3}\Lambda\big) }\,(r-r_0)^2  +\ldots\,,
	\label{[0,0]_OmegaFULL}
	\\
	\H (r) \rovno
	\frac{\Lambda}{3}-r^2-\Big(\frac{\Lambda}{3}-r_h^2\Big)\frac{r^3}{r_h^3}
	+ (b_1-b_2)\,r_0(r-r_0) -3b_2\,(r-r_0)^2
	\nonumber\\
	&& +\frac{(b_2-b_1)\big(1+\nu+\frac{1}{2kr_0^2}\big)
      -2(2+3 \nu)b_2+3b_2^2}{(1+3\nu +b_1-b_2)\,r_0}  \,(r-r_0)^3 + \ldots\,,
	\label{[0,0]_HFULL}
	\end{eqnarray}
 where
\be
\nu\equiv\oo -1-\frac{\Lambda}{3}\frac{\oo^3}{r_0^2}\,,\qquad
\oo\equiv\frac{r_0}{r_h}\,,
\label{[0,0]_DEF c0}
\ee
 and where \emph{two independent Bach parameters}, denoted as $b_1$ and $b_2$, are  proportional to the values of the two components of the Bach tensor at $r_0$
(see \eqref{[0,0]_BachInvariantb1b2B1B2},
\eqref{[0,0]_BachInvariantb1b2}). For ${b_1=0=b_2}$, the solution reduces to the Schwarzschild--(A)dS solution.

To find the coefficients $a_i$ and $c_i$ in \eqref{rozvojomeg0}--\eqref{DElta},
we start with Eqs.~\eqref{KeyEq1} and~\eqref{KeyEq3} that give
 (after relabeling ${l \to l-1}$) for $ l\ge 1$
\bea
c_{l+3} \rovno \frac{3}{k\,(l+3)(l+2)(l+1)l}\,\sum^{l}_{i=0}
a_i \,a_{l+1-i}(l+1-i)(l-3i)
\,,
\label{[0,0]initcondc}\\
a_{l+1}\rovno \frac{1}{l(l+1)c_0}
\bigg\{\tfrac{2}{3}\Lambda\! \sum^{l-1}_{j=0}\sum^{j}_{i=0}a_ia_{j-i}a_{l-j-1}-\tfrac{1}{3}\,a_{l-1}
-\sum^{l+1}_{i=1}c_ia_{l-i+1}\big[l(l-i+1)+\tfrac{1}{6}i(i-1)\big]\! \bigg\}, \label{[0,0]initconda}
\eea
respectively.
An  additional constraint follows from the lowest
nontrivial order ${l=0}$ of Eq. \eqref{KeyEq2},
\begin{equation}
c_3 =\frac{1}{6kc_1}\big[9a_1^2c_0+2k(c_2^2-1)+3a_0(a_0+a_1c_1-\Lambda a_0^3)\big] \,. \label{[0,0]initcond2}
\end{equation}

Therefore, this solution  admits five free initial parameters ${a_0,\, a_1,\, c_0,\, c_1,\, c_2}$. The remaining coefficients $a_{l+1}$, $c_{l+3}$ in
\eqref{[0,0]_OmegaFULL} and \eqref{[0,0]_HFULL}, respectively,
 are  given  by the recurrent relations \eqref{[0,0]initconda}, \eqref{[0,0]initcondc}, respectively,
\bea
&& a_2 = -\frac{a_0 + 3 a_1 c_1 + a_0 c_2 - 2 a_0^3 \Lambda}{6 c_0}\,,\ldots\,,\\
&& c_4 = -\frac{6 a_1^2 c_0 + a_0 (a_0 + 3 a_1 c_1 + a_0 c_2 - 2 a_0^3 \Lambda)}{24 c_0 k}\,,  \ldots\,.
\label{[0,0]initcond3}
\eea

Now we will show that this class of solutions contains the Schwarzschild--(A)dS solution as a special subcase.

\subsubsection{{Identification of the Schwarzschild--(A)dS black hole }}
\label{sec_Sch[0,0]}

Let us employ the scalar invariant \eqref{invB},
 with \eqref{B1}, \eqref{B2}, which
at ${r=r_0}$ read
\begin{align}
& B_{ab}\, B^{ab} (r_0) =  \frac{1}{72\,a_0^8} \big[ (\B_1)^2 + 2(\B_1+\B_2)^2\big]\,,\nonumber\\
& \hbox{where}\quad \B_1 (r_0)= 24 c_0c_4  \,,\qquad  \B_2 (r_0)= 2(3 c_1 c_3  - c_2^2 +  1)\,.
\label{[0,0]_BachInvariant}
\end{align}
The Schwarzschild--(A)dS solution  is uniquely identified by vanishing of the Bach tensor, ${B_{ab}=0}$ $\Leftrightarrow$ ${\B_1=0=\B_2}$, which implies ${c_4=0}$ and ${3 c_1 c_3  - c_2^2 +  1 =0}$. Together  with \eqref{[0,0]initcond3}, \eqref{[0,0]initcond2}, this yields two necessary conditions
\be
c_1=-\frac{a_0}{a_1}\Big(1-\Lambda a_0^2+3\frac{a_1^2}{a_0^2}c_0\Big)\,,\qquad
c_2=2-\Lambda a_0^2+3\frac{a_1^2}{a_0^2}c_0\,.
\label{[0,0]BachzeroA}
\ee
Then  the recurrent relations \eqref{[0,0]initconda}, \eqref{[0,0]initcondc} reduce to
\be
a_i=a_0\,\Big(\frac{a_1}{a_0}\Big)^i\quad \hbox{for all}\ i \ge 0 \,,\qquad
c_3=-\frac{a_1}{a_0}\Big(1+\frac{a_1^2}{a_0^2}c_0-\tfrac{1}{3}\Lambda a_0^2\Big)\,, \quad c_i=0\quad \hbox{for all}\ i \ge 4\,,
\ee
where the first sequence is  a geometrical series, while the second series is truncated to
the 3rd-order polynomial. The metric functions thus take the following closed form
\begin{eqnarray}
\Omega(r)      \rovno  a_0 \,\sum_{i=0}^\infty  \,\Big(\frac{a_1}{a_0}\Delta\Big)^i
=\frac{a_0^2}{a_0-a_1\Delta}=\frac{a_0^2}{(a_0+a_1r_0)-a_1r}\,, \label{[0,0]Omega}\\
\H (r) \rovno  c_0+c_1(r-r_0)+c_2(r-r_0)^2+c_3(r-r_0)^3 \,. \label{[0,0]H0}
\end{eqnarray}
Using the gauge  \eqref{scalingfreedom}, we can set
\be
a_0=-\frac{1}{r_0}\,,\qquad a_1=\frac{1}{r_0^2}
\label{[0,0]_a0a1}\,,
\ee
so that the metric functions read
\be
{\bar r}=\Omega(r) = -\frac{1}{r}\,, \qquad
\H (r) =
\big(r-r_0\big)\frac{r^2}{r_0}+\frac{c_0-\tfrac{1}{3}\Lambda}{r_0^3}\,r^3 +\frac{\Lambda}{3}
\,.
\label{[0,0]_Schw}
\ee
Such $\H $ can be rewritten as
\be
\H (r) = \frac{\Lambda}{3}-r^2-\Big(\frac{\Lambda}{3}-r_h^2\Big)\frac{r^3}{r_h^3}\,, \label{[0,0]_SchwSHIFT}\\
\ee
and $c_0$,  $r_h$ are related by
\be
	c_0=\frac{\Lambda}{3}(1-\oo^3)+r_0^2 (\oo-1)
	= r_0^2\nu +\frac{\Lambda}{3}\,. \label{[0,0]_DEF c0a}
\ee
This is the Schwarzschild--(A)dS solution \eqref{IIbH0Schw} (see also \eqref{SchwAdS}) with the black hole horizon  at $r=r_h$.

Therefore, for vanishing Bach tensor, the class ${[n,p]=[0,0]}$ corresponds to the Schwarzschild--(A)dS black hole solution.

\subsubsection{{More general solution with nontrivial Bach tensor}}

Now, let us return to the
 generic case of ${[n,p]=[0,0]}$ with a nonvanishing Bach tensor \eqref{[0,0]initcondc}--\eqref{[0,0]initcond3}
 with the free parameters  ${a_0,a_1,c_0,c_1,c_2}$. To simplify the two components
$\B_1(r_0)$ and $\B_2(r_0)$ of the Bach tensor \eqref{B1} and \eqref{B2} evaluated at $r_0$,
we introduce new more physical dimensionless  Bach parameters $b_1$ and $b_2$,
corresponding to $\B_1(r_0)$ and $\B_2(r_0)$,
respectively,
\be
b_1 \equiv  \tfrac{1}{3}\big(-1-6\nu -c_2+3\frac{c_1}{r_0}\big) \,,\qquad
b_2 \equiv  \tfrac{1}{3}\big(2+3 \nu -c_2\big) \,,
\label{Bach_b1b2}
\ee
so that
\be
c_1 = (1+3 \nu +b_1-b_2)\,r_0 \,,\qquad
c_2 =  2+3 \nu -3b_2 \,,
\label{Bach_c1c2}
\ee
where $\nu$ has been introduced in  \eqref{[0,0]_DEF c0} and the gauge \eqref{[0,0]_a0a1} has been used.
Then using \eqref{[0,0]initcond3} and \eqref{[0,0]initcond2},   $\B_1(r_0)$ and $\B_2(r_0)$ are related to the Bach parameters $b_1,b_2$ by
\be
b_1=\tfrac{1}{3}kr_0^2\,\B_1(r_0) \,,\qquad b_2=\tfrac{1}{3}kr_0^2\,\big(\B_1(r_0)+\B_2(r_0)\big)\,,
\label{[0,0]_BachInvariantb1b2B1B2}
\ee
and the Bach invariant at $r_0$ is
\be
B_{ab}\, B^{ab} (r_0)= \frac{r_0^4}{8k^2}\,\big( b_1^2 + 2\, b_2^2\,\big) \,.
\label{[0,0]_BachInvariantb1b2}
\ee

The coefficients $a_i$, $c_i$ take now the form
\begin{align}
&
a_0 = -\frac{1}{r_0}\,,\quad
a_1 =\frac{1}{r_0^2}\,,\quad
a_2 = -\frac{1}{r_0^3}-\frac{b_1}{2c_0 r_0}\,, \ldots\,,\label{[0,0]_expansionaINFa} \\
& c_0 = \nu\,r_0^2 +\tfrac{1}{3}\Lambda \,,\quad
c_1 =(1+3\nu)r_0 +(b_1-b_2)\,r_0 \,,\quad
c_2 = 2+3\nu -3b_2 \,,\nonumber\\
&
c_3 = \frac{(1+\nu)(1+3 \nu)-2(2+3 \nu)b_2+3b_2^2+(b_2-b_1)\frac{1}{2kr_0^2}}{(1+3 \nu +b_1-b_2)\,r_0}  \,,\ \
c_4 = \frac{b_1}{8kc_0 r_0^2}\,, \ldots\,.
\label{[0,0]_expansioncINFc}
\end{align}
For ${b_1=0=b_2}$, the Bach tensor vanishes and thus the solution reduces to the Schwarzschild--(A)dS solution (see Sec.~\ref{sec_Sch[0,0]}).
Thus the more general solution \eqref{[0,0]initcondc} and \eqref{[0,0]initconda} includes
 a modification of
the Schwarzschild--(A)dS solution  admitting a nonvanishing Bach tensor
(see also
\eqref{[0,0]_expansionaINFa}, \eqref{[0,0]_expansioncINFc} or
\eqref{[0,0]_OmegaFULL}, \eqref{[0,0]_HFULL}).

Notice that as ${\Delta\to 0}$, the functions $\bar r$, $f$, and $h$ approach \emph{constants}
(cf. \eqref{to static}, \eqref{rcehf}).

\subsubsection{{Identification of the Schwa--Bach--(A)dS black hole ${[0,1]}$ in the class ${[0,0]}$ }}

Since the generic  class ${[0,0]}$
contains also  the Schwarzschild--Bach--(A)dS black hole solution \eqref{Omega_[0,1]}, \eqref{H_[0,1]}, expressed around the horizon $r_h$ in the class ${[0,1]}$, one should be able to express the five free parameters ${a_0,\, a_1,\, c_0,\, c_1,\, c_2}$ of the ${[0,0]}$ class in terms of the free parameters of the ${[0,1]}$ class.
This can be done (within the convergence radius of~\eqref{Omega_[0,1]},~\eqref{H_[0,1]}) by evaluating the functions \eqref{Omega_[0,1]}, \eqref{H_[0,1]} and their derivatives at ${r=r_0}$, and comparing them with the expansions
\eqref{rozvojomeg0} and \eqref{rozvojcalH0}
with ${n=0,p=0}$:
\begin{align}
a_0 & = -\frac{1}{r_0}-\frac{b}{r_h}\sum_{i=1}^\infty \alpha_i\,\Big(\frac{r_h-r_0}{\X r_h}\Big)^i \,, \label{[0,0]a0}\\
a_1 & = \frac{1}{r_0^2}+\frac{b}{\X r_h^2}\sum_{i=1}^\infty i\,\alpha_i\,\Big(\frac{r_h -r_0}{\X r_h}\Big)^{i-1} \,, \\
c_0 & = (r_0-r_h)\bigg[\frac{r_0^2}{r_h}-\frac{\Lambda}{3r_h^3}(r_0^2+r_0 r_h+r_h^2 )
+3b\X \,r_h\sum_{i=1}^\infty \gamma_i\,\Big(\frac{r_0-r_h}{\X r_h}\Big)^i\, \bigg] \,, \\
c_1 & = (3r_0-2r_h)\frac{r_0}{r_h}-\frac{\Lambda r_0^2}{r_h^3}+3b\X \,r_h\sum_{i=1}^\infty (i+1)\, \gamma_i\,\Big(\frac{r_0-r_h}{\X r_h}\Big)^i \,, \\
c_2 & = (3r_0-r_h)\frac{1}{r_h}-\frac{\Lambda r_0}{r_h^3}+\frac{3}{2}b\,\sum_{i=1}^\infty i(i+1)\,\gamma_i\,\Big(\frac{r_0-r_h}{\X r_h}\Big)^{i-1} \,.\label{[0,0]c2}
\end{align}
Thus we obtain an expansion  of the Schwarzschild--Bach--(A)dS black hole \emph{around any point} $r_0$ in terms of just one Bach parameter $b$ (which determines the value of the Bach tensor on the horizon $r_h$) in the recurrent relations \eqref{[0,0]initcondc}--\eqref{[0,0]initcond2}.

\subsubsection{{Formal limit  ${r_0 \to r_h}$ }}

Finally, let us  perform a  ``consistency check'' between the two series corresponding to the Schwarzschild--Bach--(A)dS black hole solution
in the class $[0,1]$ (see \eqref{IIbOmegaFULL}, \eqref{IIbH0FULL})  and  in the class ${[0,0]}$
(see \eqref{[0,0]_OmegaFULL}, \eqref{[0,0]_HFULL}).
Here we temporarily denote the coefficients in the class ${[0,0]}$ by ${\hat c}_i$ and ${\hat a}_i$.
In the limit ${r_0 \to r_h}$ in \eqref{[0,0]a0}--\eqref{[0,0]c2}, using also
\eqref{IIb_a0} and the first relations in \eqref{IIb_expansionaLambda}, \eqref{IIb_expansioncLambda},  we obtain the following relations
\bea
{\hat a}_0 \rovno -\frac{1}{r_h} \equiv a_0\,,\qquad
{\hat a}_1 = \frac{1}{r_h^2}\Big(1+\frac{b}{\rho}\Big) \equiv a_1\,,\\
{\hat c}_0 \rovno 0\,,\qquad
{\hat c}_1 = r_h -\frac{\Lambda}{r_h}=r_h \rho\equiv c_0\,,\qquad
{\hat c}_2 = 2-\frac{\Lambda}{r_h^2}+3b \equiv c_1\,.
\eea
Since ${\hat c}_{j+1}$  and~$c_j$ satisfy the same recurrent relations (cf. \eqref{nonSchwinitcondc} and \eqref{[0,0]initcondc}), the functions $\H$ agree. In addition, the condition
${\hat c}_0=0$ in  \eqref{[0,0]initconda} implies
\be
0=\tfrac{1}{3}\,\hat a_{l-1} +l^2\, {\hat c}_1\,{\hat a}_l
+\sum^{l+1}_{i=2}\hat c_i\,\hat a_{l+1-i}\left[l(l+1-i)+\tfrac{1}{6}i(i-1)\right]
-\tfrac{2}{3}\Lambda \sum^{l-1}_{j=0}\sum^{j}_{i=0}a_ia_{j-i}a_{l-j-1}\,,
\ee
i.e.,
\be
{\hat a}_l =-\frac{1}{l^2\, {\hat c}_1} \Big[\tfrac{1}{3}\,\hat a_{l-1}
+\sum^{l}_{i=1}\hat c_{i+1}\, \hat a_{l-i}\left[l(l-i)+\tfrac{1}{6}i(i+1)\right]
-\tfrac{2}{3}\Lambda \sum^{l-1}_{j=0}\sum^{j}_{i=0}a_ia_{j-i}a_{l-j-1}\Big],
\ee
which is the same as \eqref{nonSchwinitconda} for $a_{l+1}$ (with the identification ${\hat c_{i+1}=c_i}$) and thus the functions $\Omega$ also agree.
Therefore, in the limit ${r_0\to r_h}$ we obtain
\be
{\hat c}_0 \to 0\,,\ \  \
{\hat c}_{j+1} \to c_j\,,\ \ \
{\hat a}_j \to a_j \quad \hbox{for all}\ j \geq 0\,,
\ee
which shows the consistency of the two expressions for the Schwarzschild--Bach--(A)dS black hole in the $[0,0]$ and $[0,1]$ classes.

\vspace{5mm}

\noindent
\textbf{To conclude}: The class $[0,0]$, expressed in terms of the series \eqref{[0,0]_OmegaFULL} and \eqref{[0,0]_HFULL} around an arbitrary point $r_0$, represents the spherically symmetric Schwarzschild--Bach--(A)dS black hole. However, without using the specific choice of parameters  \eqref{[0,0]_a0a1}, it may represent any of the other spherical solutions as well.

\vspace{5mm}


\subsection{Solutions with extreme double horizon in the class ${[0,2]}$}
\label{sec_[0,2]}

The ${[n,p]=[0,2]}$ class of solutions  corresponds to an expansion  around an extreme (double degenerate) horizon $r_0$ since for ${p=2}$ the key metric function reads ${\H(r) = (r-r_0)^2\Big[c_0 + c_1 \,(r-r_0)+\cdots\Big]}$. For a generic ${\Lambda}$, it contains only the extreme Schwarzschild--de~Sitter spacetime. However, for certain special values of ${\Lambda}$, other solutions with a nonvanishing Bach tensor also exist.

Indeed, for ${l\geq 0}$,  Eq. (\ref{KeyEq1}) gives
\begin{align}
\tfrac{1}{3}k c_{l}(l+2)(l+1)l(l-1)=(l-1)la_0 a_{l}+\sum^{l-1}_{i=1}a_i a_{l-i}(l-i)(l-3i-1) \,,\label{IIIb_rr}
\end{align}
while Eq.~(\ref{KeyEq3}) yields
\begin{equation}
c_0 =2\Lambda a_0^2 -1\,,\qquad a_1=\frac{3a_0 c_1}{2(3-4\Lambda a_0^2)}\,,\label{[0,2]c0}
\end{equation}
and
\begin{align}
a_{l}\left[l(l+1)(2a_0^2\Lambda-1)-\tfrac{4}{3}a_0^2\Lambda\right]+&\tfrac{1}{6}a_0c_{l}(l+1)(l+2)
+\sum^{l-1}_{i=1}c_i a_{l-i}\left[(l-i)(l+1)+\tfrac{1}{6}(i+1)(i+2)\right]\nonumber\\
&
=\tfrac{2}{3}\Lambda\bigg[
\sum^{l-1}_{i=1}a_0a_i a_{l-i}
+\sum^{l-1}_{j=1}\sum^{j}_{i=0}a_i a_{j-i} a_{l-j}\bigg]\,.\label{IIIb_trace}
\end{align}
Interestingly,  Eq.~(\ref{KeyEq2}) implies 
\begin{align}
\hbox{either \quad a)}&\ \ \Lambda a_0^2=1\,,\quad \Lambda>0\,, \label{[0,2]1}\\
\hbox{or     \quad b)}&\ \ \Lambda=\frac{3}{8k}\,. \label{[0,2]2}
\end{align}
Let us discuss these two distinct cases separately.

\subsubsection{Case a)  $\Lambda a_0^2=1$: extreme	Schwarzschild(--Bach)--dS black hole}
\label{sec_case[0,2]a}

In this case, it follows from~\eqref{[0,2]c0} that ${c_0=1}$ while $c_1$ is arbitrary.
From \eqref{IIIb_rr} and \eqref{IIIb_trace}, we infer that for the first coefficients
${a_i=a_0 \left(-\tfrac{3}{2}c_1\right)^i \equiv \pm \tfrac{1}{\sqrt{\Lambda}}
\left(-\tfrac{3}{2}c_1\right)^i}$ and ${c_j=0}$ for ${j>1}$.
Let us thus assume that ${a_i=a_0 \left(-\tfrac{3}{2}c_1\right)^i\equiv\pm \tfrac{1}{\sqrt{\Lambda}}
\left(-\tfrac{3}{2}c_1\right)^i}$ for \emph{all} ${0\leq i\leq l-1}$, and $c_j=0$ for $2\leq j\leq l-1$. Then
Eq.~\eqref{IIIb_rr} gives
\be
a_l=\frac{k}{3a_0}\,c_{l}(l+2)(l+1) +a_0 \left(-\tfrac{3}{2}c_1\right)^l
\equiv\pm \Big(\tfrac{1}{3} k\sqrt{\Lambda}\,c_{l}(l+2)(l+1) + \frac{1}{\sqrt{\Lambda}}
\left(-\tfrac{3}{2}c_1\right)^l\Big)\,,
\label{[0,2]pomal}
\ee
and Eq.~\eqref{IIIb_trace}  implies (note that this relation is valid even for $l=2$)
\be
c_{l}\big[ 2k\Lambda\left[3l(l+1)-4\right]+3\big]=0\,.\label{[0,2]condcl}
\ee
Now we have to distinguish two subcases:
\vspace{5mm}

\noindent
$\bullet$ Using the mathematical induction, the only solution of this system  with a \emph{generic value} of ${\Lambda>0}$ is
\be
a_{i}=a_0 \left(-\tfrac{3}{2}c_1\right)^i
  \equiv \pm \frac{1}{\sqrt{\Lambda}}\left(-\tfrac{3}{2}c_1\right)^i\,,   \quad  \forall i\geq 0\,,
  \qquad\hbox{and}\qquad
c_{j}=0\,,\ \forall j\geq 2\,, \label{[0,2]extSchw-ca}
\ee
with a free parameter $c_1$.
Thus
\be
\H(r) =\Delta^2+c_1\Delta^3\,,
\ee
and $\Omega$  is a sum of a geometric series \eqref{[0,2]extSchw-ca} which can be expressed in the closed form as
\be
\Omega(r)=\frac{\pm 2}{\sqrt{\Lambda}\,(2+3c_1\Delta)}\,.
\ee
Using the gauge
\be
c_1=\mp \frac{2}{3\sqrt{\Lambda}}\,,\qquad r_0=\mp\sqrt{\Lambda}\,,\label{gauge[0,2]}
\ee
 the metric functions are simplified to
\be
\Omega=-\frac{1}{r}\,,\qquad
\H = \Delta^2(1+c_1\Delta)
   \equiv\frac{(r\pm\sqrt{\Lambda})^2(\sqrt{\Lambda}\mp 2r)}{3\sqrt{\Lambda}}\,.\label{[0,2]extSchw-OmH}
\ee
It is the extreme  Schwarzschild--de~Sitter black hole \eqref{SchwAdS} characterised by the condition ${9\Lambda m^2=1}$ (see e.g., Section 9.4.2 in~\cite{GriffithsPodolsky:2009}). The double degenerate Killing horizon is located at ${r_h\equiv\mp\sqrt{\Lambda}}$, that is at ${\bar r_h \equiv\Omega(r_h) = \pm 1/\sqrt{\Lambda} = 3m}$.

\vspace{5mm}

\noindent
\textbf{To conclude}: The class $[0,2]$ with ${\Lambda a_0^2=1}$ and arbitrary ${\Lambda>0}$, with the metric functions expressed in terms of  \eqref{[0,2]extSchw-OmH}, represents  the extreme Schwarzschild--de~Sitter black hole solution.

\vspace{5mm}

\noindent
$\bullet$ The second branch of the ${[0,2]}$ class of solutions obeying \eqref{[0,2]1} exists only for \emph{discrete values} of the cosmological constant~${\Lambda>0}$, restricted by \eqref{[0,2]condcl} as
\be
\Lambda=-\frac{3}{2k\left[3L(L+1)-4\right]}\,,\qquad\hbox{where}\quad   L\in\mathbb{N} \,,\ \ L\geq 2\,, \ \ k<0\,,
\label{[0,2]lambda}
\ee
or equivalently
\be
\Lambda=-\frac{3\gamma}{[6L(L+1)-8]\alpha + 24\beta}\,.
\ee
In this class, all $c_i$ vanish for ${2\leq i\leq L-1}$, while $a_j$ are determined by
\eqref{[0,2]extSchw-ca} for ${0\leq j\leq L-1}$.  Eq.~\eqref{IIIb_rr} implies
\be
a_L =  \pm \Big(\tfrac{1}{3} k\sqrt{\Lambda}\,\,\tilde{b}_L(L+1)(L+2)
   + \frac{1}{\sqrt{\Lambda}} \left(-\tfrac{3}{2}c_1\right)^L\Big)\,,
\ee
cf. \eqref{[0,2]pomal}, where
\be
\tilde{b}_L\equiv c_L
\ee
is \emph{a new free ``Bach'' parameter}. The remaining coefficients $c_i$, $a_i$ (for ${i>L}$) are obtained  from \eqref{IIIb_rr} and \eqref{IIIb_trace} by the recurrent relations\footnote{Since these relations are valid also in the Case b), see Sec.~\ref{sec_case[0,2]b}, we do not use \eqref{[0,2]1} to simplify them.}
\bea
c_{l} \rovno 6\frac{\left[3l(l+1)(2a_0^2\Lambda-1)-4a_0^2\Lambda\right]
	U_{l-1}-3a_0 l(l-1)V_{l-1}} {(l-1)l(l+1)(l+2)[3a_0^2-8ka_0^2\Lambda+6k l(l+1)(2a_0^2\Lambda-1)]} \,,
\label{[0,2]clal}\\
a_l \rovno -3\frac{a_0 U_{l-1}+2k l(l-1)V_{l-1}}
{(l-1)l[3a_0^2-8ka_0^2\Lambda+6k l(l+1)(2a_0^2\Lambda-1)]} \,,\nonumber
\eea
where
\bea
U_{l-1} \rovno \sum^{l-1}_{i=1}a_i a_{l-i}(l-i)(l-3i-1)\,,\label{[0,2]S}\\
V_{l-1}\rovno \sum^{l-1}_{i=1}c_i a_{l-i}\left[(l-i)(l+1)+\tfrac{1}{6}(i+1)(i+2)\right]
-\tfrac{2}{3}\Lambda\Big(\sum^{l-1}_{i=1}a_0a_i a_{l-i}+\sum^{l-1}_{j=1}\sum^{j}_{i=0}a_i a_{j-i} a_{l-j}\Big)\,.\nonumber
\eea
Recall that in this case
\be
a_0=\pm\frac{1}{\sqrt{\Lambda}}\,,\qquad  c_0=1\,,\label{gauge[0,2]S}
\ee
see \eqref{[0,2]1}, \eqref{[0,2]c0}, respectively.
Therefore, in general the metric functions take the form
\bea
 \Omega(r)  \rovno \sum_{i=0}^\infty a_i\,(r-r_0)^{i} \equiv
   \frac{a_0}{1+\frac{3}{2}c_1(r-r_0)}+\sum_{i=L}^\infty \Big[a_i-a_0
 	\left(-\tfrac{3}{2}c_1\right)^i \,\Big](r-r_0)^{i}\,,\label{rozvoj[0,2]Omega}\\
 \H(r)      \rovno  (r-r_0)^2\Big[1  +c_1{(r-r_0)}+\sum_{i=L}^\infty c_i\,(r-r_0)^i\,\Big] \,.
 \label{rozvoj[0,2]}
\eea

The corresponding leading terms in the Bach and Weyl invariants \eqref{invB}, \eqref{invC} read
\bea
B_{ab}\,B^{ab} \rovno \frac{(3 L^2 + 8 L + 8)(L-1)^2(L+1)^2(L+2)^2}{72 a_0^8}\,\,\tilde{b}_L^2\,(r-r_0)^{2L}+\cdots \,,\\
C_{abcd}\,C^{abcd} \rovno \frac{16}{3 a_0^4}+\cdots\,.
\eea
Interestingly, unlike for the Schwarzschild--Bach--(A)dS black hole discussed in Sec.~\ref{SchwaBach_[n,p]=[0,1]}, for these black holes the (in general nonzero) \emph{Bach tensor vanishes on the extreme horizon} localized at $r_0$ since ${(r-r_0)^{2L} \to 0}$.

Close to this extreme horizon, that is for ${r\rightarrow r_0}$, Eqs.~\eqref{to static}, \eqref{rcehf}
imply that in the standard spherically symmetric coordinates
\be
{\bar r}\ \ \rightarrow \  a_0 \,,\qquad
h\ \ \sim \ \ ({\bar r}-a_0)^2\,,\qquad
f\ \  \sim \ \ ({\bar r}-a_0)^2\,,\label{[0,2]asymp3}
\ee
where ${a_0=\pm 1/\sqrt{\Lambda}\equiv {\bar r}_h}$ denotes the position of the extreme horizon.

\begin{figure}[t!]
	\includegraphics[scale=0.44]{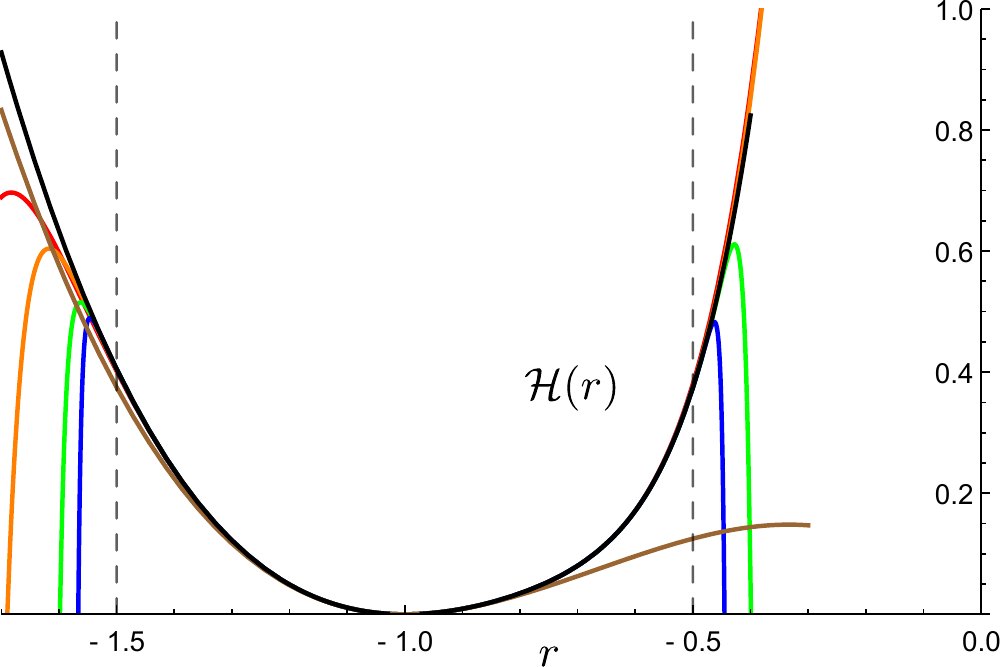}
	\hspace{8mm}
	\includegraphics[scale=0.44]{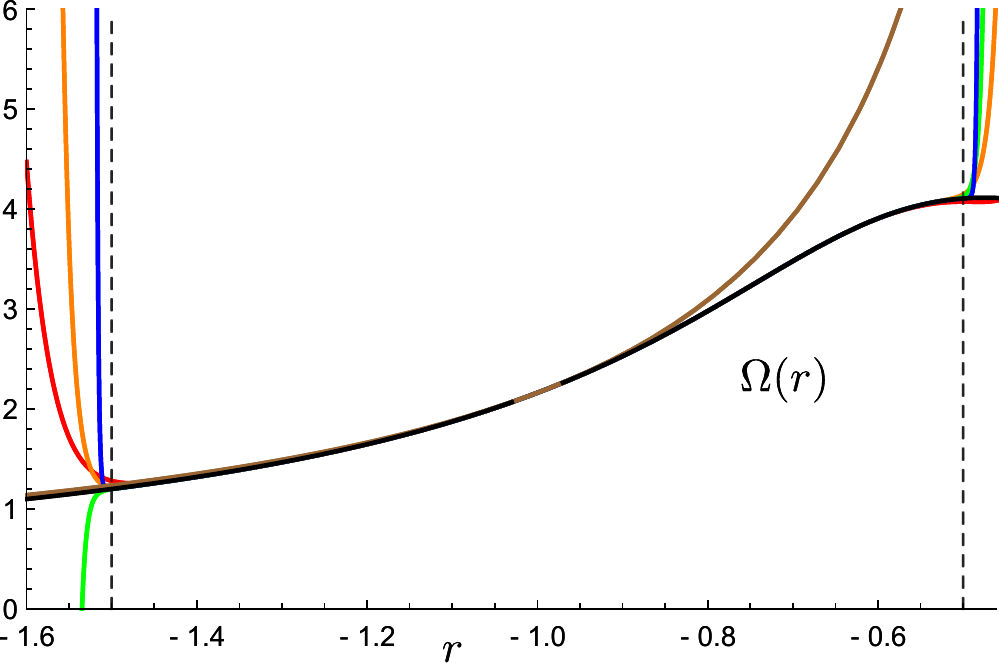}
	\caption{   The metric functions $\H(r)$ (left) and $\Omega(r)$ (right) for the extreme higher-order (discrete) Schwarzschild--Bach--de~Sitter solution $[0,2]$ with ${L=2}$, ${k=-1/2}$, ${\Lambda=3/14}$, ${c_1=-1}$, ${\tilde{b}_2=1}$, ${r_0=-1}$ and ${a_0=1/\sqrt{\Lambda}}$. First 10 (red), 20 (orange), 50 (green), and 100 (blue) terms in the expansions are	used. Furthermore, the corresponding Schwarzschild--de~Sitter solution obtained by setting ${\tilde{b}_2=0}$ is shown (brown). A numerical solution with initial values taken from the first 100 terms of the metric functions at ${r=-1.15}$ and ${r=-0.85}$ is also given (black). The numerics breaks down on the extreme horizon. Vertical dashed lines indicate the radius of the convergence obtained using the same method as in Fig.~\ref{fig_H_Omega}. 
	} \label{fig:20}
\end{figure}

These solutions exist only for  ${k<0}$ and ${\Lambda>0}$, coupled by~\eqref{[0,2]lambda}, and represent a \emph{new family of black holes} with
an extreme horizon and nonvanishing Bach tensor, with the physical Bach parameter~$\tilde{b}_L$ and two gauge parameters $c_1$, $r_0$. The  Bach tensor  vanishes only in the case of the extreme Schwarzschild--de~Sitter black hole, see~\eqref{[0,2]extSchw-OmH}, which can be obtained  by setting ${\tilde{b}_L=0}$ (in the ${L\rightarrow\infty}$ limit, the solution approaches Schwarzschild with $\Lambda=0$). Therefore, we may call these solutions \emph{extreme higher-order (discrete) Schwarzschild--Bach--de~Sitter black holes}. Typical behaviour of the metric functions is shown in Fig.~\ref{fig:20}

The first representative of this new class is given by  $L=2$. Its initial coefficients in~\eqref{rozvoj[0,2]Omega} and \eqref{rozvoj[0,2]} read
\begin{align}
&a_2-a_0\left(-\tfrac{3}{2}c_1\right)^2
=\pm 6\,\sqrt{\frac{|k|}{21}}\,\tilde b_2\,,\quad
a_3-a_0\left(-\tfrac{3}{2}c_1\right)^3
 =\pm 29\,\sqrt{\frac{|k|}{21}}\,c_1\tilde b_2\,, \dots \,,\nonumber\\
&c_2=\tilde{b}_2\,,\quad
c_3
=-2c_1\tilde b_2\,, \dots \,.
\end{align}
With the natural gauge choice \eqref{gauge[0,2]}, the explicit solution takes the form
\bea
 \Omega(r)
  \rovno -\frac{1}{r}+\tilde b_2 (r\pm\sqrt{\Lambda})^2 \bigg[ \pm 2\,\sqrt{\frac{3}{7}|k|} +\frac{116}{9}k\, (r\pm\sqrt{\Lambda})+\cdots\bigg] \,,\nonumber\\
 \H(r)
   \rovno (r\pm\sqrt{\Lambda})^2
   	\bigg[ 1\mp \frac{4}{3}\sqrt{\frac{7}{3}|k|}\,  (r\pm\sqrt{\Lambda})+\tilde b_2 (r\pm\sqrt{\Lambda})^2\bigg(1
   	\pm \frac{8}{3}\sqrt{\frac{7}{3}|k|}\,(r\pm\sqrt{\Lambda})\cdots \bigg)\bigg] \,.
 \label{rozvoj[0,2]L=2}
\eea
It reduces to the extreme Schwarzschild--de~Sitter black hole~\eqref{[0,2]extSchw-OmH} when the Bach tensor vanishes, that is for ${\tilde{b}_2=0}$.

\vspace{5mm}

\noindent
\textbf{To conclude}: The class $[0,2]$ with ${\Lambda a_0^2 =1}$ and the discrete values of ${\Lambda>0}$ given by \eqref{[0,2]lambda}, with the metric functions expressed in terms of the series~\eqref{rozvoj[0,2]Omega}, \eqref{rozvoj[0,2]} around the horizon at $r_0$, represents the extreme higher-order (discrete) Schwarzschild--Bach--de~Sitter black holes with vanishing Bach tensor on the horizon.

\vspace{5mm}

\subsubsection{Case b) ${\Lambda=\frac{3}{8k}\equiv \frac{3\gamma}{8(\alpha-3\beta)}\,}$: another extreme Bachian--dS  black hole}
\label{sec_case[0,2]b}

In a given QG theory with fixed values of the parameters $\alpha$, $\beta$, and $\gamma$, there exists a \emph{unique} value of the cosmological constant such that ${\Lambda=\frac{3}{8k}}$, that is ${\Lambda= \frac{3\gamma}{8(\alpha-3\beta)}\,}$.\footnote{In the particular theory with ${\alpha=3\beta}$, necessarily ${\gamma=0}$ but $\Lambda$ remains arbitrary.}

In this case, ${a_0, c_1}$ are free parameters. The coefficients ${c_0,a_1}$ are given by~\eqref{[0,2]c0}, while the
remaining coefficients ${c_l,a_l}$ for all ${l\geq 2}$ are again determined by the recurrent relations
\eqref{[0,2]clal} with  \eqref{[0,2]S}. The first few such coefficients explicitly read
\begin{align}
&
c_0 = \frac{3 a_0^2}{4k} -1 \,,\ \
c_2 = \frac{a_0^2 c_1^2 (3 a_0^2 - 8 k)k}{24 (3 a_0^2 - 4 k) (a_0^2 - 2 k)^2}\,,\ \
c_3 =-\frac{a_0^2 c_1^3 (7 a_0^2 - 12k)( 3a_0^2 -8k)k^2 }{60 (3 a_0^2 - 4 k)^2 (a_0^2 - 2 k)^3}\,, \dots \,, \label{[0,2]specLci}\\
&
a_1 = -\frac{a_0 c_1 k}{a_0^2 - 2 k}\,,\ \
a_2 = \frac{a_0 c_1^2 (21 a_0^2 - 32 k)k^2  }{6 (3 a_0^2 - 4 k) (a_0^2 - 2 k)^2}\,,\ \
a_3 = -\frac{a_0 c_1^3(231 a_0^4 - 724 a_0^2 k + 576 k^2) k^3 }{18 (3 a_0^2 - 4 k)^2 (a_0^2 - 2 k)^3}\,, \dots \,.\label{[0,2]specLai}
\end{align}
The metric functions thus take the form
\bea
	\Omega(r)  \rovno
	a_0\bigg[\, 1
	-\frac{c_1 k}{a_0^2 - 2 k}\,(r-r_0)
	+\Big(\frac{c_1 k}{a_0^2 - 2 k}\Big)^2\,\Big(1+\frac{3a_0^2-8k}{18a_0^2-24k}\Big)\,(r-r_0)^2+\cdots
\,,\qquad\label{rozvoj[0,2]bOmega}\\
	\H(r)      \rovno (r-r_0)^2 \bigg[\, 2 a_0^2 \Lambda - 1 + c_1(r-r_0)+	\frac{a_0^2 c_1^2 k (3 a_0^2 - 8 k)}
	 {24 (a_0^2 - 2 k)^2(3 a_0^2 - 4 k)}(r-r_0)^2+\cdots \bigg]	\,.
	\label{rozvoj[0,2]b}
\eea

Note that for a special case $a_1=0$, it follows that $c_1=0$  and subsequently $a_2$, $a_3$, $c_2$, $c_3$, $U_1$, $U_2$, $U_3$, $V_1$, $V_2$, $V_3$ also vanish (see \eqref{[0,2]specLci}, \eqref{[0,2]specLai}, \eqref{[0,2]S}). The relations \eqref{[0,2]clal} with  \eqref{[0,2]S} then determine $a_4=c_4=0$ which implies $U_4=V_4=0$, etc. Thus, all $a_i=0$ and $c_i=0$, $i\geq 1$, and the metric functions reduce to   $\Omega=a_0$ and
	$\H=(r-r_0)^2 ( 2\Lambda a_0^2 - 1)$. This  Kundt metric is a Bachian generalization of the Nariai spacetime \eqref{Nariaimetric}, cf. \eqref{anothergenNariaimetric}, with an extreme horizon.

 In the limit ${r\rightarrow r_0}$,  the relations~\eqref{[0,2]asymp3} hold, and the values of the Bach and Weyl invariants \eqref{invB}, \eqref{invC} on the horizon are
\be
{B}_{ab}\,{B}^{ab}(r_0) = {\frac{1}{4 k^2}\Big(\,\frac{3}{8 k}-\frac{1}{a_0^2}\,\Big)^2}\,,\qquad
{C}_{abcd}\,{C}^{abcd}(r_0) = \frac{3}{4 k^2}\,.
\ee
Therefore, the \emph{Bach tensor is nonvanishing on the horizon}, unless ${a_0^2= \tfrac{8}{3}k}$.

The \emph{extreme Schwarzschild--de~Sitter black hole} \eqref{[0,2]extSchw-OmH} is recovered for ${B_{ab}\,B^{ab}=0}$, i.e., for
\be
a_0^2=\frac{8k}{3}=\frac{1}{\Lambda}>0\,.
\ee
Thus ${a_0^2\Lambda=1}$ as in  Case a), cf.~\eqref{[0,2]1}. Moreover, ${c_0=1}$, ${c_i=0}$ for all ${i\geq 2}$, and ${a_i=(-\tfrac{3}{2}c_1)^i a_0}$, so that Eqs.~\eqref{[0,2]extSchw-ca} hold, leading to the solution~\eqref{[0,2]extSchw-OmH}.

In general, it is natural to introduce a dimensionless Bach parameter $b_e$ by
\be
b_e \equiv \frac{3 a_0^2}{8 k}-1\,,
\qquad\hbox{so that}\quad a_0^2=\frac{8k}{3}(b_e+1)
\quad\hbox{and}\quad
{B}_{ab}\,{B}^{ab}(r_0) = \bigg(\frac{3}{16 k^2}\frac{b_e}{b_e+1}\bigg)^2 \,, \label{be_definice}
\ee
which vanishes for the extreme Schwarzschild--de~Sitter black hole. The metric function $\Omega$ can then be rewritten as
\be
	\Omega(r)  = a_0 \,\sum_{i=0}^\infty  \,\Big(-\frac{c_1 k}{a_0^2 - 2 k}\,(r-r_0)\Big)^i
  +\Big(\frac{c_1 k}{a_0^2 - 2 k}\Big)^2\,\frac{8ka_0}{18a_0^2-24k}\,b_e\,(r-r_0)^2+\cdots
\,.\qquad\label{rozvoj[0,2]bOmegabe}
\ee
The first term in $\Omega(r)$ is a geometric series which can by summed up to
\be
	 a_0 \,\sum_{i=0}^\infty  \,\Big(-\frac{c_1 k}{a_0^2 - 2 k}\,(r-r_0)\Big)^i = \frac{a_0(a_0^2 - 2 k)}{(a_0^2 - 2 k -c_1 k\,r_0)+c_1 k\,r}= -\frac{1}{r}\,,
\label{rozvoj[0,2]bOmegasum}
\ee
when the unique gauge
\be
a_0=-\frac{1}{r_0}\,,\qquad
c_1=\frac{1-2kr_0^2}{kr_0^3}
\ee
is used, so that
\be
 b_e = \frac{3}{8 k r_0^2}-1  \equiv \frac{\Lambda}{r_0^2}-1\,.\label{[0,2]be}
\ee
The explicit solution thus reads
\bea
\Omega(r) \rovno -\frac{1}{r}-\frac{4k\,b_e}{3r_0(3-4kr_0^2)}(r-r_0)^2  + \dots \,, \label{rozvoj[0,2]bOmegaFinal}\\
\H(r)\rovno (r-r_0)^2 \bigg[ (1+2b_e)+\frac{1}{3r_0} (2+8b_e)(r-r_0)+
\frac{b_e}{3r_0^2(3-4kr_0^2)}(r-r_0)^2 + \dots \bigg] 	\label{rozvoj[0,2]bFinal}\,.
\eea
Note that this solution has only one free parameter since $r_0$ and $b_e$ are related via \eqref{[0,2]be}.

This class of spacetimes describes a black hole with the (double degenerate) \emph{extreme horizon located at $r=r_0$ and a nonvanishing Bach tensor}. Its value on the horizon is given by the invariant (\ref{be_definice}),
	\be
	{B}_{ab}\,{B}^{ab}(r_0) = \Big(\frac{r_0^2}{2k}\,b_e\Big)^2 \,.
	\label{BCInvariants_[0,2]final}
	\ee
For ${b_e=0}$, i.e.,  ${r_0=\mp\sqrt{\Lambda}\equiv\mp\sqrt{\frac{3}{8k}}}$, this solution reduces to the extreme Schwarzschild--de~Sitter solution (\ref{[0,2]extSchw-OmH}).

\vspace{5mm}

\noindent
\textbf{To conclude}: The class $[0,2]$ with $\Lambda=3/(8k)$, expressed in terms of the series \eqref{rozvoj[0,2]bOmegaFinal},	 \eqref{rozvoj[0,2]bFinal} around the double degenerate horizon at $r_0$, represents  another extreme  Bachian--de~Sitter  black hole generalizing the extreme Schwarzschild--dS black hole with nonvanishing Bach tensor on the horizon.

\vspace{5mm}


\subsection{Higher-order Schwarzschild--Bach--(A)dS black holes  in the class ${[-1,0]}$}
\label{IIIa}

We will now prove that for a \emph{generic} value of the cosmological constant $\Lambda$, the class
${[n,p]=[-1, 0]}$ admits only the Schwarzschild--(A)dS solution. However, for \emph{specific} values
${ \Lambda=-\frac{3}{2k(L+3)(L+2)}}$, where ${L\in\mathbb{N}}$, ${L\geq0}$, it represents    ``higher--order discrete Schwarzschild--Bach--(A)dS black holes''.

\subsubsection{Uniqueness of the Schwarzschild--(A)dS black hole for a generic value of $\Lambda$}
\label{IIIa1}

For ${l\geq 0}$, Eq.
(\ref{KeyEq1}) gives
\begin{align}
\tfrac{1}{3}kc_{l+4}(l+4)(l+3)(l+2)(l+1)=(l+4)(l+5)a_0 a_{l+4}+\sum^{l+3}_{i=1}a_i a_{l-i+4}(l-i+3)(l-3i+4) \,,\label{IIIa_rr}
\end{align}
and
\begin{align}
a_1=a_2=a_3=0\,.
\end{align}
Eq.~(\ref{KeyEq3})  specifies
\begin{equation}
c_0 =\tfrac{1}{3}\Lambda a_0^2\,,\ \ c_1=0\,,\ \ c_2=-1\,,
\end{equation}
and
\begin{align}
&a_{l+4}[c_0(l+2)(l+3)-2\Lambda a_0^2]+\tfrac{1}{6}a_0c_{l+4}l(l+1)
+\tfrac{1}{3}a_{l+2} +\sum^{l+3}_{i=1}c_i a_{l+4-i}[(l+3-i)(l+2)+\tfrac{1}{6}i(i-1)]\nonumber\\
& =\tfrac{2}{3}\Lambda\bigg[\sum^{l+3}_{i=1}a_0a_ia_{l+4-i}
+\sum^{l+3}_{j=1}\sum^{j}_{i=0}a_ia_{j-i}a_{l+4-j}\bigg]\,.\label{IIIa_trace}
\end{align}
Finally, Eq.
(\ref{KeyEq2}) implies
\begin{align}
&a_0^2c_{l+4}(l+1) +2\Lambda a_0^3 a_{l+4}(l+5)=\sum^{l+3}_{j=1}\sum^{j}_{i=0}a_i a_{j-i}c_{l-j+4}(j-i-1)(l-j+3i+1)\nonumber\\
& +\sum^{l+2}_{i=0}a_ia_{l-i+2}+3c_0\sum^{l+3}_{i=1}a_ia_{l+4-i}(l+3-i)(i-1)
-\tfrac{1}{3}k\sum^{l+3}_{i=1}c_ic_{l-i+4}i(l+4-i)(l+3-i)\left(l+\tfrac{1}{2}(5-3i)\right)
\nonumber\\
& -\Lambda
\bigg[2a_0^2\sum^{l+3}_{i=1}a_ia_{l-i+4}+\sum^{l+3}_{m=1}
\Big(\sum^{m}_{i=0}a_ia_{m-i}\Big)\Big(\sum^{l-m+4}_{j=0}a_ja_{l-m-j+4}\Big)\bigg]
\,, \
\ \ l\geq 1\,. \label{IIIa_ru}
\end{align}

Therefore, the functions $\Omega(r)$ and $\H (r)$  have the form
\be
\Omega = \frac{a_0}{\Delta}+a_{\ell+1}\Delta^{\ell}+\cdots \,,\qquad
\H  = \frac{\Lambda}{3}a_0^2-\Delta^2 +c_{3}\,\Delta^{3}+c_{\k}\,\Delta^{\k}+\cdots \,,
\ee
where ${\ell\equiv l+3 \geq 3}$, ${\k\geq 4}$.

Assuming that there exists $l$ such that $a_i=0$ for all $1\leq i\leq l+3$ and $c_j=0$ for all $4\leq j\leq l+3$ (the following equations hold also for $l=0$),
equations \eqref{IIIa_rr} and \eqref{IIIa_ru} simplify to
\bea
2\Lambda(l+5)\, a_0\, a_{l+4} \rovno -(l+1)\,c_{l+4}\,, \label{caseIIIaeq2}\\
(l+5)\,a_0\,a_{l+4} \rovno {\textstyle\frac{1}{3}} k\,(l+3) (l+2)(l+1)\,c_{l+4}\,.\label{caseIIIaeq1}
\eea
For \emph{generic values} of $\Lambda$ (i.e., other than those given by condition \eqref{caseIIIaL}), necessarily
${a_{l+4}=0=c_{l+4}}$ and thus ${a_i=0=c_{i+3}}$ \emph{for all} ${i\geq 1}$ (otherwise, the two equations \eqref{caseIIIaeq2}, \eqref{caseIIIaeq1} are not compatible).
Thus, for a generic~$\Lambda$, the  only  free parameters are $a_0$ and $c_3$, and the metric functions become
\be
\Omega=\frac{a_0}{\Delta}\,,\qquad
\H =\frac{\Lambda}{3}a_0^2-\Delta^2+c_{3}\,\Delta^{3}\,.\label{metrH[-1,0]}
\ee
Using the remaining coordinate freedom, we can set ${a_0=-1}$ and ${r_0=0}$ (implying ${\Delta=r}$), which
gives the Schwarzschild--(A)dS solution \eqref{SchwAdS} with the identification ${c_3=-2m}$.

\vspace{5mm}

\noindent
\textbf{To conclude}: The class $[-1,0]$ with an arbitrary value of $\Lambda$, expressed in terms of the metric functions \eqref{metrH[-1,0]}, represents the spherically symmetric Schwarzschild--(A)dS black hole.

\vspace{5mm}

\subsubsection{Higher-order Schwarzschild--Bach--(A)dS black holes for special values of $\Lambda$ }
\label{IIIa2}

Interestingly, for \emph{ special} (discrete) values of the cosmological constant~$\Lambda$ given by
\be
\Lambda=-\frac{3}{2k(L+2)(L+3)}\,,\qquad\hbox{where}\quad
L\in\mathbb{N}_0
\label{caseIIIaL}
\ee
or equivalently, using \eqref{fieldeqsEWmod}, by
\be
\Lambda=-\frac{3\gamma}
{24\beta+2\alpha(L+2)(L+3)}\,,
\ee
the system of field equations \eqref{caseIIIaeq2}, \eqref{caseIIIaeq1} also admits another class of solutions in the form
\be
a_{L+4}=k\,\frac{(L+1) (L+2)(L+3)}{3(L+5)\,a_0}\,b_L \,, \qquad L\in\mathbb{N}_0 
\ee
where
\be
b_L\equiv c_{L+4}
\ee
is \emph{an additional free ``Bach'' parameter}. All the other coefficients for ${l>L}$  are determined by \eqref{IIIa_rr} and \eqref{IIIa_trace} as
\be
\tfrac{1}{3}kc_{l+4}(l+4)(l+3)(l+2)(l+1)=(l+4)(l+5)a_0 a_{l+4}+\sum^{l+3}_{i=1}a_i a_{l-i+4}(l-i+3)(l-3i+4) \,,
\label{[-1,0]_cl+4}
\ee
and
\begin{align}
a_{l+4}&\frac{a_0^2l(l+5)}{6(l+2)(l+3)}\Big[2\Lambda(l+2)(l+3)+\frac{3}{k}\,\Big]
\nonumber\\
&=-\frac{a_0 l}{2k(l+4)(l+3)(l+2)}
\sum^{l+3}_{i=1}a_i a_{l-i+4}(l-i+3)(l-3i+4)-\tfrac{1}{3}a_{l+2}
\nonumber\\
&
-\sum^{l+3}_{i=1}c_i a_{l+4-i}[(l+3-i)(l+2)+\tfrac{1}{6}i(i-1)]
+\tfrac{2}{3}\Lambda\bigg[
\sum^{l+3}_{i=1}a_0a_ia_{l+4-i}
+\sum^{l+3}_{j=1}\sum^{j}_{i=0}a_ia_{j-i}a_{l+4-j}\bigg],
\label{[-1,0]_al+4}
\end{align}
where $\Lambda$ is given by (\ref{caseIIIaL}),
so that \eqref{[-1,0]_al+4} gives $a_{l+4}$ and then \eqref{[-1,0]_cl+4} $c_{l+4}$.

In the limit ${\Delta\to 0}$, the dominant terms of the metric functions become
${\Omega=a_0\,\Delta^{-1}+\cdots}$ and ${\H=c_0+\cdots=\tfrac{1}{3}\Lambda a_0^2+\cdots}$, similarly as in~\eqref{SchwAdS}, in which case  the relations
\eqref{to static}, \eqref{rcehf}
give
\bea
&& {\bar r}=\frac{a_0}{\Delta}\ \rightarrow\ \infty
\qquad (\hbox{so that}\
r\ \to\  \frac{a_0}{\bar r}+r_0)\,,\\
&& h \ \to\ -\frac{\Lambda}{3}a_0^2\,{\bar r}^2\,,\qquad
f \ \to\ -\frac{\Lambda}{3}\,{\bar r}^2\,.
\eea
With the natural gauge choice ${a_0=-1}$, this is exactly the asymptotic behavior of the
Schwarzschild--(A)dS solution in the canonical form \eqref{Einstein-WeylBH}, \eqref{SchwarzschildAdSBH} with
${\,h=f=1-2m/\bar{r}-\tfrac{\Lambda }{3}\,\bar{r}^2\,}$ as ${{\bar{r}\to\infty}}$.
Therefore, \emph{these solutions asymptotically approach the Schwarzschild--(A)dS solution}.

In this case, the Bach and Weyl invariants \eqref{invB}  and \eqref{invC} read
\bea
{ B}_{ab}\,{B}^{ab} \rovno\frac{ [\,(L+1)(L+2)(L+3)(L+4)\,\Lambda\,]^2}{216 a_0^4} \,
b_L^ 2 \,\Delta^{2L+8} +\cdots\,,
\label{Bachinv1}
\\
{ C}_{abcd}\, {C}^{abcd} \rovno \frac{12}{{a_0}^4 }\,c_3^2\,\Delta^6 +\cdots
\,,\label{Weylinv1}
\eea
respectively. For  ${b_L \not=0\not = c_3}$, they are \emph{both nonvanishing}, but
for large values of ${\bar r}$, the invariants \emph{approach zero asymptotically} for  ${\bar r}\rightarrow \infty$ (that is for ${\Delta\rightarrow 0}$).

Note that the new free (non-Schwarzschild)  \emph{Bach parameter} $b_L$  can be chosen to be \emph{arbitrarily small} and thus (assuming analyticity) these solutions include arbitrarily small perturbations of the Schwarzschild--(A)dS solution.

\begin{figure}[t!]
		\includegraphics[scale=0.44]{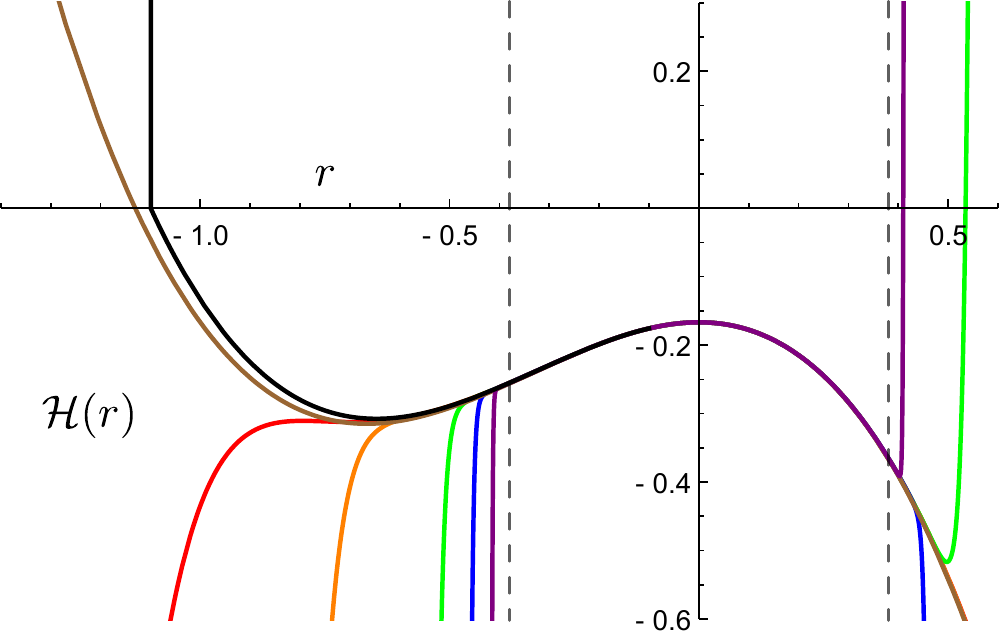}
		\hspace{8mm}
		\includegraphics[scale=0.44]{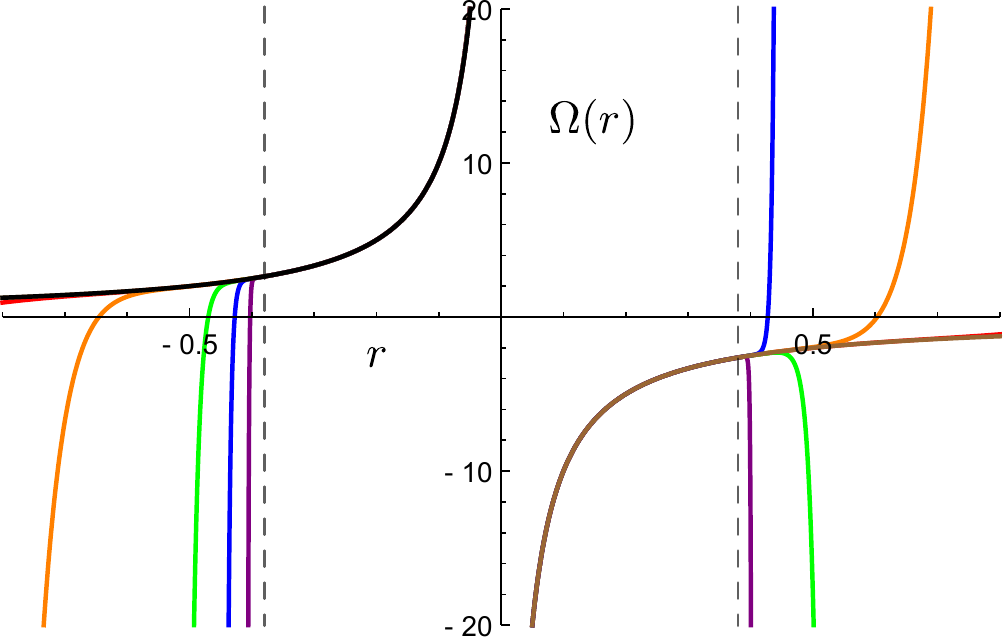}
		\caption{   The metric functions $\H(r)$ (left) and $\Omega(r)$ (right) for the non-Schwarzschild  (Bachian) solution $[-1,0]$ with the parameters ${L=0}$, ${k=1/2}$, ${\Lambda=-1/2}$, ${r_0=0}$, ${a_0=-1}$, ${c_3=-1}$ and ${b_0\equiv c_4=1/20}$. First 10 (red), 20 (orange), 50 (green), 100 (blue), and 300 (purple) terms in the expansions have been used. Furthermore, we plot the corresponding Schwarzschild--AdS solution obtained by setting ${b_0\equiv c_4=0}$ (brown). A numerical solution with initial values taken from the first 100 terms of the metric functions at $r=-0.2$ is also plotted (black). Note that the numerics breaks down on the horizon given by ${\H=0}$. Vertical dashed lines indicate the	radius of the convergence. From the behaviour of the metric function $\Omega$ we can read off that, in the standard spherical coordinates $\bar r$, the metric functions converge from ${\bar r\gtrapprox 2.63}$. (In contrast with the $[0,1]$ case, the series in the $[-1,0]$ class do not approach geometric series asymptotically. Thus we have estimated the radius of convergence by searching for which values of $r$ these series can be bounded by appropriately chosen geometric series.)  All the plotted functions visually overlap within the radius of convergence, validating the results.} \label{fig:2}
	\end{figure}

The presence of a horizon is indicated by numerical calculations shown in Fig.~\ref{fig:2} (however, note that the numerical calculation breaks down at the horizon).
Recall that the condition ${\H=0}$ implies ${f(\bar r)=0=h(\bar r)}$, see \eqref{rcehf}.
This  suggests that, at least for some values of $b_L$,
these solutions represent a \emph{new family of black holes with}~$\Lambda$.
Such solutions can naturally be considered as a generalization of the Schwarzschild--(A)dS family
since, in addition to mass encoded by the parameter $c_3$ and a cosmological constant~$\Lambda$,
they contain further physical/geometrical parameter~$b_L$.
With this parameter, the Bach tensor $B_{ab}$ becomes non-zero, see \eqref{Bachinv1} with \eqref{Bach=0iffINV=0}, and due to \eqref{fieldeqsEWmod}, the Ricci tensor $ R_{ab}$ is also non-trivial.
Since they are vacuum solutions, the Birkhoff theorem is clearly violated in QG.
Such solutions are not possible in the Einstein theory since ${k=0}$ formally corresponds to infinite value of $\Lambda$ in \eqref{caseIIIaL}.

For a general value of integer $L$ (${L\in\mathbb{N}_0}$), 
the metric functions read
\bea
\Omega(r)\rovno\frac{a_0}{\Delta}+a_{L+4}\,\Delta^{L+3}+\cdots \,,\label{metO2[-1,0]}\\
\H (r)\rovno\frac{\Lambda}{3}\,a_0^2-\Delta^2 +c_{3}\,\Delta^{3}+c_{L+4}\,\Delta^{L+4}+\cdots
\,.\label{metH2[-1,0]}
\eea
Let us present additional terms in the following special cases for the lowest orders of $L$:
\begin{itemize}	
\item
	For ${L=0}$, Eq.~\eqref{caseIIIaL} gives ${\displaystyle \Lambda=-\frac{1}{4k}}$.
    Then, the first coefficients in the  series
	\eqref{rozvojomeg0} and \eqref{rozvojcalH0} are
    \begin{align}
	&a_1=a_2=a_3=0\,,\quad
	a_4=\frac{2k}{5a_0}\,b_0\,,\quad
	a_5=0\,,\quad
	a_6=-\frac{288k^2}{49a_0^3}\,b_0\,,\quad
	a_7=\frac{4k^2}{a_0^3}\,c_3b_0\,,\dots\,,\\
	&c_0=\frac{\Lambda}{3}\,a_0^2\,,\quad
	c_1=0\,,\quad
	c_2=-1\,,\quad
	c_5=0\,,\quad
	c_6=-\frac{72k}{35a_0^2}\,b_0\,,\quad
	c_7=\frac{4k}{5a_0^2}\,c_3b_0\,,\dots\,,
	\end{align}
	with the free parameters $a_0$, $c_3$, and ${b_0\equiv c_4}$. Choosing ${a_0=-1}$, ${r_0=0}$ and denoting ${c_3=-2m}$, the metric functions become
    \be
    \Omega=-\frac{1}{r}-\frac{2}{5}k\,b_0\,r^{3}+\cdots\,,\qquad
    \H =\frac{\Lambda}{3}-r^2 -2m\,r^{3}+b_0\,r^{4}+\cdots\,,
    \ee
    which reduces to the Schwarzschild--(A)dS solution \eqref{SchwAdS} when ${b_0=0}$.
\item
	For ${L=1}$, equation~\eqref{caseIIIaL} gives ${\displaystyle \Lambda=-\frac{1}{8k}}$.
    The coefficients are
    \begin{align}
	&a_1=a_2=a_3=a_4=0\,,\quad
	a_5=\frac{4k}{3a_0}\,b_1\,,\quad
	a_6=0\,, \quad
	a_7= -\frac{400 k^2}{9 a_0^3}\, b_1\,,\dots\,,\\
	&c_0=\frac{\Lambda}{3}\,a_0^2\,,\quad
	c_1=0\,,\quad
	c_2=-1\,,\quad
	c_4=0\,,\quad
	c_6=0\,,\quad
	c_7= -\frac{80k}{9 a_0^2}\,b_1 \,,\dots\,,
    \end{align}
	with free parameters $a_0$, $c_3$, and ${b_1 \equiv c_5}$. Choosing ${a_0=-1}$, ${r_0=0}$ and denoting ${c_3=-2m}$,
    \be
    \Omega=-\frac{1}{r}-\frac{4}{3}k\,b_1\,r^{4}+\cdots\,,\qquad
    \H =\frac{\Lambda}{3}-r^2 -2m\,r^{3}+b_1\,r^{5}+\cdots\,,
    \ee
    which also reduces to the Schwarzschild--(A)dS solution when ${b_1=0}$.
	\item
	Explicit expressions for ${L\geq 2}$ can be obtained analogously. Notice that as ${L \rightarrow\infty}$,
    the	cosmological constant ${\Lambda\to 0}$, and the solution  approaches the Schwarzschild black hole metric.
\end{itemize}

\vspace{5mm}

\noindent
\textbf{To conclude}: The class $[-1,0]$ with the discrete values of $\Lambda$ given by \eqref{caseIIIaL}, with the metric functions expressed in terms of the series \eqref{metO2[-1,0]} and \eqref{metH2[-1,0]} around an arbitrary point $r_0$ corresponding to the asymptotic physical region ${\bar r \rightarrow \infty}$, represents the family of spherically symmetric higher-order discrete Schwarzschild--Bach--(A)dS black holes with an additional Bach parameter $b_L$ (where ${L=0, 1, 2, \dots}$).

\vspace{5mm}


\subsection{Bachian singularity in the class ${[n,p]=[1,0]}$ }
\label{SchwaBach_[n,p]=[1,0]}

Now let us investigate the case ${[1,0]}$. Since the Bach tensor is always nonvanishing (see below), this class does not contain the Schwarzschild--(A)dS as a special case. Moreover, it possesses a curvature
singularity in both the Bach and Weyl tensors at $r=r_0$ corresponding to ${\bar r =0}$. Thus it can be nicknamed as \emph{Bachian singularity}.

After relabeling ${l \to l-3}$, Eq. \eqref{KeyEq1} for the case ${[n,p]=[1,0]}$ yields
\be
c_{l+1}=\frac{3}{k\,(l+1)l(l-1)(l-2)}\,\sum^{l-3}_{i=0}
a_i \,a_{l-3-i}(l-2-i)(l-5-3i)\,,  \qquad \forall\ l\ge 3\,.
\label{(2,2)initcondc} \ee
The lowest order ${l=0}$ of  \eqref{KeyEq3} gives
\be
a_1=-\frac{a_0c_1}{2c_0}\,,
\label{(2,2)initcond3}
\ee
while higher orders imply for ${l\geq 1}$ (the first sum is empty for ${l=1,2}$)
\begin{align}
a_{l+1}\!=\!\frac{1}{(l+1)(l+2)\,c_0}
\bigg[\tfrac{2}{3}\Lambda \sum^{l-3}_{j=0}\sum^{j}_{i=0}a_ia_{j-i}a_{l-j-3}-\tfrac{1}{3}a_{l-1}
-\sum^{l+1}_{i=1}c_ia_{l-i+1}\big[(l-i+2)(l+1)+\tfrac{1}{6}i(i-1)\big] \bigg].
\label{(2,2)initconda}
\end{align}
Finally, the lowest
 order ${l=0}$ of  \eqref{KeyEq2} leads to
\be
c_3=\frac{1}{6kc_1}\big[9a_0^2c_0+2k(c_2^2-1)\big]\,.
\label{(2,2)initcond2}
\ee
All the coefficients $a_{l+1}$, $c_{l+1}$ can be determined from the recurrent relations \eqref{(2,2)initcondc},
\eqref{(2,2)initconda}, which give
\be
\Omega(r)  = (r-r_0)\Big[a_0 + \sum_{i=1}^\infty a_i \,(r-r_0)^{i}\Big]\,,\qquad
\H(r)      = c_0 + \sum_{i=1}^\infty c_i \,(r-r_0)^{i}\,,
\label{rozvoj[1,0]}
\ee
where  the coefficients $a_1, c_3$ are given  by \eqref{(2,2)initcond2}
and \eqref{(2,2)initcond3}, respectively, while
\bea
a_2 \rovno -\frac{a_0}{18c_0^2}\big[c_0(1+7c_2)-6c_1^2\big]\,, \nonumber\\
a_3 \rovno -\frac{a_0 }{36 k c_0^3 c_1}\big[18 a_0^2 c_0^3 + k[4 c_0^2 (c_2^2-1) - 2 c_0 c_1^2 (1 + 10 c_2)
+9 c_1^4]\big]\,, \ldots \,,\nonumber\\
c_4 \rovno -\frac{a_0^2}{4 k}\,, \qquad
c_5 = \frac{3 a_0^2 c_1}{40 k c_0}\,, \ldots\,.
\label{coef_[1,0]}
\eea
Here ${a_0,\, c_0,\, c_1,\, c_2}$ are four free parameters (using the gauge freedom \eqref{scalingfreedom}, one can fix some of these parameters, e.g., ${a_0=1}$ and ${r_0=0}$).
The coefficients \eqref{coef_[1,0]}
coincide with those of \cite{PodolskySvarcPravdaPravdova:2020}  since the cosmological constant~$\Lambda$ enters only the coefficients $a_i$ with ${i\geq 4}$ and $c_i$ with ${i\geq 9}$.

From the scalar invariants \eqref{invB}, \eqref{invC},
\be
B_{ab}\, B^{ab}(r)  = \frac{3 c_0^2}{4 a_0^4 k^2} \frac{1}{(r-r_0)^8}+\ldots \,,\qquad
C_{abcd}\, C^{abcd}(r) = \frac{4}{3 a_0^4}\frac{(1 + c_2)^2}{(r-r_0)^4}+\ldots \,,\label{(2,2)BInv2}
\ee
 it follows that the Bach tensor $B_{ab}$ cannot be set to zero since, by definition,  ${c_0\not= 0}$. Consequently, the Schwarzschild--(A)dS solution does not belong to this class and  the Bach invariant always diverges at ${r=r_0}$. Notice that there is also a Weyl curvature singularity at ${r=r_0}$ (in the special case  ${c_2 =-1}$, ${C_{abcd}\, C^{abcd} \propto (r-r_0)^{-2}}$).

For \eqref{rozvoj[1,0]}, the limit ${r\rightarrow r_0}$  implies
\be
 {\bar r}=\Omega(r) \to 0\,\label{(2,2)horizon}\,,
\ee
and thus the Bach/Weyl curvature \emph{singularity is located at the origin}, which is also indicated by the behaviour of the
 metric functions  \eqref{to static}, \eqref{rcehf}  in terms of the physical radial coordinate $\bar r$,
\be
h \sim  -c_0\,{\bar r}^2 \to   0\,, \qquad
f \sim  -a_0^2c_0\, ({\bar r})^{-2} \to   \infty\,. \label{(2,2)horizonC}
\ee
Notice also that in the  ${\Lambda =0}$ case, this ${[n,p]=[1,0]}$ class of solutions corresponds to the ${(s,t)=(2,2)}$ family  \cite{Stelle:1978,LuPerkinsPopeStelle:2015b, HoldomRen:2017,PerkinsPhD,PodolskySvarcPravdaPravdova:2020}.

\vspace{5mm}

\noindent
\textbf{To conclude}: The class $[1,0]$, with the metric functions expressed in terms of the series \eqref{rozvoj[1,0]}  around an arbitrary point $r_0$ corresponding to the physical/geometrical origin ${\bar r =0}$, represents a spherically symmetric spacetime with Bachian singularity.

\vspace{5mm}


\subsection{Empty class ${[n,p]=[0,>2]}$}
\label{sec_[0,>2]}

In what follows we show that, in fact, this class is empty.

For ${p\not\in \mathbb{N}}$, Eq.~(\ref{KeyEq1}) implies $c_0=0$, which must be nonzero by definition.
For all integers ${p>2}$ at the order $\Delta^0$, Eqs.~\eqref{KeyEq2} and~\eqref{KeyEq3}
give the conditions ${3 a_0^2(1  -  \Lambda a_0^2) = 2 k}$ and $ 2\Lambda a_0^2 =1$, respectively, which together imply
\be
\Lambda=\frac{3}{8k}\,,\qquad  a_0=\pm\sqrt{\frac{4k}{3}}\,. \label{Iaa0}
\ee

First, let us discuss the case $[0,3]$. At the order $\Delta^0$, Eq.~\eqref{KeyEq1}  gives
${a_2 = (a_1^2 + 4 k c_1)/a_0}$.
At the order $\Delta^1$, Eq.~(\ref{KeyEq3}) gives ${a_1=\tfrac{3}{2}a_0 c_0 }$, while Eq.~(\ref{KeyEq1})
gives ${a_3=27 a_0 c_0^3/8 + 6 k c_0 c_1/a_0  + 20k c_2 /(3a_0)}$.
Finally, at the order $\Delta^2$, Eq.~(\ref{KeyEq3}) implies
${c_0=0}$, and thus this case cannot occur.

Let us show that this is also the case for all integers ${p>3}$.
For  ${p>3}$ at the order $\Delta^1$,  Eq.~(\ref{KeyEq3}) implies ${a_1=0}$. Using the mathematical induction, we show that all $a_i$ for ${1\leq i\leq p-1}$ vanish.
So let us assume that all $a_i$ for ${1\leq i\leq j-1}$, ${j< p-2}$, vanish.
Then
Eq. (\ref{KeyEq1})  at the order $\Delta^{j-2}$  gives ${a_0 a_{j}(j-1)=0}$ which implies that also ${a_j=0}$.
 Thus the second nonvanishing coefficient (after ${a_0}$) is $a_{p-2}$ which is determined by the order $\Delta^{p-4}$ of Eq. (\ref{KeyEq1}) as
\be
a_{p-2}=kp(p-1)\frac{c_0}{3a_0}\,.\label{Iaapm2}
\ee
Then using \eqref{Iaa0}, Eqs.~(\ref{KeyEq1}) and~(\ref{KeyEq3})
at the orders $\Delta^{2p-6}$ and $\Delta^{2p-4}$  yield
\begin{align}
2(2p-5)a_0 a_{2p-4}
-(p-1)a_{p-2}^2
-\tfrac{4}{3}kc_{p-2}(p-1)(2p-3)(2p-5)&= 0\,,\\
-2a_{2p-4}
+c_0 a_{p-2}\left[3(p-2)(2p-3)+\tfrac{1}{2}p(p-1)\right]
-6\Lambda a_0 a_{p-2}^2
+a_0 c_{p-2}(p-1)(2p-3) &= 0\,,
\end{align}
respectively. A linear combination of these two equations, using also~\eqref{Iaa0} and~\eqref{Iaapm2}, leads to
\be
p(p-1) (p-2)(5p^2-21p+20) c_0^2=0\,.
\ee
This, for integers ${p>3}$, implies ${c_0=0}$ which is a contradiction.

\vspace{5mm}

\noindent
\textbf{To conclude}: The class $[0,>2]$ is empty and thus, under our assumptions, there are no black hole solutions to QG field equations with  a higher-than-doubly degenerate horizon.

\vspace{5mm}


\subsection{Solutions with regular Bachian infinity in the class ${[n,p]=[<0,2n+2]}$}
\label{sec_[n,2n+2]}

Finally, we will show that there are \emph{infinitely many vacuum solutions} in the class ${[n,p]=[<0,2n+2]}$, namely ${[n,p]=[-J/2,2-J]}$, where
${J\in \mathbb{N}}$, ${J>2}$.
Each of these solutions contains an asymptotic region ${{\bar r}\rightarrow\infty}$ where both the Bach and Weyl invariants approach a nontrivial constant. Therefore, these solutions \emph{do not admit the classical Schwarzschild--(A)dS limit}.

In Eq.~(\ref{KeyEq3}), the first and last terms start with the power ${\Delta^{3n}}$, while the second one with  ${\tfrac{1}{3}a_0\Delta^{n}}$. Since in all sums there are integer steps,  to allow ${a_0\not= 0}$, the expression ${3n-n}$ has to be a (negative) integer.
Thus,
\be
	{n=-J/2}\,, \ \ \mbox{where}\ \
	{J\in \mathbb{N} }\,,\ \
	{J\geq 3}\,,
\ee
and the coefficient $p$ is ${p=2n+2=-J+2}$. Then from Eq.~\eqref{CaseIII_summary3} we obtain a \emph{unique value of the cosmological constant},
\be
\Lambda = \frac{3}{32k}\frac{11J^2-12J+4}{1-J^2}\,,\quad
|\Lambda|\geq \frac{201}{256 |k|}\,,\qquad\hbox{and}\quad
c_0 = \frac{3a_0^2}{4k(1-J^2)}\,. \label{[mJpul,]L}
\ee
The subleading terms in Eqs.~(\ref{KeyEq1}) and~(\ref{KeyEq3}) give
\bea
c_1\rovno-\frac{3a_0 a_1 (n+2)}{2k(n+1)(2n+1)(2n+3)}\,,
\nonumber\\
0\rovno
c_0 a_1 (11 n^2 +18n+7) +a_0 c_1
(11n^2+11n+3)-6\Lambda a_0^2 a_1\,,
\eea
respectively. Their combination yields
\be
a_1\,(55 J^4-335 J^3+554 J^2-412J +120)=0\,,
\ee
which cannot be fulfilled for ${J\in \mathbb{N} }$, ${J>2}$ unless
\be
a_1=0=c_1\,.
\ee

As an illustration of such solutions, let us present two explicit examples. The highest possible value of $n$ is obtained for ${J=3}$, namely ${n=-3/2}$,  ${p=-1}$. For ${J=4}$, we obtain ${n=-2}$, ${p=-2}$:

\begin{itemize}	
\item In the case ${[-3/2,-1]}$ we get
\begin{align}
& \Lambda = -\frac{201}{256 k}\,,\\
& a_1 = a_2=a_3=a_4=0\,,\ \ a_5=\frac{16 k c_5}{67 a_0}\,,\ \ a_6=\frac{43520k^2}{117117 a_0^3}\,,\ \
a_7=0\,, \dots \,,\\
& c_0 = -\frac{3 a_0^2}{32 k}\,,\ \
c_1=c_2=0\,, \ \ c_3=-\frac{4}{13}\,,\ \ c_4=0\,,\ \
c_6=\frac{16864 k}{39039 a_0^2}\,,  \ \ c_7=0\,, \dots \,,
\end{align}
and thus
\begin{align}
\Omega(r) & = a_0 (r-r_0)^{-3/2}+
	\frac{16 k c_5}{67 a_0}(r-r_0)^{7/2}
	+\frac{43520k^2}{117117 a_0^3}(r-r_0)^{9/2}+ \cdots\,,
	\label{[mJ]O}\\
\H(r)  &=  -\frac{3 a_0^2}{32 k} (r-r_0)^{-1}
	-\frac{4}{13}(r-r_0)^{2}+c_5 (r-r_0)^{4}+
	\frac{16864 k}{39039 a_0^2}(r-r_0)^{5}+\cdots\,, \label{[mJ]H}
\end{align}

with two free parameters $a_0,c_5$.

\item In the case ${[-2,-2]}$ we get
\begin{align}
& \Lambda = -\frac{33}{40 k}\,,\\
& a_1 = a_2=a_3=a_4=a_5=a_6=0\,,\ \ a_7=\frac{20 k c_7}{33 a_0}\,,\ \ a_8=\frac{400k^2}{441 a_0^3}\,,\ \
a_9=a_{10}=0\,, \dots \,,\\
& c_0 = -\frac{ a_0^2}{20 k}\,,\ \
c_1=c_2=c_3=0\,, \ \ c_4=-\frac{1}{7}\,,\ \
c_5=c_6=0\,,\ \
c_8=\frac{40 k}{63 a_0^2}\,,\ \
c_9=c_{10}=0 \,, \dots\,,
\end{align}
with two free parameters $a_0,c_7$.
\end{itemize}

For all admitted values of $n$, the invariants \eqref{invB}, \eqref{invC} approach  a  constant as ${r\rightarrow r_0}$, namely
\bea
B_{ab}\, B^{ab} \rovno p^2(p-1)^2(p-3)^2(11p^2-32p+24)\frac{c_0^4}{144a_0^8}+\cdots\,,\label{invBB[<0,p]}\\
C_{abcd}\, C^{abcd}  \rovno  p^2(p-1)^2\frac{c_0^2}{3a_0^4}+\cdots\,. \label{invCC[<0,p]}
\eea
Note that for the permitted values of $p$, the invariant \eqref{invBB[<0,p]} \emph{cannot vanish}, and therefore this class does not admit the Schwazschild--(A)dS solution as a limit.

Moreover, since ${n<0}$, it follows from \eqref{to static} that
\be
{\bar r}=\Omega=a_0\,\Delta^n+\cdots\ \rightarrow\ \infty
\ee
as ${r\rightarrow r_0}$. Thus this limit  corresponds to an \emph{asymptotic region}, where the standard metric functions \eqref{rcehf} behave as
\bea
h  \rovno -a_0^2 c_0\,\Delta^{4n+2}+\cdots\ \sim\  {\bar r}^{\,4+2/n}+\cdots\rightarrow\ \infty\,,\\
f  \rovno -n^2c_0\,\Delta^{2n}+\cdots \quad\, \sim\  {\bar r}^{\,2}+\cdots\rightarrow\ \infty \,.
\eea
Note that, in the notation of \cite{Stelle:1978,LuPerkinsPopeStelle:2015b}, these solutions would be described as families ${(s,t)=\left(-2,4-\frac{4}{J}\right)_\infty}$ with ${J\in\mathbb{N}}$, ${J\geq 3 }$, i.e., with the parameter ${t\in [ \frac{8}{3},4)}$.

\vspace{5mm}

\noindent
\textbf{To conclude}: The class $[-J/2,2-J]$ with $\Lambda$ uniquely determined by \eqref{[mJpul,]L} is, in fact, an infinite discrete family of metrics parametrized by an integer ${J\geq 3 }$. They all have a regular Bachian infinity because both the Bach and Weyl invariants approach a finite nonzero value \eqref{invBB[<0,p]}, \eqref{invCC[<0,p]} in the asymptotic physical region ${\bar r \rightarrow \infty}$. In particular, for ${J=3}$ the metric functions are  expressed in terms of the series \eqref{[mJ]O} and \eqref{[mJ]H}.

\newpage
\section{Discussion of solutions using the expansion in powers of~$r^{-1}$}
\label{expansiont_INF}

Now let us study and classify all possible solutions to the field equations of QG in the case of an asymptotic expansion as ${r\rightarrow \infty}$. Here we assume that the metric functions $\Omega(r)$,  $\H (r)$  of (\ref{BHmetric}) can be expanded in (negative) powers of $r$ as  (\ref{rozvojomegINF}), (\ref{rozvojcalHINF}), that is
\begin{equation}
\Omega(r)    = r^N   \sum_{i=0}^\infty A_i \,r^{-i}\,, \qquad
\H (r) = r^P \,\sum_{i=0}^\infty C_i \,r^{-i}\,.\label{rozvojomagAcalHINF}
\end{equation}
Employing these expansions in the field equation
 (\ref{Eq1}), we obtain
\begin{align}
&\sum_{l=-2N+2}^{\infty}r^{-l}\sum^{l+2N-2}_{i=0}A_i\,A_{l-i+2N-2}\,(l-i+N-2)(l-3i+3N-1) \nonumber \\
& \hspace{45.0mm}=\tfrac{1}{3}k\sum^{\infty}_{l=-P+4}r^{-l}\,C_{l+P-4}\,(l-4)(l-3)(l-2)(l-1) \,.
\label{KeyEq1INF}
\end{align}
The second field equation (\ref{Eq2}) gives
\begin{align}
&\sum_{l=-2N-P+2}^{\infty}r^{-l}\sum^{l+2N+P-2}_{j=0}\sum^{j}_{i=0}A_i\,A_{j-i}\,C_{l-j+2N+P-2}\,(j-i-N)(l-j+3i-N-2) \nonumber \\
& \hspace{10.0mm} +\sum_{l=-2N}^{\infty}r^{-l}\sum^{l+2N}_{i=0}A_i\,A_{l-i+2N}
-\Lambda \sum_{l=-4N}^{\infty}r^{-l}\sum^{l+4N}_{m=0}\Big(\sum^{m}_{i=0}A_i\,A_{m-i}\Big)\Big(\sum^{l-m+4N}_{j=0}A_j\,A_{l-m-j+4N}\Big) \nonumber \\
& = \tfrac{1}{3}k \bigg[2+\sum^{\infty}_{l=-2P+4}r^{-l}\sum^{l+2P-4}_{i=0}C_{i}\,C_{l-i+2P-4}\,(i-P)(l-i+P-4)(l-i+P-3)
(l-\tfrac{3}{2}i+\tfrac{3}{2}P-\tfrac{5}{2})\bigg],
\label{KeyEq2INF}
\end{align}
and finally, the trace equation (\ref{trace}) leads to
\begin{align}
&\sum_{l=-N-P+2}^{\infty}r^{-l}\sum^{l+N+P-2}_{i=0}C_i\,A_{l-i+N+P-2}\,\big[(l-i+P-2)(l-1)+\tfrac{1}{6}(i-P)(i-P+1)\big] \nonumber \\
& \hspace{50mm}+\tfrac{1}{3}\sum^{\infty}_{l=-N}r^{-l}\,A_{l+N} =\tfrac{2}{3}\Lambda \sum^{\infty}_{l=-3N}r^{-l}\sum^{l+3N}_{j=0}
\sum^{j}_{i=0}A_i\,A_{j-i}\,A_{l-j+3N} \,.
\label{KeyEq3INF}
\end{align}

Expressions for the coefficients $C_j$  in terms of $A_j$s
can be obtained by comparing coefficients appearing at the same powers
of $r^{-l}$ on both sides of  (\ref{KeyEq1INF}).
The terms with the lowest order imply that there are three distinct cases:
\begin{itemize}
	\item Case I$^\infty$: ${\ \ -2N+2<-P+4}$\,, \ i.e.,  ${\ P<2N+2}$\,,
	\item Case II$^\infty$: ${\ -2N+2>-P+4}$\,,  \ i.e.,  ${\ P>2N+2}$\,,
	\item Case III$^\infty$: ${-2N+2=-P+4}$\,,   \ i.e.,  ${\ P=2N+2}$\,.
\end{itemize}
In what follows, we derive all possible solutions in these cases.

\subsection{\textbf{Case I}$^\infty$ }

For  ${-2N+2<-P+4}$, the {highest} order (namely $r^{-l}$,  ${-l=2N-2}$) in (\ref{KeyEq1INF}) gives
\begin{equation}
N(N+1)=0 \,,
\label{KeyEq1CaseIINF}
\end{equation}
and thus only the cases  ${N=0}$ and ${N=-1}$ are allowed.

Since the leading orders appearing in (\ref{KeyEq3INF}), \begin{equation}
\big[6N(N+P-1)+P(P-1)\big]C_0\,r^{N+P-2}+\cdots=-2\,r^N+\cdots +4\Lambda A_0^2 r^{3N}+\dots\,,
\label{KeyEq3CaseIINF}
\end{equation}
for  ${N=0}$  are ${r^{P-2}}$, ${r^0}$, and ${r^0}$, respectively, the condition ${P-2<2N=0}$ implies that the highest order is ${0=2(-1+\Lambda A_0^2 )r^0}$, leading to ${A_0^2=1/\Lambda}$.
Then Eq.~(\ref{KeyEq3INF}) implies ${\Lambda=3/(8k)}$.

In contrast, the case ${N=-1}$ cannot occur since the leading powers are ${r^{P-3}}$, ${r^{-1}}$,
and ${r^{-3}}$, respectively, however, the condition ${P-3<2N-1=-3<-1}$ implies that the highest order in \eqref{KeyEq3CaseIINF} is  ${0=-2r^{-1}}$, which is a contradiction.
\vspace{5mm}

\noindent
\textbf{To summarize}: The only possible solutions in Case~I$^\infty$ are
\begin{align}
\ [N,P] &= [0,<2]^\infty  \quad \hbox{with}\quad A_0^2=\frac{1}{\Lambda} \quad\hbox{and}\quad
 \Lambda=\frac{3}{8k}  \quad \Big(\hbox{so that}\ A_0^2=\frac{8k}{3}\Big)\,.
\label{CaseI_summaryINF}
\end{align}

\subsection{\textbf{Case II}$^\infty$ }

The condition ${-2N+2>-P+4}$ implies that  the {highest} order (namely $r^{-l}$, ${l=-P+4}$) in  (\ref{KeyEq1INF}) is on the {right hand} side, which gives
\begin{equation}
P(P-1)(P-2)(P-3)=0 \,,
\label{KeyEq1CaseIIINF}
\end{equation}
leading to four possible cases ${P=0}$, ${P=1}$, ${P=2}$, and ${P=3}$.
For these four cases, the leading orders of \eqref{KeyEq3INF}, i.e., \eqref{KeyEq3CaseIINF},  read
\begin{align}
\hbox{for}& \quad P=0,\ N<-1:
 &&
[6N(N-1)]C_0\,r^{N-2}+\cdots=-2\,r^N+\cdots+4\Lambda r^{3N}+\cdots
\,,\label{eq0m1}\\
\hbox{for}& \quad P=1,\ N<-1/2:
&&
[6N^2] C_0\,r^{N-1}+\cdots=-2\,r^N+\cdots+4\Lambda r^{3N}+\cdots
\,,\label{eq1m12}\\
\hbox{for}&\quad P=2,\ N<0:
 &&
\big[6N(N+1)+2\big]C_0\,r^N+\cdots=-2\,r^N+\cdots+4\Lambda r^{3N}+\cdots
\nonumber \\
& \quad && \Rightarrow\ \hbox{necessarily}\quad (3N^2+3N+1)C_0=-1\,,\label{contrp=2c0INF}
\\
\hbox{for}&\quad P=3,\ N<1/2:
&&
\big[6N(N+2)+6\big]C_0\,r^{N+1}+\cdots=-2\,r^N+\cdots+4\Lambda r^{3N}+\cdots\nonumber \\
&
 && \Rightarrow\ \hbox{necessarily}\quad N=-1\,.
\end{align}
Eqs. \eqref{eq0m1}, \eqref{eq1m12} do not admit solutions for ${N<-1}$ and ${N<-1/2}$, respectively.
For the case ${P=2}$, implying ${N<0}$, we employ the leading orders of Eq.~\eqref{KeyEq2INF},
\be
3A_0^2\,[N(3N+2)C_0+1]\,r^{2N} +2k(C_0^2 -1)-3\Lambda A_0^4 r^{4N}
+ \cdots =0 \,, \label{eq2rozvoj0omegINF}
\ee
which requires ${(3N^2+2N)C_0=-1}$. In combination with  \eqref{contrp=2c0INF}, we obtain ${N=-1}$, ${C_0=-1}$.
\vspace{5mm}

\noindent
\textbf{To summarize}: The only possible two classes of solutions in Case~II$^\infty$ are
\bea
\ [N,P] \rovno [-1,3]^\infty \,,  \label{CaseII_summaryINF-a}\\
\ [N,P] \rovno [-1,2]^\infty \,. \label{CaseII_summaryINF-b}
\eea

\subsection{\textbf{Case III}$^\infty$ }
\label{sec_IIIinf}

For ${P=2N+2}$, the {highest} order (which is $r^{-l}$,  ${l=2-2N}$) in  (\ref{KeyEq1INF}) appears {on both sides} and implies
\be
P(P-2)\big[3A_0^ 2+4kC_0(P-1)(P-3)\big]=0\,,
\label{KeyEq1CaseIIIINF}
\ee
leading to three possible subcases ${P=0}$,  ${P=2}$, and ${3A_0^2=-4kC_0 (P-1)(P-3)}$ with ${P\not= 0,1,2,3}$,
corresponding to
${N=-1}$,  ${N=0}$, and  ${3A_0^2=-4kC_0(4N^2-1)}$ with ${N\not= -1, -1/2, 0, 1/2}$, respectively.

For these three cases, the leading orders of  \eqref{KeyEq3INF},
\bea
\left[(11N^2+6N+1) C_0-2\Lambda A_0^2\right]\,r^{3N}  +\cdots \rovno - r^N+\cdots  \,, \label{eqtr00omegIIIINF}
\eea
read (note that necessarily $N\geq 0$):
\begin{align}
&\hbox{for}\quad N=-1\Leftrightarrow P=0:\quad &
(6C_0-2\Lambda A_0^2)\,r^{-3}+\cdots=-\,r^{-1}+\cdots\quad& \hbox{not compatible}\,,\label{contrp=2c0IIIaINF}\\
&\hbox{for}\quad N=0\Leftrightarrow P=2:\quad & C_0-2\Lambda A_0^2+\cdots=-1+\cdots\quad& \Rightarrow C_0=-1+2\Lambda A_0^2\,,\label{contrp=2c0IIIbINF}\\
&\hbox{for}\quad 3A_0^2=4kC_0(1-4N^2): & (11N^2+6N+1)C_0-2\Lambda A_0^2+\cdots=0&\,. \label{contrp=2c0IIIcINF}
\end{align}

Eq.~\eqref{KeyEq2INF} for the case \eqref{contrp=2c0IIIbINF} requires  ${3A_0^2(1-\Lambda A_0^2)+2k(C_0^2-1)=0}$, which after substituting for $C_0$ from \eqref{contrp=2c0IIIbINF}
gives ${(1-\Lambda A_0^2)(8k\Lambda-3)=0}$.
Thus, there are two possibilities:
\be
\Lambda A_0^2=1\quad\Rightarrow\quad C_0=1\,,
\ee
and
\be
\Lambda=\frac{3}{8k}\quad\Rightarrow\quad C_0=\frac{3}{4k}A_0^2-1\,.
\ee

In the case \eqref{contrp=2c0IIIcINF}, the two conditions, namely
 ${3A_0^2=4kC_0(1-4N^2)=-4kC_0 (P-3)(P-1)}$, with ${P\not= 0,1,2,3}$ (and thus ${N=P/2-1\not=-1,-1/2,0,1/2}$), and ${(11N^2+6N+1)C_0=2\Lambda A_0^2}$ (for ${N>0}$,  ${N\not=0, 1/2}$)  imply
\be
\Lambda =\frac{3}{8k}\frac{11N^2+6N+1}{1-4N^2}\quad\Rightarrow\quad C_0=\frac{3}{4k}\frac{A_0^2}{1-4N^2}\,.\label{IIIcLC0}
\ee
The highest order of Eq.~\eqref{eq2rozvoj0omegINF} is then identically satisfied.
\vspace{5mm}

\noindent
\textbf{To summarize}: In Case~III$^\infty$, there are three possible classes of solutions
\bea
\ [N,P] \rovno [0,2]^\infty  \hspace{5mm} \hbox{with} \quad
\Lambda A_0^2=1\,,\quad C_0=1\,, \label{CaseIII_summaryINF-a}\\
\ [N,P] \rovno [0,2]^\infty  \hspace{5mm} \hbox{with} \quad
\Lambda=\frac{3}{8k}\,,\quad C_0=\frac{3}{4k}A_0^2-1\,, \label{CaseIII_summaryINF-b}\\
\ [N,P] \rovno [>0,2N+2]^\infty \quad  \hbox{with} \quad
\Lambda =\frac{3}{8k}\frac{11N^2+6N+1}{1-4N^2}\,,\ C_0=\frac{3}{4k}\frac{A_0^2}{1-4N^2}
\,,\ N\not=\frac{1}{2}\,. \label{CaseIII_summaryINF-c}
\eea

\section{Description and study of all possible solutions in powers of~$r^{-1}$}
\label{description_INF}

In this section, by solving  Eqs.~(\ref{KeyEq1INF}), (\ref{KeyEq2INF}), and the trace (\ref{KeyEq3INF}), we will study all spherically symmetric solutions contained in Cases~I$^\infty$, II$^\infty$, III$^\infty$ in the asymptotic region in the coordinate $r$, i.e., as $r\to\infty$.
As follows from the previous section, there are {six classes} of solutions to be discussed, namely (\ref{CaseI_summaryINF}), (\ref{CaseII_summaryINF-a}), (\ref{CaseII_summaryINF-b}), (\ref{CaseIII_summaryINF-a}), (\ref{CaseIII_summaryINF-b}), and (\ref{CaseIII_summaryINF-c}).

\subsection{Schwarzschild--Bach--(A)dS black hole in the class ${[N,P]=[-1,3]^\infty}$: near the singularity }
\label{Schw_[N,P]=[-1,3]}

The expansions \eqref{rozvojomegINF}, \eqref{rozvojcalHINF}  in negative powers of~$r$  for ${N=-1}$, ${P=3}$, which is the class (\ref{CaseII_summaryINF-a}), give
\begin{eqnarray}
\Omega(r)     \rovno  -\frac{1}{r}
+\frac{B}{r}\,\bigg(\frac{2}{9}\,\frac{1}{C_0^3 r^3}
+\frac{1}{6}\,\frac{1}{C_0^4 r^4}
+\frac{2}{15}\,\frac{1}{C_0^5 r^5} +\ldots\bigg)\,, \label{IIdOmegaFULL}\\
\H (r) \rovno
\frac{\Lambda}{3} - r^2 - \Big(\frac{\Lambda}{3}-r_h^2 \Big)\frac{r^3}{r_h^3}
+ B\,\bigg( \frac{1}{C_0^2}-\frac{1}{90k}\,\frac{1 }{C_0^3 r^3}
-\frac{1}{140k}\,\frac{1}{C_0^4r^4}
-\frac{1}{210k}\,\frac{1}{C_0^5 r^5}+ \ldots \bigg) \,, \label{IIdH0FULL}
\end{eqnarray}
which represents the Schwarzschild--Bach--(A)dS black hole in QG, and for ${B=0}$ reduces to the Schwarzschild--(A)dS solution~\eqref{IIbH0Schw} with the horizon located at~$r_h$.

Since in the limit ${r\to\infty}$ we obtain ${{\bar r}=\Omega(r)\sim -1/r \to 0 }$ and ${\H \to \infty}$, the physical origin ${\bar r = 0}$ represents the curvature singularity, cf. \eqref{BInv2INFfinal}. Note also that  ${h({\bar r})\sim 1/( r_h \, \bar r) \to \infty}$ and ${f({\bar r})\sim h({\bar r})}$, see \eqref{rcehf}. In what follows, we  derive this class of solutions.

Eq.~\eqref{KeyEq1INF} gives  (relabeling  ${l \to l+2 }$) all $C_{l+1}$ in
	terms of ${A_0,\ldots, A_{l-2}}$, starting from ${C_4=0}$:
\be
C_{l+1}=\frac{3}{k\,(l-2)(l-1)l(l+1)}\,\sum^{l-2}_{i=0}
A_i\,A_{l-2-i}(l-1-i)(l-2-3i)\,,  \qquad \forall\ l\ge 3\,.
\label{nonSchwinitcondcINF} \ee
	Eq. \eqref{KeyEq3INF} determines all
$A_{l}$
in terms of ${A_0,\ldots, A_{l-1}}$ and ${C_1,\ldots, C_{l}}$, starting from $A_1$:
\begin{align}
l^2C_0A_l=&
\tfrac{2}{3}\Lambda \sum^{l-3}_{j=0}\sum^{j}_{i=0}A_iA_{j-i}A_{l-j-3}-\tfrac{1}{3}\,A_{l-1}  -\sum^{l}_{i=1}C_iA_{l-i}
\,\big[\,l(l-i)+\tfrac{1}{6}i(i+1)\big]
\,,
\qquad \forall\ l\ge 1\,.
\label{nonSchwinitcondaINF}
\end{align}
Finally, the additional constraint, namely
\be
C_2=\frac{C_1^2-1}{3C_0}\,,
\label{nonSchwinitcond2INF}
\ee
follows from the
lowest nontrivial order ${l=0}$ of the field equation \eqref{KeyEq2INF}.

Thus, this class has four initial parameters ${A_0, C_0, C_1, C_3}$. The constant  ${C_2}$ is given by \eqref{nonSchwinitcond2INF}, and $A_l$, $C_{l+3}$  for all ${l\ge 1}$  by the recurrent relations
\eqref{nonSchwinitcondaINF}, \eqref{nonSchwinitcondcINF}, namely
\bea
&&
C_4=0\,,\ \
C_5=0\,,\ \
C_6=A_0^2 \frac{(C_1+1)^2(C_1-2)+9C_0^2(A_0^2\Lambda -3 C_3)}{2430 C_0^3 k}\,,\dots\,,\
\label{[-1,3]infC6}\\
&&
A_1=-A_0\frac{(C_1+1)}{3 C_0}\,,\ \
A_2=A_0\frac{(C_1 +1)^2}{9 C_0^2}\,,\\
&&
A_3= -A_0 \frac{(C_1+1)^2 (13 + 7 C_1) + 54 C_0^2 C_3 - 18 \Lambda A_0^2 C_0^2}{243 C_0^3}\,, \dots\,.
\eea

\subsubsection{Identification of the Schwarzschild--(A)dS black hole }

For this class of solutions of the form \eqref{rozvojomagAcalHINF}, the curvature invariants \eqref{invB}, \eqref{invC} read
\be
B_{ab}\, B^{ab}(r \to \infty) = \Big(45 \frac{C_0}{A_0^4}\, C_6\Big)^2  \,,
\qquad
C_{abcd}\, C^{abcd}(r \to \infty) \sim 12 \frac{C_0^2}{A_0^4}\,r^6  \,.
\label{BInv2INF}
\ee
To identify the Schwarzschild--(A)dS solution, we have to set the Bach tensor to zero, which is achieved by setting
\be
C_6 = 0\,.
\label{C_6=0}
\ee
This is equivalent to ${ 27 C_0^2 C_3 = (C_1+1)^2(C_1-2)  + 9\Lambda A_0^2 C_0^2 }$.
Then the sequence  $A_i$  reduces to a geometrical series, while the sequence $C_i$  truncates to a 3rd-order polynomial
\bea
&& A_i=A_0\,\Big(\!\!-\frac{C_1+1}{3 C_0}\,\Big)^i\quad \hbox{for all}\ i \ge 0 \,,\\
&& C_2=\frac{C_1^2-1}{3C_0}\,, \quad C_3=\frac{(C_1+1)^2(C_1-2)+9\Lambda A_0^2  C_0^2}{27 C_0^2 }\,, \quad C_i=0\quad
\hbox{for all}\ i \ge 4\,,
\eea
and the metric functions can be expressed in the closed form
\begin{eqnarray}
\Omega(r)      \rovno  \frac{A_0}{r}   \,\sum_{i=0}^\infty \,\Big(\!\!-\frac{C_1+1}{3C_0\,r}\Big)^i
=\frac{A_0}{r+(C_1+1)/(3C_0)}\,, \label{IIdOmega}\\
\H (r) \rovno  C_0\,r^3+C_1\,r^2+\frac{C_1^2-1}{3C_0}\, r+\frac{(C_1+1)^2(C_1-2)+9\Lambda A_0^2  C_0^2 }{27 C_0^2 } \,. \label{IIdH0}
\end{eqnarray}
Choosing a gauge \eqref{scalingfreedom} such that
\be
A_0=-1\,,\qquad C_1=-1\,,
\label{IId_a0}
\ee
the metric functions simplify to
\be {\bar r}=\Omega(r) = -\frac{1}{r}\,,
\qquad \H (r) = \frac{\Lambda}{3}-r^2+C_0\,r^3 \,,
\label{IIdH0Schw}
\ee
which is the \emph{Schwarzschild--(A)dS black hole metric}~\eqref{SchwAdS},
where $C_0$ is given by the horizon position~$r_h$ and the cosmological constant $\Lambda$ as
\be
C_0=\Big(r_h^2-\frac{\Lambda}{3}\Big)\frac{1}{r_h^3}\,,
\ee
see (\ref{IIbH0Schw}).

\subsubsection{More general Schwarzschild--Bach--(A)dS black hole}

For a more general solution with a nonvanishing Bach tensor, see \eqref{BInv2INF}, it is convenient to introduce  a \emph{dimensionless Bach parameter}~$B$ proportional to $C_6$. Employing \eqref{IId_a0}, ${C_6= -(C_3-\tfrac{\Lambda}{3}) / (90k C_0)}$, and thus we choose
\be
B \equiv C_0^2\left(C_3-\frac{\Lambda}{3}\right) \,.
\label{b_definiceINF}
\ee
Using \eqref{b_definiceINF} and the gauge \eqref{IId_a0},  the recurrent
	relations \eqref{nonSchwinitcondaINF}, \eqref{nonSchwinitcondcINF}
	simplify to
\begin{align}
& A_0 = -1\,,\qquad
A_1 = 0\,,\qquad
A_2 = 0\,,
\qquad
 A_3 = \frac{2B}{9C_0^3}\,, \qquad
A_4 = \frac{B}{6C_0^4}\,, \nonumber \\
&A_5 = \frac{2B}{15C_0^5} \,,
\qquad
 A_6=-\frac{B(10B-9)}{81C_0^6}
-\frac{7B(3+40k\Lambda)}{9720kC_0^4}
\,, \ldots\,,\label{IId_expansionaINFa} \\
&  C_0=\Big(r_h^2-\frac{\Lambda}{3}\Big)\frac{1}{r_h^3}
 \,,\qquad
C_1 = -1 \,,\qquad
C_2 = 0 \,,\qquad
C_3 = \frac{B}{C_0^2} +\frac{\Lambda}{3}\,,\qquad
C_4 = 0 \,, \nonumber \\
& C_5 = 0 \,, \qquad
C_6 = -\frac{B}{90kC_0^3}  \,,\qquad
C_7 = -\frac{B}{140kC_0^4}  \,,\qquad
C_8 = -\frac{B}{210kC_0^5}\,, \ldots\,,
\label{IId_expansioncINFc}
\end{align}
 giving the explicit expansions \eqref{IIdOmegaFULL}, \eqref{IIdH0FULL}.

Finally, the corresponding curvature invariants \eqref{BInv2INF} at ${\bar r = 0}$ read
\be
B_{ab}\, B^{ab}(r \to \infty) = \frac{1}{4k^2C_0^4}\,B^2  \,,\qquad
C_{abcd}\, C^{abcd} (r \to \infty) \sim 12C_0^2\,r^6 \to \infty \,,
\label{BInv2INFfinal}
\ee

\vspace{5mm}

\noindent
\textbf{To conclude}: The class $[-1,3]^\infty$, with the metric functions expressed in terms of the series \eqref{IIdOmegaFULL} and \eqref{IIdH0FULL} around the physical origin
	${\bar r=0}$, represents the Schwarzschild--Bach--(A)dS  black hole. As was already pointed out in \cite{PodolskySvarcPravdaPravdova:2020} for the ${\Lambda=0}$ case, this metric may, in fact, represent a distinct Schwarzschild--Bach--(A)dS black hole from the one in the $[0,1]$ class discussed in Sec.~\ref{SchwaBach_[n,p]=[0,1]}.

\vspace{5mm}

\subsection{Bachian--(A)dS vacuum in the class ${[N,P]=[-1,2]^\infty}$ }
\label{Schw_[N,P]=[-1,2]}

Now we will analyze the second possibility \eqref{CaseII_summaryINF-b}  in  Case~II$^\infty$. After relabeling ${l \to l+2}$, Eq. \eqref{KeyEq1INF} for ${N=-1}$, ${P=2}$ yields
\be
C_l=\frac{3}{k\,(l-2)(l-1)l(l+1)}\,\sum^{l-2}_{i=0}
A_i \,A_{l-2-i}(l-1-i)(l-2-3i)  \qquad \forall\ l\ge 3\,.
\label{[-1,2]initcondc}
\ee
Eq. \eqref{KeyEq3INF} in its lowest orders ${r^{-1}, r^{-2} }$ determines $A_1$ and $C_0$ as
\begin{equation}
A_1=\tfrac{1}{2}A_0C_1 \,, \qquad  C_0=-1\,,
\end{equation}
while for higher orders gives
\begin{align}
l(l-1)C_0A_{l-1}=\tfrac{2}{3}\Lambda \sum^{l-3}_{j=0}\sum^{j}_{i=0}A_iA_{j-i}A_{l-j-3}
-\sum^{l-1}_{i=1}C_iA_{l-i-1}\left[(l-i)(l-1)+\tfrac{1}{6}(i-2)(i-1)\right]\ \ \forall\ l\ge 3\,.
\label{[-1,2]initconda}
\end{align}
There are no additional constraints following from Eq. \eqref{KeyEq2INF}. There are thus three free parameters,  ${A_0, C_1, C_2}$, and all other
coefficients are determined by the recurrent relations \eqref{[-1,2]initcondc}, \eqref{[-1,2]initconda}, e.g.,
\begin{align}
& A_2 = \frac{A_0}{3}(C_1^2 + C_2-\tfrac{1}{3} \Lambda A_0^2 )\,,\qquad
A_3 = \frac{ A_0 }{4}C_1 (C_1^2 + 2C_2-\tfrac{2}{3} \Lambda A_0^2 )\,,\nonumber\\
& A_4 = \frac{A_0 }{5}\left[C_1^4 + 3 C_1^2 C_2 + C_2^2 +\frac{A_0^2}{192 k}\big[C_1^2 + 4 C_2- 8 (23 C_1^2 + 12 C_2) k \Lambda\big]+\frac{A_0^4}{144 k}  \Lambda (8 k \Lambda-1) \right],  \ldots\,,
\label{[-1,2]_expansionaINFa} \\
& C_3 = 0 \,,\qquad
C_4 = \frac{A_0^2 }{240 k}(C_1^2 + 4C_2- \tfrac{4}{3} \Lambda A_0^2 ) \,,\qquad
C_5 = \frac{A_0^2}{240 k} C_1 (C_1^2 + 4C_2- \tfrac{4}{3} \Lambda A_0^2 ) \,,\nonumber\\
& C_6 = \frac{A_0^2 }{67200 k^2}\left[ A_0^2(3-\tfrac{32}{3}k\Lambda)+ 4k(59 C_1^2 + 26 C_2)\right](C_1^2 + 4C_2- \tfrac{4}{3} \Lambda A_0^2 )  \,, \ldots\,.
\label{[-1,2]_expansioncINFc}
\end{align}

\subsubsection{Identification of Minkowski and (A)dS spaces}

First, let us observe that in the limit ${r\to\infty}$, the scalar invariants \eqref{invB}, \eqref{invC} remain finite,
\be
B_{ab}\, B^{ab}(r\to\infty) = \frac{300}{A_0^8}\,C_4^2  \,,\qquad
C_{abcd}\, C^{abcd}(r\to\infty) \sim \frac{12}{A_0^4\,r^4}\,C_4^2  \,,
\label{[-1,2]_Inv_INF}
\ee
suggesting that there is no
	physical singularity there. Both the Bach and Weyl tensor invariants  vanish iff ${C_4=0}$, i.e.,
${C_2=-\frac{1}{4} C_1^2+\tfrac{1}{3} \Lambda A_0^2 }$. Then all the coefficients
\eqref{[-1,2]_expansionaINFa}, \eqref{[-1,2]_expansioncINFc} simplify  to
${A_i=A_0\,(\frac{1}{2}C_1)^i}$ for all $i$, and ${C_i=0}$ for all ${i\ge3}$, respectively, so that the metric functions reduce to
\BE
\Omega(r)  =  \frac{A_0}{r}   \,\sum_{i=0}^\infty \,\Big(\frac{C_1}{2\,r}\Big)^i
=\frac{A_0}{r-\tfrac{1}{2}C_1}\,, \qquad
\H (r) = \frac{1}{3} \Lambda A_0^2 - \left(r-\frac{1}{2}C_1\right)^2   \,. \label{[-1,2]_Omega_H0}
\EE
Employing the gauge freedom \eqref{scalingfreedom}, we may set
\be
A_0=-1\,,\qquad C_1=0\,.
\label{[-1,2]_a0}
\ee
Then the metric functions simplify to
\be {\bar r}=\Omega(r) = -\frac{1}{r} \,,
\qquad \H (r) = \frac{\Lambda}{3} - r^2 \,.
\label{[-1,2]_solution}
\ee
Comparing with \eqref{SchwAdS}, this case ${C_4=0}$ corresponds to \emph{Minkowski} or \emph{(anti-)de~Sitter} space with vanishing Bach and Weyl tensors. For ${\Lambda>0}$, there is a cosmological horizon at $r_{cosm}= \sqrt{\tfrac{\Lambda}{3}}$.

\subsubsection{Bachian--(A)dS vacuum}
\label{sec_[-1,2]inf}

Let us return to the class ${[N,P]=[-1,2]^\infty}$ with a general Bach tensor \eqref{[-1,2]_Inv_INF}. Using the gauge \eqref{[-1,2]_a0}, we introduce a dimensionless Bach parameter
\be
B_v \equiv \Big(C_2-\frac{\Lambda}{3}\Big)k
\,
\label{beta_definiceINF}
\ee
so that ${ C_4 = (C_2-\tfrac{1}{3}\Lambda )/(60 k)=B_v/(60 k^2)}$, cf. \eqref{[-1,2]_expansioncINFc}.

The coefficients \eqref{[-1,2]_expansionaINFa}, \eqref{[-1,2]_expansioncINFc} then take the form
\begin{align}
& A_0 = -1\,,\quad
A_1 = 0\,,\quad
A_2 = -\frac{B_v}{3k}\,,\quad
A_3 = 0\,, \nonumber\\
& A_4 =-\frac{B_v}{240 k^2}\left(1+8k\Lambda + 48 B_v \right)
\,, \quad
A_5 = 0 \,, \ldots \,,   \label{[-1,2]_expansionaINFaa} \\
& C_0 = -1 \,,\quad
C_1 = 0 \,,\quad
C_2 = \frac{B_v}{k}+\frac{\Lambda}{3} \,,\quad
C_3 = 0 \,,\nonumber\\
& C_4 = \frac{B_v}{60k^2}  \,,\quad
C_5 = 0  \,,\quad
C_6 = \frac{B_v}{16800k^3}\left(3+24k\Lambda+104B_v \right)
\,, \ldots \,,
\label{[-1,2]_expansioncINFcc}
\end{align}
leading to the metric functions
\begin{eqnarray}
\Omega(r)     \rovno  -\frac{1}{r}
- \frac{B_v}{k\,r^3}\,\bigg(\frac{1}{3}
+\frac{1}{5kr^2}\,\Big(\frac{1}{48}+\frac{1}{6}k\Lambda+B_v \Big) +\ldots\bigg)\,, \label{[-1,2]_OmegaFULL}\\
\H (r) \rovno \frac{\Lambda}{3} -r^2
+ \frac{B_v}{k}\,
\bigg( 1+\frac{1}{60kr^2} +\frac{1}{700k^2r^4}\,\Big(\frac{1}{8}+k\Lambda+\frac{13}{3}B_v \Big)
+ \ldots \bigg) \,, \label{[-1,2]_H0FULL}
\end{eqnarray}
and the curvature invariants \eqref{[-1,2]_Inv_INF}
\be
B_{ab}\, B^{ab} (r\to\infty) = \frac{B_v^2}{12k^4}  \,,\qquad
C_{abcd}\, C^{abcd} (r\to\infty) \sim \frac{B_v^2}{300k^4r^4} \to 0  \,.
\label{[-1,2]_Inv_INF_beta}
\ee

In the limit ${r\to\infty}$, the metric functions behave as ${\Omega\sim -1/r}$ and ${\H \sim  \frac{\Lambda}{3} -r^2 + \frac{B_v}{k}}$. From \eqref{to static}, \eqref{rcehf} we thus obtain
\bea
&& {\bar r}=\Omega(r) \to 0 \,, \qquad h \sim 1\,,\qquad f \sim 1\,, \label{[-1,2]C}
\eea
i.e., at the physical origin ${\bar r=0}$, both metric functions $h$ and $f$ remain nonzero and finite, and there is no  horizon, nor  singularity therein.

\vspace{5mm}

\noindent
\textbf{To conclude}: The class ${[-1,2]^\infty}$, with the metric functions expressed in terms of the series \eqref{[-1,2]_OmegaFULL} and \eqref{[-1,2]_H0FULL} around the physical origin ${\bar r =0}$, describes a one-parameter Bachian generalization of  Minkowski or (A)dS spacetime with a nonzero Bach tensor whose magnitude is determined by the parameter $B_v$. This Bachian--(A)dS vacuum is a ``massless limit'' of the class ${[-1,3]^\infty}$, corresponding to ${C_0=0}$ in (\ref{IIdH0Schw}), cf.  Sec. \ref{sec_consistency [-1,3] [-1,2]}.

\vspace{5mm}

\subsubsection{The class $[-1,2]^\infty$ as a limit of the $[-1,3]^\infty$ class}
\label{sec_consistency [-1,3] [-1,2]}

A limiting procedure between the class of solutions $[-1,2]^\infty$, described by 	 \eqref{[-1,2]initcondc}--\eqref{[-1,2]initconda} with the coefficients denoted here by hats, and the class
$[-1,3]^\infty$, described by \eqref{nonSchwinitcondcINF}--\eqref{nonSchwinitcond2INF},  requires
\be
C_0\rightarrow\ 0\,,\qquad
C_i\rightarrow\ \hat C_{i-1},\ \ \ i\geq 1\,,\qquad
A_i\rightarrow\ \hat A_i,\ \ i\geq 0\,.
\ee
Then the relation \eqref{nonSchwinitcondaINF} for ${l=1}$, i.e., $3C_0A_1=-A_0(1+C_1)$, gives
\be
C_1\rightarrow\ -1,\ \ \ \mbox{i.e.,}\ \ \ \hat C_0=-1 \,.
\ee
The relations \eqref{nonSchwinitcondcINF} for $C_{l+1}$ and \eqref{[-1,2]initcondc} for $\hat C_l$
	are the same, and the relation \eqref{nonSchwinitcondaINF} for~$A_l$
\begin{align}
l^2C_0A_l=
\tfrac{2}{3}\Lambda \sum^{l-3}_{j=0}\sum^{j}_{i=0}A_iA_{j-i}A_{l-j-3}-\tfrac{1}{3}\,A_{l-1}
 -\sum^{l}_{i=1}C_iA_{l-i}\,\big[\,l(l-i)+\tfrac{1}{6}i(i+1)\big]\,,
\ \ \forall\ l\ge 1\,,
\end{align}
for $C_0=0$  leads to
\be
\hat A_{l-1}
=\frac{1}{l(l-1)}\bigg\{
\tfrac{2}{3}\Lambda \sum^{l-3}_{j=0}\sum^{j}_{i=0} \hat A_i \hat A_{j-i} \hat A_{l-j-3}
+\sum^{l-1}_{i=1} \hat C_i\,\hat A_{l-1-i}\,\big[(l-1)(l-i)+\tfrac{1}{6}(i-2)(i-1)\big]
\bigg\},
\quad \forall\ l\ge 2\,,
\ee
which is exactly \eqref{[-1,2]initconda}.

Note that the four free parameters ${A_0,C_0,C_1,C_3}$ of the $[-1,3]^\infty$ family reduce to three free parameters
${\hat A_0,\hat C_1, \hat C_2}$	of the $[-1,2]^\infty$ family since two parameters
become fixed (${C_0\rightarrow0}$,	${C_1\rightarrow\hat C_0=-1}$)  and one parameter ${C_2\rightarrow\hat C_1}$ becomes free	 ($3C_0C_2=C_1^2-1\rightarrow 0$).

\vspace{5mm}

\subsection{Nariai--Bach solutions in the class $[N,P]=[0,2]^\infty$  }
\label{sec_[0,2]infty}

In this section, we will show that the class $[0,2]^\infty$ contains \emph{static  spherically symmetric Nariai spacetime and its Bachian generalizations}. Only
the non-Kundt solutions with  \emph{discrete values} of $\Lambda$ given by Eq.~\eqref{[0,2]inf_L}  can be transformed into the standard static spherically symmetric form
\eqref{Einstein-WeylBH}.

For ${l\geq 3}$, Eq.~\eqref{KeyEq1INF} gives
\be
A_0 A_{l-2} (l-2)(l-1)=\tfrac{1}{3}k\,C_{l-2} (l-4)(l-3)(l-2)(l-1)
-\sum^{l-3}_{i=1}A_iA_{l-i-2}(l-i-2)(l-3i-1)\,,\label{rr_III_infty}
\ee
which for ${l=3,4}$ yields
\be
A_1=0\,, \qquad A_2=0\,,\label{[0,2]infA1A2}
\ee
respectively. For ${l\geq 1}$, the trace equation \eqref{KeyEq3INF} implies
\bea
&&A_l\left[ 2\Lambda A_0^2\left(l(l-1)-\tfrac{2}{3}\right)-l(l-1)\right]
+ \tfrac{1}{6} C_l A_0 (l-2)(l-1)
\label{trace_III_infty}\\
&&=-\sum^{l-1}_{i=1}C_i A_{l-i}\left[(l-i)(l-1)+\tfrac{1}{6}(i-2)(i-1)\right]
+\tfrac{2}{3}\Lambda \bigg[\sum^{l-1}_{i=1}A_0 A_{i}A_{l-i}
+\sum^{l-1}_{j=1}\sum^{j}_{i=0}A_i A_{j-i}A_{l-j}\bigg],\nonumber
\eea
where we employed the relation
\be
C_0=2\Lambda A_0^2-1\,,
\label{[0,2]infC0}
\ee
which follows from the leading order of Eq.~\eqref{KeyEq3INF}.
Finally, the leading term of Eq.~\eqref{KeyEq2INF} gives
\be
(\Lambda A_0^2 - 1)(8k\Lambda-3)=0\,.
\ee
Similarly as in the class $[0,2]$ investigated in Section~\ref{sec_[0,2]},
we have two cases to consider, namely
\begin{align}
\hbox{ a)}&\ \ \Lambda A_0^2=1\,,\quad \Lambda>0\,, \label{[0,2]infpodmA0}\\
\hbox{ b)}&\ \ \Lambda=\frac{3}{8k}\,, \label{[0,2]binfL}
\end{align}
see Eqs.~\eqref{CaseIII_summaryINF-a} and~\eqref{CaseIII_summaryINF-b}. Let us discuss these two distinct cases separately.

\subsubsection{Case a)  ${\Lambda A_0^2=1}$: Nariai(--Bach)	spacetime}
\label{sec_Nariai-Bach}

Recall that ${A_1=0=A_2}$, see \eqref{[0,2]infA1A2}, and from \eqref{[0,2]infC0} with \eqref{[0,2]infpodmA0} we obtain
\be
C_0=1\,.
\ee
Assuming ${A_i=0}$ for all ${1\leq i\leq l-1}$, and ${C_i=0}$ for ${3\leq i\leq l-1}$, ${l\geq 4}$,
Eqs.~\eqref{rr_III_infty} and~\eqref{trace_III_infty}
give (also for $l=3$) the relations
\bea
 A_{l}\,A_0-C_{l}\, \tfrac{1}{3}k\, (l-1)(l-2) \rovno 0\,,\label{IIIb_infty_eqs1}\\
 A_l\left[ 2\Lambda A_0^2\left(l(l-1)-\tfrac{2}{3}\right)-l(l-1)\right]
+C_l\, A_0 \tfrac{1}{6}(l-1)(l-2) \rovno 0\,.\label{IIIb_infty_eqs2}
\eea
This is a system of two linear equations for two unknowns $A_l,C_l$ with the determinant
\be
\tfrac{1}{6}(l-1)(l-2) \Big[ A_0^2 + 2k\left[2\Lambda A_0^2 \left(l(l-1)-\tfrac{2}{3}\right)-l(l-1)\right] \Big] \,,
\label{[0,2]detA0}
\ee
which, after substituting from \eqref{[0,2]infpodmA0}, reduces to
\be
\tfrac{1}{6}(l-1)(l-2) \Big[\,\frac{1}{\Lambda} +2k\left(l(l-1)-\tfrac{4}{3}\right) \Big]\,, \qquad l\geq 3\,.
\label{[0,2]det}
\ee
Therefore, we have to distinguish two subcases:

\vspace{5mm}

\noindent
$\bullet$ For a \emph{generic value} of ${\Lambda>0}$ this determinant is nonvanishing, and necessarily
\be
C_l=0=A_l\qquad \hbox{for all} \quad l\geq 3 \,.
\ee
The only such solution is thus
\be
\Omega = A_0 = \frac{1}{\sqrt{\Lambda}}\,,\qquad
\H(r) = r^2 + C_1\,r + C_2\,,\label{[0,2]infty-Nariai}
\ee
so that the metric (\ref{BHmetric}) reads
\be
\dd s^2 = A_0^2 \big(\dd \theta^2+\sin^2\theta\,\dd \phi^2\big)
+ A_0^2 \big[ (r^2 + C_1\,r + C_2)\,\dd u^2 -2\,\dd u\,\dd r \big]\,.
\label{genNariaimetric}
\ee
After performing the transformation
\bea
  r\rovno \tfrac{1}{4}(4C_2-C_1^2+\Lambda^2 \tilde u^2)\,\tilde r + \tfrac{1}{2}(\Lambda \tilde u-C_1)\,,\\
  u\rovno \frac{4}{\sqrt{4C_2-C_1^2}} \arctan{\bigg(\frac{\Lambda\tilde u}{\sqrt{4C_2-C_1^2}}\bigg)}\,,
\eea
(and dropping tilde symbols), or  for ${C_1^2=4C_2}$ using the gauge freedom (\ref{scalingfreedom}) and the scaling ${A_0^2\, u \to u}$, the solution reduces to
\be
\dd s^2 = \frac{1}{\Lambda} \big(\dd \theta^2+\sin^2\theta\,\dd \phi^2\big)
  -2\,\dd u\,\dd r + \Lambda\,r^2 \dd u^2 \,.
\label{Nariaimetric}
\ee

This is the metric of the \emph{Nariai spacetime} \cite{Nariai:1951} which in Einstein's theory is a vacuum solution of algebraic type~D. In fact, it belongs to the class of direct-product geometries, see Chapter~7 of \cite{GriffithsPodolsky:2009}. In particular, it has the Kundt form (18.49) for ${\alpha=\Lambda/2}$ therein (see also \cite{MO:02,PodolskyOrtaggio:2003, KadlecovaZelnikovKrtousPodolsky:2009}). It is a nonsingular, homogeneous and spherically symmetric Einstein space. Interestingly, since the conformal factor ${\Omega=A_0}$ is constant, and thus ${{\bar r}=\Omega=\,}$const., this solution \emph{cannot be transformed into the standard static spherically symmetric coordinates} \eqref{Einstein-WeylBH}.

\vspace{5mm}

\noindent
\textbf{To conclude}: The class $[0,2]^\infty$ with ${\Lambda A_0^2=1}$ and arbitrary ${\Lambda>0}$, with the metric functions expressed in terms of  \eqref{[0,2]infty-Nariai}, represents  the direct-product (${S^2 \times \dd S_2}$) Nariai solution~\eqref{Nariaimetric}.

\vspace{5mm}

\noindent
$\bullet$ The second branch of the $[0,2]^\infty$ class of solutions obeying \eqref{[0,2]infpodmA0} exists only for special \emph{discrete values} of the cosmological constant~${\Lambda>0}$, such that
\be
\Lambda=-\frac{3}{2k\left[3L(L-1)-4\right]}\,,\qquad \hbox{where}\quad {L\in\mathbb{N}} \,,\ \  L\geq 3\,, \ \ \ k<0
\,.\label{[0,2]inf_L}
\ee
The determinant~\eqref{[0,2]det} vanishes,
in which case $C_1,C_2$, together with a new \emph{Bach parameter}
\be
\tilde{B}_L\equiv C_L \,,
\ee
 are three free parameters. Recall that
\be
A_0 = \frac{1}{\sqrt{\Lambda}}\,,\qquad C_0=1\,,
\ee
but ${A_i=0}$ for ${1\leq i\leq L-1}$, and also ${C_i=0}$ for ${3\leq i\leq L-1}$. The first nontrivial coefficient $A_L$ is given by~\eqref{IIIb_infty_eqs1},
\bea
A_{L} \rovno  \tfrac{1}{3}k\,\sqrt{\Lambda}\,  (L-1)(L-2)\, \tilde{B}_L \,,\label{IIIb_infty_eqs1AL}
\eea
and all the subsequent coefficients $C_i$, $A_i$ for ${i>L}$ are determined by~\eqref{rr_III_infty}, \eqref{trace_III_infty} as
 \bea
C_{l} \rovno 6
\frac{\left[3l(l-1)-4\right]
	U_{l-1} -3A_0 l(l+1)V_{l-1}} {(l+1)l(l-1)(l-2)[3 A_0^2 -8k + 6kl(l-1)]}
\,,\nonumber\\
A_l  \rovno -3
\frac{A_0 U_{l-1} +2k l(l+1)V_{l-1}}{(l+1)l[3A_0^2 -8k + 6kl(l-1)]} \,, \label{[0,2]ClAl_inf}
\eea
  where
\bea
U_{l-1} \rovno
\sum^{l-1}_{i=1}A_i A_{l-i}(l-i)(l-3i+1)\,,\label{[0,2]S_inf}\\
V_{l-1} \rovno
\sum^{l-1}_{i=1}C_i A_{l-i}\left[(l-i)(l-1) +\tfrac{1}{6}(i-1)(i-2)\right]
-\tfrac{2}{3}\Lambda
\Big( \sum^{l-1}_{i=1}A_0A_i A_{l-i} +\sum^{l-1}_{j=1}\sum^{j}_{i=0} A_i A_{j-i} A_{l-j}\Big)\,.
\nonumber
\eea
Thus, the metric functions  take the explicit form
\bea
\Omega(r) \rovno  \frac{1}{\sqrt{\Lambda}}+A_L\, r^{-L}+\sum_{i=L+1}^\infty A_i\, r^{-i}\,,\label{rozvoj[0,2]infOmega}\\
\H(r) \rovno r^2+C_1\,r+C_2+\tilde{B}_L\, r^{2-L}+\sum_{i=L+1}^\infty C_i\, r^{2-i}\,,\label{rozvoj[0,2]infH}
\eea
where, $A_L$
and $A_i, C_i$ are determined by~\eqref{IIIb_infty_eqs1AL} and \eqref{[0,2]ClAl_inf}.

This is clearly a \emph{Bachian generalization of the Nariai metric}~\eqref{genNariaimetric}, to which it reduces when~${\tilde{B}_L=0}$. Since now ${\bar{r}=\Omega(r)}$ is \emph{not a constant}, it is possible to transform these Nariai--Bach   spacetimes to standard static spherically symmetric coordinates~\eqref{Einstein-WeylBH}.

For ${r\rightarrow \infty}$,
the leading terms of the curvature invariants
\eqref{invB} and \eqref{invC} read
\bea
B_{ab}\,B^{ab}(r\rightarrow \infty) \rovno \frac{1}{72}\,(3 L^2 - 8 L + 8)(L-2)^2(L-1)^2(L+1)^2  \, \Lambda^4\,\tilde{B}_L^2\, r^{-2L}+\cdots \,,
\\
C_{abcd}\,C^{abcd}(r\rightarrow \infty) \rovno \frac{16}{3}\, \Lambda^2+\cdots\,.
\eea
 Also, from \eqref{to static}, \eqref{rcehf} we obtain
\bea
&& {\bar r}=\Omega(r) \to A_0 = \frac{1}{\sqrt{\Lambda}} \,, \label{[0,2]A}\\
&& h \sim -\frac{(A_L)^{2/L}} {\Lambda(\bar r -A_0)^{2/L}}\rightarrow \ \infty\,,\qquad
f \sim -L^2\Lambda(\bar r -A_0)^2\ \rightarrow\ 0\,. \label{[0,2]C}
\eea

Although the discrete-$\Lambda$ spectrum of possible solutions~\eqref{[0,2]inf_L} in this case $[0,2]^\infty$ is the same as the spectrum~\eqref{[0,2]lambda} in the case $[0,2]$, these two metrics do not describe the same solution since only the $[0,2]$ case (the extreme higher-order (discrete) Schwarzschild--Bach--dS black holes)
contains the Schwarzschild--de~Sitter limit for vanishing Bach parameter. The Robinson--Trautman-type metric $[0,2]^\infty$ given by \eqref{rozvoj[0,2]infOmega}, \eqref{rozvoj[0,2]infH} is a generalization of the Nariai metric (of the Kundt type) to a nonvanishing Bach tensor.

\vspace{5mm}

\noindent
\textbf{To conclude}: The class $[0,2]^\infty$ with ${\Lambda A_0^2=1}$ and the discrete values of $\Lambda$ given by \eqref{[0,2]inf_L}, for which the metric functions are expressed in terms of the series \eqref{rozvoj[0,2]infOmega}, \eqref{rozvoj[0,2]infH} around a finite point ${{\bar r}= \frac{1}{\sqrt{\Lambda}}}$, represents a spherically symmetric generalization of the Nariai metric to a nonzero Bach tensor, which can be called Nariai--Bach spacetimes.

\vspace{5mm}

\subsubsection{Case b) ${\Lambda=\frac{3}{8k}\equiv \frac{3\gamma}{8(\alpha-3\beta)}\,}$: another Bachian generalization of the Nariai solution}
\label{sec_[0,2]nekb}

Let us now investigate the second distinct case of possible solutions in the class $[0,2]^\infty$, namely~\eqref{[0,2]binfL}.

Equations~\eqref{rr_III_infty} and~\eqref{trace_III_infty} lead, as in  previous Case~a), to the system \eqref{IIIb_infty_eqs1}, \eqref{IIIb_infty_eqs2} with the determinant \eqref{[0,2]detA0}, which for
\eqref{[0,2]binfL} is proportional to
\be
3A_0^2-4k\,.
\ee
Therefore, either the determinant vanishes for ${A_0^2=\frac{4}{3}k \Leftrightarrow 2\Lambda A_0^2=1}$, which using \eqref{[0,2]infC0} implies ${C_0=0}$, i.e., a contradiction with our assumptions, or the determinant is nonvanishing, leading to trivial solutions ${C_i=0}$ for all ${i\geq 3}$, and ${A_i=0}$ for all ${i\geq 1}$.
This solution
thus has only three free parameters ${A_0, C_1, C_2}$, while $C_0$ is determined as
\be
C_0 = 2 \Lambda A_0^2-1 = \frac{3}{4k}A_0^2-1\,,
\ee
see~\eqref{[0,2]infC0} and \eqref{CaseIII_summaryINF-b}.
The explicit metric functions read
\be
\Omega = A_0 \,,\qquad
\H(r)  = C_0\,r^2 + C_1\,r + C_2\,,\label{[0,2]infty-caseb-Nariai}
\ee
and lead to a direct-product metric
\be
\dd s^2 = A_0^2 \big(\dd \theta^2+\sin^2\theta\,\dd \phi^2\big)
+ A_0^2 \Big[ \big((2\Lambda A_0^2-1)\,r^2 + C_1\,r + C_2\big)\,\dd u^2 -2\,\dd u\,\dd r \Big]\,.
\label{anothergenNariaimetric}
\ee
whose invariants \eqref{invB}, \eqref{invC} are \emph{constants}
\bea
B_{ab}\, B^{ab} \rovno \left( \frac{4\Lambda}{3 A_0^2}(1-\Lambda A_0^2)\right)^2 =
\left( \frac{8 k - 3 A_0^2}{16 k^2 A_0^2 }\right)^2\,,\\
C_{abcd}\, C^{abcd} \rovno  \frac{16}{3}\Lambda^2 =\frac{3}{4 k^2}
\,.
\eea
Clearly, the Bach tensor vanishes and the solution~\eqref{anothergenNariaimetric} becomes the Nariai spacetime~\eqref{Nariaimetric}, which is an Einstein space (with ${B_{ab}=0}$),  if and only if ${  A_0^2=1/\Lambda \Leftrightarrow A_0^2=\frac{8}{3}k }$. 

The direct-product solution~\eqref{anothergenNariaimetric} is of the Kundt type, and cannot be transformed into the standard static spherically symmetric coordinates~\eqref{Einstein-WeylBH} since
${{\bar r}=\Omega=\,}$const.

\vspace{5mm}

\noindent
\textbf{To conclude}: The class $[0,2]^\infty$ with $\Lambda=3/(8k)$ necessarily leads to the metric \eqref{anothergenNariaimetric}. It represents another Bachian generalization of the Nariai solution~\eqref{Nariaimetric} which belongs to the Kundt family.

\vspace{5mm}

\subsection{Pleba\'{n}ski--Hacyan spacetime in the class ${[N,P]=[0,<2]^\infty}$ }
\label{sec_[0,m2]infty}

In what follows we will show that, in fact, this class admits only the Kundt-type spacetimes $[0,1]^\infty$ and $[0,0]^\infty$.

First, for ${P\not\in \mathbb{N}}$, Eq.~\eqref{KeyEq1INF} implies ${C_0=0}$ which must be nonzero by definition. Therefore, $P$ {is necessarily an integer}.

For all $P<2$, Eqs.~\eqref{KeyEq2INF} and \eqref{KeyEq3INF} at the order $r^{0}$
give the conditions  ${3 A_0^2(1-\Lambda A_0^2 )- 2k=0}$ and ${2\Lambda A_0^2-1=0}$, respectively, which imply
\be
\Lambda =\frac{3}{8k}>0\,,\qquad\hbox{and}\qquad
A_0 = \pm\sqrt{\frac{4k}{3}}=\pm \frac{1}{\sqrt{2\Lambda}}\,. \label{Iaa0inf}
\ee
Let us now investigate the following three distinct possibilities, namely ${P=1}$, ${P=0}$, and ${P<0}$.
\vspace{5mm}

\noindent
$\bullet$ In the case $[0,1]^\infty$, Eq.~\eqref{KeyEq1INF} at the orders $r^{-3}$ and $r^{-4}$  gives
${A_1=A_2=0}$, respectively, while at the orders $r^{-5}$, $r^{-6}$ and  $r^{-7}$ it yields ${A_3=\tfrac{2 kC_2}{3 A_0}}$, ${A_4=\tfrac{2k C_3}{A_0}}$, and ${A_5=\tfrac{4kC_4}{A_0}}$, respectively.
At the order $r^{-4}$, Eq.~\eqref{KeyEq3INF} implies ${C_2=0}$. Let us use the mathematical induction to {prove that all} $C_\ell,\ A_{\ell-1}$ vanish for ${\ell\geq 2}$. So, let us assume that ${A_1=A_2=\dots=A_j=0}$,  ${C_2=C_3=\dots=C_{j-1}=0}$, and ${A_{l+1}=\tfrac{k}{3A_0}(l-1)l\, C_l}$ for ${l\in \{j,\, j+1\}}$. We prove that then ${C_{j}=0=A_{j+1}}$ and also that $A_{j+3}=\tfrac{k}{3A_0}(j+1)(j+2)\,C_{j+2}$. Indeed, Eq.~\eqref{KeyEq1INF} at the order $r^{-(j+5)}$ gives
\be
A_{j+3}=\frac{k}{3A_0}(j+1)(j+2)\,C_{j+2}\,,
\ee
while
\eqref{KeyEq3INF}  at the order $r^{-(j+2)}$, using \eqref{Iaa0inf} and also the expressions for $A_{j+1},\, A_{j+2}$, yields
\be
C_0 \, C_j=0\,,
\ee
which implies ${C_j=0=A_{j+1}}$, and thus ${C_\ell,\, A_{\ell-1}}$ vanish for  all $\ell\geq 2$.
Therefore,
\be
{\bar r}= \Omega = A_0 \,,\qquad
\H(r)  = C_0\,r + C_1\,.\label{[0,2]infty-caseb-PlebHacyan}
\ee
The metric~\eqref{BHmetric} with such functions can further be simplified using the coordinate transformation
\be
u = \frac{2}{C_0}\,\ln\bigg( \frac{\tilde u}{A_0^2} \bigg) \,,\qquad
r = \frac{\tilde u}{A_0^2} \bigg(\frac{C_0}{2}\,\tilde{r} + \frac{1}{C_0}\bigg) - \frac{C_1}{C_0}\,,
\label{[0,2]infty-caseb-transformation}
\ee
to the form
\be
\dd s^2 = A_0^2 \big(\dd \theta^2+\sin^2\theta\,\dd \phi^2\big)-2\,\dd u\,\dd r\,.
\label{PlebnHacyanmetric}
\ee
This is a simple direct-product ${S^2 \times M_2}$ geometry. Actually, this is the  \emph{Pleba\'{n}ski--Hacyan spacetime} with ${\Lambda>0}$, see the metric~(7.19) in \cite{GriffithsPodolsky:2009}. Interestingly, it is a type~D electrovacuum spacetime in Einstein's theory, while here it is a spherically symmetric \emph{vacuum solution to QG}, with the radius $A_0$ of the sphere ${S^2}$ given by
\be
A_0^2 = \frac{4k}{3}=\frac{1}{2\Lambda}\,. \label{Iaa0infradius}
\ee
The curvature invariants are
\begin{align}
R_{ab}\, R^{ab} =  8\Lambda^2 \,, \qquad C_{abcd}\, C^{abcd} = \frac{16}{3}\Lambda^2 \,,\qquad
B_{ab}\, B^{ab} =  \frac{16}{9}\Lambda^4 \,,\label{invPlebHacyan}
\end{align}
i.e., the curvature invariants are constant, and the nonvanishing Bach tensor is uniform. Notice again that, due to ${{\bar r}=A_0}$, this Kundt-type solution cannot be put into the standard static spherically symmetric coordinates.
\vspace{5mm}

\noindent
$\bullet$ In the case $[0,0]^\infty$, Eq.~\eqref{KeyEq1INF} at the orders $r^{-3}$ and $r^{-4}$ again gives
${A_1=A_2=0}$, respectively, while at the orders $r^{-5}$, $r^{-6}$ and  $r^{-7}$, ${A_3=\tfrac{2 kC_1}{3 A_0}}$, ${A_4=\tfrac{2k C_2}{A_0}}$, and ${A_5=\tfrac{4kC_3}{A_0}}$, respectively.
At the orders  $r^{-5}$, $r^{-6}$, $r^{-7}$, Eq.~\eqref{KeyEq3INF} implies
${C_1=C_2=C_3=0}$, respectively. Using the mathematical induction we again prove that all ${C_\ell=0=A_{\ell}}$ for ${\ell\geq 1}$. We assume that ${A_1=A_2=\dots=A_{j+1}=0}$,  ${C_1=C_2=\dots=C_{j-1}=0}$, and ${A_{l+2}=\tfrac{k}{3A_0}(l+1)l\, C_l}$ for ${l\in\{j,\ j+1,\ j+2\}}$
and prove that then ${C_{j}=0=A_{j+2}}$ and ${A_{j+5}=\tfrac{k}{3A_0}(j+4)(j+3)C_{j+3}}$.
Indeed, Eq.~\eqref{KeyEq1INF} at the order $r^{-(j+7)}$ gives
\be
A_{j+5}=\frac{k}{3A_0}(j+3)(j+4) C_{j+3}\,,
\ee
while
\eqref{KeyEq3INF}  at the order $r^{-(j+4)}$, using \eqref{Iaa0inf} and the expressions for $A_{j+2},\ A_{j+4}$, yields
\be
C_0\, C_j=0\,,
\ee
which implies ${C_j=0=A_{j+2}}$ and thus  $C_\ell,\ A_{\ell}$ vanish for all ${\ell\geq 1}$.
Therefore,
\be
{\bar r}= \Omega = A_0 \,,\qquad
\H(r)  = C_0\,,\label{[0,2]infty-caseb-PlebHacyan-b}
\ee
which is the special case of \eqref{[0,2]infty-caseb-PlebHacyan}. With the coordinate transformation
\be
r = \frac{\tilde r}{A_0^2} + \frac{C_0}{2}\,u \,,
\label{[0,2]infty-caseb-transformationnext}
\ee
we again obtain the metric~\eqref{PlebnHacyanmetric} of the Pleba\'{n}ski--Hacyan direct-product spacetime.
\vspace{5mm}

\noindent
$\bullet$ Finally, the case $[0,<0]^\infty$ with ${P<0}$ is, in fact, \emph{empty}. Indeed, using the mathematical induction we show that  $A_i$ for ${1\leq i\leq 1-P}$ vanish.
For all ${P<0}$, Eq.~\eqref{KeyEq1INF} at the order $r^{-3}$ implies ${A_1=0}$. Let us assume that all $A_i$ for ${1\leq i\leq j-1}$, ${j< 2-P}$, vanish. Then Eq.~\eqref{KeyEq1INF}  at the order $r^{-j-2}$  gives ${A_0 A_{j}(j+1)=0}$, which implies that also ${A_j=0}$. Thus the second nonvanishing coefficient (after $A_0$) is $A_{2-P}$ which is determined by the order $r^{P-4}$ of Eq.~\eqref{KeyEq1INF},
\be
A_{2-P}=kP(P-1)\frac{C_0}{3A_0}\,.\label{Iaapm2inf}
\ee
Then, using \eqref{Iaa0inf}, Eqs.~\eqref{KeyEq1INF} and \eqref{KeyEq3INF}
at the orders $r^{2P-6}$ and $r^{2P-4}$ yield
\bea
2(5-2P)A_0 A_{4-2P} +(P-1)A_{2-p}^2-\tfrac{4}{3}kC_{2-P}(1-P)(3-2P)(5-2P)\rovno 0\,,\nonumber\\
-2A_{4-2P}+C_0 A_{2-P}\left[3(2-P)(3-2P)+\tfrac{1}{2}P(P-1)\right]
-6\Lambda A_0 A_{2-P}^2+A_0 C_{2-P}(1-P)(3-2P) \rovno 0\,,\nonumber
\eea
respectively. A linear combination of these two equations, using also  \eqref{Iaa0inf} and \eqref{Iaapm2inf}, leads to
\be
P(P-1) (P-2)(5P^2-21P+20) C_0^2=0\,,
\ee
which for integers ${P<0}$ would imply ${C_0=0}$, a contradiction. Therefore,  this class is indeed empty.

\vspace{5mm}

\noindent
\textbf{To conclude}: The class $[0,<2]^\infty$ includes only the  ${S^2 \times M_2}$ direct-product Pleba\'{n}ski--Hacyan spacetime~\eqref{PlebnHacyanmetric}, \eqref{Iaa0infradius} which is the spherically symmetric vacuum solution to  QG with ${\Lambda =3/(8k)>0}$ and a nonvanishing Bach tensor.  As in the previous case, this Kundt metric cannot be put into the standard static spherically symmetric coordinates.

\vspace{5mm}

\subsection{Solutions with regular Bachian infinity in the class ${[N,P]=[>0,2N+2]^\infty}$ }
\label{sec_[N,2N+2]infty}

Let us investigate the last possible class~\eqref{CaseIII_summaryINF-c}.
Similarly as in Section~\ref{sec_[n,2n+2]}, in Eq.~(\ref{KeyEq3INF}) the first and last terms start with the power ${r^{3N}}$, while the second one with  ${\tfrac{1}{3}A_0r^{N}}$. Since in all sums there are integer steps, to allow ${A_0\not= 0}$, the expression ${3N-N}$ has to be a positive integer.
Thus, $N$ must have the form
\be
N=J/2\,, \qquad\hbox{where}\quad J\in \mathbb{N}\,, \ \  J \ge 2\,,
\ee
(recall that ${N=1/2}$ is not admitted, see \eqref{CaseIII_summaryINF-c}). Consequently, ${N\geq 1}$ and  ${P=2N+2=J+2\geq 4}$. Then from Eq.~\eqref{IIIcLC0} we obtain \emph{discrete values of the cosmological constant}~$\Lambda$ for ${J=2,3,\ldots}$ as
\bea
\Lambda\rovno \frac{3}{32k}\frac{11J^2+12J+4}{1-J^2}\,, \label{[J,]L}\\[2mm]
&& \hbox{so that}\ \
\Lambda\in \big[ -\tfrac{9}{4k}, -\tfrac{33}{32k} \big) \ \mbox{for}\ k>0\,,\ \hbox{and}\ \
\Lambda\in \big( \tfrac{33}{32|k|},\tfrac{9}{4|k|}\big] \ \mbox{for}\ k<0\,, \nonumber \\
C_0\rovno \frac{3}{4k}\frac{A_0^2}{1-J^2} \,.
\eea
Notice that the product $k\Lambda$ is \emph{always negative}.
From the subleading term of~\eqref{KeyEq3INF} we get
\be
A_1=-k\frac{2N(2N+1)(2N-1)}{3(N+1)}\frac{C_1}{A_0}\,,
\ee
and similarly we can express further coefficients for any ${N=J/2}$. Let us  present two examples for  ${J=2}$ and ${J=3}$.
\vspace{5mm}

\noindent
$\bullet$ In the case $[1,4]^\infty$, there are three free parameters $A_0,C_1,C_5$ with the remaining coefficients determined by  Eqs.~\eqref{KeyEq1INF}--\eqref{KeyEq3INF} as
\bea
\Lambda \rovno -\frac{9}{4k}\,, \nonumber\\
A_1 \rovno -k\frac{C_1}{A_0}\,,\ \
A_2=A_3=A_4=0\,,\ \ A_5=k\frac{2C_5}{9A_0}\,,\ \ \dots\,,\\
C_0 \rovno -\frac{A_0^2}{4k}\,,\ \
C_2=-\frac{1}{7}-\frac{3k C_1^2}{2A_0^2}\,,\ \
C_3=\frac{2kC_1}{7A_0^2}+\frac{k^2C_1^3}{A_0^4}\,,\ \
C_4=\frac{32k}{147A_0^2}-\frac{k^2C_1^2}{7A_0^4}-\frac{k^3C_1^4}{4A_0^6}\,,\ldots \nonumber
\eea
and thus the metric functions take the form
\bea
\Omega\rovno A_0\, r-k\frac{C_1}{A_0} + k\frac{2C_5}{9A_0}\, r^{-4}+\cdots\,,\nonumber\\
\H\rovno -\frac{A_0^2}{4k}\,r^4 +C_1\, r^3 - \frac{1}{7}\bigg(1+\frac{21k C_1^2}{2A_0^2}\bigg)\, r^2
+\frac{2kC_1}{7A_0^2}\bigg( 1 + \frac{7kC_1^2}{2A_0^2}\bigg)\, r +
\cdots\,.\label{rozvoj[J,]}
\eea

\vspace{5mm}

\noindent
$\bullet$ In the case $\left[\frac{3}{2},5\right]^\infty$, there are three free parameters $A_0,C_1,C_7$, and
\bea
\Lambda \rovno  -\frac{417}{256 k}\,,\nonumber\\
A_1 \rovno  -\frac{16 k C_1}{5 A_0}\,,\ \
A_2 =\frac{128 k^2 C_1^2}{75 A_0^3}\,,\ \
A_3=\frac{2048 k^3 C_1^3}{3375 A_0^5}\,,\ \
A_4=\frac{8192 k^4 C_1^4}{16875 A_0^7}\,,\ \
A_5 =\frac{131072 k^5 C_1^5}{253125 A_0^9}\,, \ldots\,,
\nonumber\\
C_0 \rovno  -\frac{3 A_0^2}{32 k}\,, \ \
C_2=-\frac{64 k C_1^2}{15 A_0^2}\,,\ \
C_3= -\frac{4}{11025 A_0^4} (225 A_0^4 - 25088 k^2 C_1^3)\,,
\\
C_4\rovno \frac{256k C_1}{165375 A_0^6} (225 A_0^4 - 6272 k^2 C_1^3)\,,\ \
C_5 = -\frac{4096 k^2 C_1^2}{12403125 A_0^8} (1125 A_0^4 - 12544 k^2 C_1^3)\,,
\nonumber\\
C_6\rovno \frac{8480 k}{38073 A_0^2}\,,
\ \
\ldots \,, \nonumber
\eea
with $C_7$ appearing in ${A_7, A_8, \ldots}$ and ${C_8, C_9, \ldots\,}$.

For all allowed values of $N$, the curvature invariants \eqref{invB}, \eqref{invC} \emph{approach a constant} as ${r\to\infty}$,
\bea
B_{ab}\,B^{ab}(r\to\infty) \rovno P^2(P-1)^2(P-3)^2(11P^2-32P+24)\frac{C_0^4}{144A_0^8} +\cdots\,,\label{invBB[>0,P]}\\
C_{abcd}\,C^{abcd}(r\to\infty) \rovno P^2(P-1)^2 \frac{C_0^2}{3A_0^4}+\cdots\,. \label{invCC[>0,P]}
\eea
For all the permitted values of ${P=4,5,\ldots}$, the \emph{Bach invariant \eqref{invBB[>0,P]} cannot vanish}, and therefore this class does not include the Schwazschild--(A)dS solution as a special subcase or a limit.

Since
\be
{\bar r}=\Omega(r)=A_0\,r^N+\cdots\ \to\ \infty \,,
\ee
cf.~\eqref{to static}, the limit ${r\to\infty}$  corresponds to an \emph{asymptotic region} far away from the origin at ${\bar r = 0}$. In this asymptotic region,  the  metric functions \eqref{rcehf}
behave as
\bea
h \rovno -A_0^2 C_0\,r^{4N+2}+\cdots\ \sim\ {\bar r}^{\,4+2/N}+\cdots \rightarrow \infty \,,\\
f \rovno -N^2C_0\, r^{2N}+\cdots \ \sim\  {\bar r}^{\,2}+\cdots \rightarrow \infty \,.
\eea

Finally, let us observe that in the notation of \cite{Stelle:1978,LuPerkinsPopeStelle:2015,LuPerkinsPopeStelle:2015b}, these solutions  correspond to the families ${(s,t)=(-2,4+4/J)_\infty}$ with $ J\in \mathbb{N}$, ${J\geq 2}$, i.e., with the parameter ${t\in (4,6]}$. Interestingly, together with the case of Section~\ref{sec_[n,2n+2]} with the same  spectrum of $\Lambda$ (cf. \eqref{[mJpul,]L} and \eqref{[J,]L}), they both describe solutions with regular Bachian infinity and together correspond to families
$(s,t)=(-2,4+4/J)_{\infty,{J\geq 2}}\cup
	(-2,4-4/J)_{\infty,{J\geq 3 }}	$, i.e.,
${(s,t)=(-2,t)_\infty}$, where ${t \in [\tfrac{8}{3},4) \cup (4,6]} $.

\vspace{5mm}

\noindent
\textbf{To conclude}:
The class $[J/2,J+2]^\infty$ with $\Lambda$ uniquely determined by \eqref{[J,]L} is, in fact, an infinite discrete family of metrics parameterized by an integer ${J\ge2}$. They all have a regular Bachian infinity since both the Bach and Weyl invariants approach finite nonzero values \eqref{invBB[>0,P]}, \eqref{invCC[>0,P]} in the asymptotic physical region ${\bar r \rightarrow \infty}$. In particular, for ${J=2}$, the metric functions are  expressed in terms of the series~\eqref{rozvoj[J,]}.
	
\vspace{5mm}

\section{Identifying the classes in the spherically symmetric coordinates}
\label{summary}

Up to now, we mostly used the Kundt coordinates to derive and analyze the vacuum spherically symmetric solutions to QG  with a nonvanishing cosmological constant, found in this paper as power series in powers of ${\Delta \equiv r-r_0}$ and $r^{-1}$.
In this section, we identify them in terms of the standard notation  used in the literature in the spherically symmetric coordinates. This allows us to establish the connection with the previously known classes for ${\Lambda=0}$.

In Table \ref{tbl:000}, solutions with ${\Lambda=0}$ found in \cite{Stelle:1978,LuPerkinsPopeStelle:2015b} near the origin, and  in \cite{LuPerkinsPopeStelle:2015b,PerkinsPhD} near a finite point ${{\bar r}\to {\bar r}_0 \ne0}$ are listed with ${(s,t)}$ and ${(w=-s,t)}$ denoting the powers  of the {leading terms} of a Laurent expansion of the two metric functions
in the {standard spherically symmetric form} \eqref{Einstein-WeylBH}, respectively,\footnote{Here,   the arguments of the metric functions $A(r), B(r)$ of \cite{LuPerkinsPopeStelle:2015b} are relabeled to $\bar r$.}
\bea
f^{-1}({\bar r}) \rovno A({\bar r}) \sim {\bar r}^{\,s}
\sim {\bar r}^{\, -w}
\,, \label{rcef2}\\
h({\bar r})      \rovno B({\bar r}) \sim {\bar r}^{\,t} \,.\label{rceh2}
\eea
	\begin{table}[H]
		\begin{center}
				\begin{tabular}{|c||c|}
					\hline
					Family $(s,t)_0$ near the origin&   Family $(w,t)_{\bar r_0}$ near ${\bar r_0}$\\[0.5mm]
					\hline\hline
					$(0,0)_0$&	$(1,1)_{\bar r_0}$\\[1mm]
					$(1,-1)_0$& 	$(0,0)_{\bar r_0}$
					\\[1mm]
					$(2,2)_0$&	$(1,0)_{\bar r_0}$ \\[1mm]
					\hline
				\end{tabular} \\[2mm]
					\caption{Families  of spherically symmetric solutions to QG with $\Lambda=0$  near the origin found
					in \cite{Stelle:1978,LuPerkinsPopeStelle:2015b}, and  near a finite point  ${{\bar r}\to {\bar r}_0 \ne0}$ found in \cite{LuPerkinsPopeStelle:2015b,PerkinsPhD}. The subscripts ``$\,_0$'' and ``$\,_{{\bar r}_0}$'' indicate an expansion around ${\bar r =0}$ and ${\bar r=\bar r_0}$, respectively.}
				\label{tbl:000}
		\end{center}
\end{table}

Since our calculations are performed  in the ``nonphysical'' Kundt coordinates, for the physical interpretation, it is important to identify  the corresponding points in the spherically symmetric coordinates, around which expansions are taken.  This can be obtained using
 \eqref{to static},  \eqref{rozvojomeg0}, and \eqref{rozvojomegINF}, and  is summarized in Table \ref{tbl:001}.
	\begin{table}[H]
		\begin{center}
				\begin{tabular}{|c||c|c|}
					\hline
					$\bar r\rightarrow$&   Corresponding $n$ for $r\rightarrow r_0$&  Corresponding $N$ for $r\rightarrow \infty$\\[0.5mm]
					\hline\hline
					$0$ &$>0$	& $<0$\\[1mm]
					$\bar r_0$ &$=0$&
					$=0$	
					\\[1mm]
					$\infty$ &$<0$&
					$>0$ \\[1mm]
					\hline
				\end{tabular} \\[2mm]
					\caption{
				The correspondence between the points around which expansions are being performed in the Kundt coordinate $r$ and the spherically symmetric radial coordinate $\bar r=\Omega(r)$. 	}
				\label{tbl:001}
		\end{center}
\end{table}

 Now, let us study the relation between the families $(s,t)$ or $(w,t)$, and $[n,p]$ and $[N,P]^\infty$. The cases with ${n=0}$ in \eqref{4classes} or ${N=0}$ in \eqref{2classes} have to be treated separately. Moreover, they contain special subcases that in the notation $(w,t)$ represent \emph{new classes with noninteger steps} in ${\bar\Delta={\bar r -\bar r_0}}$.

\subsection{Classes   ${[n\not=0,p]}$ and ${[N\not= 0,P]^\infty}$}

	Using the expressions \eqref{rcehf} with ${\bar{r} = \Omega(r)}$  and \eqref{rozvojomeg0}, \eqref{rozvojcalH0} for ${r \to r_0}$ when  ${n\not=0}$, and \eqref{rozvojomegINF}, \eqref{rozvojcalHINF} for	${r \to \infty}$ when ${N\not=0}$, we obtain the relations between $(s,t)$, introduced by \eqref{rcef2}, \eqref{rceh2}, and $[n,p]$, $[N,P]^\infty$ as
	\bea
	(s,t) \rovno  \left(\tfrac{2-p}{n}\,,  2+\tfrac{p}{n}\right), \label{st-np}\\
	(s,t) \rovno \left( \tfrac{2-P}{N}\,, 2+\tfrac{P}{N}\right), \label{st-NP}
	\eea
	respectively. For the solutions found in this paper, these are summarized in Table \ref{tbl:0003}.

	\begin{table}[H]
		\begin{center}
				\begin{tabular}{|c||c|c|}
					\hline
					Class $[n,p]$ &
					Class $[N,P]^\infty$ &
					Corresponding family $(s,t)$
					\\[0.5mm]
					\hline\hline
					$[n,p]$& --&	$\left(\frac{2-p}{n}\,,  2+\frac{p}{n}\right)$  \\[1mm]\hline
					$[1,0]$ &  -- &
					$(2,2)_{0}$\\[1mm]
					$[-1,2]$ &  -- &
					$(0,0)_{\infty}$\\[1mm]
					$[-1,0]$ &  -- &
					$(-2,2)_{\infty}$\\[1mm]
					$\left[-\tfrac{J}{2},2-J\right]$ & -- & $\left(-2,4\tfrac{J-1}{J}
					\right)_\infty$
					\\
					\hline\hline
					-- & $[N,P]^\infty$& $\left(\frac{2-P}{N}\,,  2+\frac{P}{N}\right)$
					\\[1mm]\hline
					-- & $[-1,3]^\infty$&
					$(1,-1)_0$\\[1mm]
					-- & $[-1,2]^\infty$&
					$(0,0)_0$\\[1mm]
					-- & $\left[\tfrac{J}{2},J+2 \right]^\infty$   & $\left(-2,4\tfrac{J+1}{J} \right)_\infty$
					\\[1mm]				
					\hline
				\end{tabular} \\[2mm]
				\caption{Values of $(s,t)$ for all solutions  $[n,p]$ with $n\not=0$, and $[N,P]^\infty$ with $N\not=0$ found in this paper. The subscripts ``$\,_0$'' and ``$\,_\infty$'' indicate the expansion for ${{\bar r}\to 0}$ and ${{\bar r}\to\infty}$, respectively. }
				\label{tbl:0003}
		\end{center}
\end{table}
Note that power series expansions with integer steps in the powers of $\Delta$ in $[n,p]$ with $n\not=0$, and in the powers of $r^{-1}$ in $[N,P]^\infty$ with $N\not=0$  correspond to integer steps in ${\bar\Delta}$ since ${\Delta\propto \bar\Delta}$ and ${r\propto \bar\Delta^{-1}}$, respectively.

Additional information about the cases discussed in this section is summarized  in subsequent Tables~\ref{tbl:011}, \ref{tbl:02}, and~\ref{tab:3}.

\subsection{Classes $[0,p]$ and $[0,P]^\infty$}
\label{sec_[0,p]}

When ${n=0}$ or ${N=0}$, there are the following subcases (recall that in the class $[0,2]^\infty$, necessarily $A_1=A_2=0$, see Section \ref{sec_[0,2]infty}):
\vspace{2mm}

\noindent
	$\bullet$ {\bf{Classes  $[0,p]$ with  ${a_1\not=0}$
		}}\\
     		In the generic case with ${a_1\not=0}$,
     		using \eqref{rcehf}, \eqref{rozvojomeg0}, \eqref{rozvojcalH0},
     		the relations are
	\be
	(w,t)=  (p,p)_{\bar r_0}\,.
	\ee
	These solutions 
	 are summarized in Table \ref{tbl:0031}.
	\begin{table}[H]
		\begin{center}
				\begin{tabular}{|c||c|}
					\hline
					Class $[n=0,p]$ &
					Corresponding family $(w,t)$ for $\bar r\rightarrow \bar r_0$
					\\[0.5mm]
					\hline\hline
					$[0,p]$& 	$(p,p)_{\bar r_0}$  \\[1mm]\hline
					$[0,0]$  &
					$(0,0)_{\bar r_0}$\\[1mm]
					$[0,1]$ &
					$(1,1)_{\bar r_0}$\\[1mm]
					$[0,2]$  &
					$(2,2)_{\bar r_0}$
					\\[1mm]				
					\hline
				\end{tabular} \\[2mm]
				\caption{Values of $(w,t)$ for all solutions  $[0,p]$ with ${a_1\not=0}$.
					 }
				\label{tbl:0031}
		\end{center}
\end{table}
Similarly as in the previous case, integer steps in $\Delta$
correspond to integer steps in  $\bar\Delta$
 since  ${{\bar\Delta \equiv \bar r -\bar r_0 \sim a_1 \Delta}}$.

\vspace{2mm}

\noindent
	$\bullet$ {\bf{Classes  $[0,p]$ with   ${a_1=0\not=a_2}$
					}}\\
In this case,
\be
(w,t) =\left(\tfrac{p+2}{2}\,,\tfrac{p}{2}\right)_{\bar r_0,1/2}\,
\ee
and these solutions  are summarized in Table \ref{tbl:0032}.

	\begin{table}[H]
		\begin{center}
				\begin{tabular}{|c||c|}
					\hline
					Class $[n=0,p]$ &
					Corresponding family $(w,t)$ for $\bar r\rightarrow \bar r_0$
					\\[0.5mm]
					\hline\hline
					$[0,p]_{a_1=0}$& 	$\left(\frac{p+2}{2}\,,\frac{p}{2}\right)_{\bar r_0,1/2}$  \\[1mm]\hline
					$[0,0]_{a_1=0}$ &
					$(1,0)_{\bar r_0,1/2}$\\[1mm]
					$[0,1]_{a_1=0}$ &
					$\left(\tfrac{3}{2},\tfrac{1}{2} \right)_{\bar r_0,1/2}$\\[1mm]
					\hline
				\end{tabular} \\[2mm]
				\caption{
				 Values of $(w,t)$ for all solutions  $[0,p]$ with $a_1=0\not=a_2$.
				 			 }
				\label{tbl:0032}
		\end{center}
\end{table}
Integer steps in $\Delta$
lead to half integer steps in $\bar \Delta$ since ${\bar\Delta\sim a_2\Delta^2}$.

\vspace{2mm}

\noindent
	$\bullet$ {\bf{Classes  $[0,p]$ with   ${a_1=a_2=0\not=a_3}$, and  $[0,P]^\infty$ with  ${{A_1=A_2=0\not=A_3}}$}}\\
In this case, using also \eqref{rozvojomegINF}, \eqref{rozvojcalHINF},
\bea
(w,t)\rovno \left(\tfrac{p+4}{3},\tfrac{p}{3}\right)\,,\\
(w,t)\rovno \left(\tfrac{8-P}{3}, -\tfrac{P}{3} \right)_{\bar r_0,1/3}\,
\eea
and these solutions  are summarized in Table \ref{tbl:0033}.
	\begin{table}[H]
		\begin{center}
				\begin{tabular}{|c|c|c|}
					\hline
					Class $[n=0,p]$ &
					Class $[N=0,P]^\infty$ &
					Corresponding family $(w,t)$ for $\bar r\rightarrow \bar r_0$
					\\[0.5mm]
					\hline\hline
					$[0,p]_{a_1=a_2=0}$& --&	$\left(\frac{p+4}{3}\,,\frac{p}{3}\right)_{\bar r_0,1/3}$  \\[1mm]\hline
					$[0,0]_{a_1=a_2=0}$ &  -- &
					$\left(\tfrac{4}{3},0 \right)_{\bar r_0,1/3}$\\[1mm]
					$[0,1]_{a_1=a_2=0}$ &  -- &
					$\left(\tfrac{5}{3},\tfrac{1}{3} \right)_{\bar r_0,1/3}$\\[1mm]
					\hline\hline
					-- & $[0,P]^\infty_{_{A_1=A_2=0}}$& $\left(\tfrac{8-P}{3},-\tfrac{P}{3} \right)_{\bar r_0,1/3}$
					\\[1mm]\hline
					-- & $[0,2]^\infty_{_{A_1=A_2=0}}$& $\left(2,-\tfrac{2}{3}\right)_{\bar r_0,1/3}$
					\\[1mm]				
					\hline
				\end{tabular} \\[2mm]
				\caption{Values of $(w,t)$ for all solutions  $[0,p]$ with $a_1=a_2=0\not=a_3$ and $[0,P]^\infty$ with $A_1=A_2=0\not=A_3$.}
				\label{tbl:0033}
		\end{center}
\end{table}
Integer steps in $\Delta$ and $r^{-1}$ lead to steps in $\bar \Delta^{1/3}$ since ${{{\bar \Delta}\sim a_3\Delta^3}}$ and ${{{\bar \Delta}\sim A_3r^{-3}}}$, respectively.

\vspace{2mm}

\noindent
	$\bullet$ {\bf{Classes  $[0,P]^\infty$ with ${A_1=A_2=\dots=A_{L-1}=0\not=A_L\,,\ L\geq 4}$}}\\
In this case,
\be
(w,t)=\left(\tfrac{2L+2-P}{L},-\tfrac{P}{L}\right)_{\bar r_0,1/L}\,,
\ee
see  Table \ref{tbl:0034}.
	\begin{table}[H]
		\begin{center}
				\begin{tabular}{|c|c|}
					\hline
					Class $[N=0,P]^\infty_{_{A_1=\dots=A_{L-1}=0}}$ &
					Corresponding family $(w,t)$ for $\bar r\rightarrow \bar r_0$
					\\[0.5mm]
					\hline\hline
					 $[0,2]^\infty_{{A_1=\dots=A_{L-1}=0}}$ & $\left(2,-\tfrac{2}{L}\right)_{\bar r_0,1/L}$
					\\[1mm]				
					\hline
				\end{tabular} \\[2mm]
				\caption{Values of $(w,t)$ for the solutions  $[0,2]^\infty$ with ${A_1=A_2=\dots=A_{L-1}=0\not=A_L}$.}
				\label{tbl:0034}
		\end{center}
\end{table}
Integer steps in $r^{-1}$ lead to steps in $\bar \Delta^{-1/L}$ since ${{{\bar \Delta}\sim A_Lr^{-L}}}$.
\vspace{8mm}

Apart from the generic cases ${[0,0]}$, ${[0,1]}$, ${[0,2]}$, ${[0,2]^\infty}$,  discussed
in Sections~\ref{SchwaBach_[n,p]=[0,0]}, \ref{SchwaBach_[n,p]=[0,1]}, \ref{sec_[0,2]}, \ref{sec_[0,2]infty}, respectively, only
the special case $[0,2]_{a_1=0}$ has been studied so far, see Sec. \ref{sec_case[0,2]b}.
For completeness, let us briefly discuss both  generic and also special cases in the $[0,p]$ and $[0,P]^\infty$ classes. Recall that the classes with ${[n=0,p>0]}$ contain horizons at $r=r_0=r_h$.
\\[2mm]
$\bullet$ Generic family $[n,p]=[0,0]$ has  the highest number of free parameters, and it seems that it can be connected  to all the other solutions. It represents an expansion around a generic point in these spacetimes;\\[1mm]
$\bullet$ family ${[0,0]_{a_1=0}}$, for which  the Bach invariant \eqref{invB} is always nonvanishing, represents a generalization of the family ${(1,0)_{\bar r_0,1/2}}$  of
\cite{LuPerkinsPopeStelle:2015b,PerkinsPhD} with $\Lambda\not=0$. It is the only family that describes a wormhole since
${f=0}$ and ${h\not=0}$  at ${{\bar r}_0}$ ($\H \not= 0$, $\Omega'=0$ implies ${n=0=p}$,  ${a_1=0}$), see \cite{LuPerkinsPopeStelle:2015b}.
 In particular, it corresponds to a wormhole with two different patches (half-integer wormhole), see \cite{PerkinsPhD};\\[1mm]
$\bullet$ family
${[0,0]_{a_1=0,E}}$
(only even powers in $\Delta$ are considered,  this is indicated by the subscript ``$\,_E$''), for which the Bach invariant \eqref{invB} is always nonvanishing, is a generalization of the family
 ${(1,0)_{\bar r_0}}$ of \cite{LuPerkinsPopeStelle:2015b,PerkinsPhD} for $\Lambda\not=0$ and
describes a   wormhole with two same patches (integer wormhole),  see \cite{PerkinsPhD};\\[1mm]
$\bullet$ family ${[0,0]_{a_1=0=a_2}}$, which can be denoted as ${\left(\tfrac{4}{3},0\right)_{\bar r_0,1/3}}$ in the notation
of \cite{LuPerkinsPopeStelle:2015b}, is a generalization of the family found in  \cite{PodolskySvarcPravdaPravdova:2020} in the case $\Lambda=0$;\\[1mm]	
 $\bullet$ family  $[0,1]=(1,1)_{\bar r_0}$ is the Schwarzschild--Bach--(A)dS black hole,
see Sec.~\ref{SchwaBach_[n,p]=[0,1]}; \\[1mm]
$\bullet$ family  ${[0,1]_{a_1=0}}={\left(\tfrac{3}{2},\tfrac{1}{2}\right)_{\bar r_0,1/2}}$ is a generalization of the black hole found in \cite{PodolskySvarcPravdaPravdova:2020}  in the case $\Lambda=0$;
\\[1mm]
$\bullet$ family  $[0,2]=(2,2)_{\bar r_0}$ represents extreme black holes (for generic $\Lambda$ the extreme Schwarzschild--dS black hole, for discrete values of $\Lambda$ the extreme higher-order discrete  Schwarzschild--Bach--dS black holes, for
$\Lambda=\tfrac{3}{8k}$ the extreme  Bachian-dS black hole), see Sec. \ref{sec_[0,2]};\\[1mm]
$\bullet$ family  $[0,2]_{a_1=0}$ with $\Lambda=\frac{3}{8k}$  represents the Bachian generalization of the Nariai spacetime with an extreme horizon, belonging to the  Kundt class, see Sec.~\ref{sec_case[0,2]b};\\[1mm]
$\bullet$ family  $[0,2]^\infty$ with a generic $\Lambda$ (necessarily ${A_i=0}$ for all ${i\geq 1}$) represents  the Nariai spacetime belonging to the Kundt family,  see Sec.~\ref{sec_Nariai-Bach};\\[1mm]
$\bullet$ families $[0,2]^\infty_{A_1=\dots=A_{L-1}=0} = \left(2,-\tfrac{2}{L}\right)_{\bar r_0,1/L}$, starting with ${L=3}$, and
$\Lambda$ given by \eqref{[0,2]inf_L} represent  the higher-order discrete
Nariai-Bach spacetimes with steps in $\bar r^{-1/L}$, see Sec.
\ref{sec_Nariai-Bach};\\[1mm]
$\bullet$ family  $[0,2]^\infty$ with ${\Lambda=\frac{3}{8k}}$
(necessarily ${A_i=0}$ for all ${i\geq 1}$) represents another Bachian generalization of the   Nariai spacetime belonging to the Kundt family,  see Sec. \ref{sec_[0,2]nekb};\\[1mm]
$\bullet$ family  $[0,<2]^\infty$: only the solutions $[0,0]^\infty$, $[0,1]^\infty$ exist, and they represent the  Pleba\'{n}ski--Hacyan 
 solution belonging to the Kundt family, see Sec.~\ref{sec_[0,m2]infty}.

\section{{Summary} }
\label{sec_sum2}

To conclude, let us summarize the solutions discussed in this paper.

First, all families compatible with the field equations \eqref{Eq1}--\eqref{Eq2} in the Kundt coordinates as ${r \to r_0}$ and ${r \to \infty}$, in terms of the series expansions \eqref{rozvojomeg0}--\eqref{DElta} and \eqref{rozvojomegINF}--\eqref{rozvojcalHINF},
are summarized in Tables~\ref{tbl:011} and \ref{tbl:02}, respectively. Their physical interpretation and the reference to the corresponding section, in which these solutions are studied, are also indicated. Note that the special subcases of the cases with ${n=0}$
are not included here, and can be found in Section~\ref{sec_[0,p]} and  in Table~\ref{tab:3}.

In Table \ref{tab:3},  all the classes and subclasses found and identified in this paper, both in the standard and Kundt coordinates, are summarized. They are arranged according to the regions, in which the expansions are taken in the ``physical'' radial coordinate~$\bar r$.

\begin{table}[t!]
	\begin{center}
		\begin{footnotesize}
			\begin{tabular}{|c||c||c|c|c|c|c|c|}
				\hline
				Class $[n,p]$ &   Family $(s,t)$ &$\Lambda$& Interpretation & Section\\[0.5mm]
				\hline\hline
				$[-1,2]$ &   $(0,0)_\infty$ & $0$& Schwarzschild black hole   (S)       & \ref{Schw_[n,p]=[-1,2]} \\[1mm]
				$[0,1]$  &   $(-1,1)_{\bar r_0}$ & any & Schwarzschild--Bach--(A)dS black hole (near the horizon) (S) & \ref{SchwaBach_[n,p]=[0,1]},
				\ref{analysisofScwaBacgAdS} \\[1mm]
				$[0,0]$  &   $(0,0)_{\bar r_0}$ &any& generic solution, including the Schwa--Bach--(A)dS black hole (S) & \ref{SchwaBach_[n,p]=[0,0]}\\[1mm]
				$[0,2]$ & $(-2,2)_{\bar r_0}$ & any& extreme Schwarzschild--dS black hole (near the horizon) (S)  & \ref{sec_case[0,2]a}
				\\[1mm]
				$[0,2]$ & $(-2,2)_{\bar r_0}$ & disc.& extreme higher-order discrete  Schwa--Bach--dS black holes (S) &
				\ref{sec_case[0,2]a} 		
				\\[1mm]
				$[0,2]$ & $(-2,2)_{\bar r_0}$ & $\tfrac{3}{8k}$&  extreme  Bachian--dS black hole (S) &
				\ref{sec_case[0,2]b}		
				\\[1mm]
					$[-1,0]$ & $(-2,2)_{\infty}$ & any&Schwarzschild--(A)dS black hole (S) &
				\ref{IIIa1}	
				\\[1mm]
				$[-1,0]$ & $(-2,2)_{\infty}$ & disc.&higher-order discrete  Schwa--Bach--(A)dS black holes (S)&
				\ref{IIIa2}		
				\\[1mm]
				$[1,0]$  &   $(2,2)_0$ &any& Bachian singularity (near the singularity) (nS)       &  \ref{SchwaBach_[n,p]=[1,0]}\\[1mm]
				$[0,>2]$& & & empty  &	  \ref{sec_[0,>2]}\\[1mm]		
				$\left[-\tfrac{J}{2},2-J\right] $, 
				& $\left(-2,\left[ \tfrac{8}{3},4\right) \right)_\infty$  &disc.& solutions with regular Bachian infinity (nS)  &
				\ref{sec_[n,2n+2]}	\\
				$J\in \mathbb{N}$,  $J\geq 3$ & & & &
				\\[1mm]
				\hline
			\end{tabular} \\[2mm]
			\caption{All solutions to QG  that can be written as the power series \eqref{rozvojomeg0}--\eqref{rozvojcalH0},
				expanded in the Kundt coordinates around any constant value~$r_0$. For some solutions, only specific discrete values of $\Lambda$ are allowed (indicated by ``disc."). The symbols ``(S)" and ``(nS)" indicate that a class contains or does \emph{not} contain the Schwarzschild--(A)dS black hole as a special case. Note that the second and fourth columns apply only to the generic cases with $a_1\not=0$ (see Section~\ref{sec_[0,p]} for the special subcases of the cases with ${n=0}$).	}
			\label{tbl:011}
		\end{footnotesize}
	\end{center}
\end{table}

\qquad

\vspace{20mm}

\begin{table}[b!]
	\begin{center}
		\begin{footnotesize}
			\begin{tabular}{|c||c||c|c|c|c|c|c|}
				\hline
				Class $[N,P]^\infty$ &   Family $(s,t)$ &$\Lambda$& Interpretation & Section\\[0.5mm]
				\hline\hline
				$[-1,3]^\infty$  &   $(1,-1)_0$ & any&	Schwa--Bach--(A)dS black hole
				(near the singularity) (S) & \ref{Schw_[N,P]=[-1,3]} \\[1mm]
				$[-1,2]^\infty$  &   $(0,0)_0$  &any& Bachian (A)dS vacuum  (near the origin)  (nS) & \ref{Schw_[N,P]=[-1,2]}\\[1mm]
				$[0,2]^\infty$  & --    & any& Nariai spacetime  (K, nS)  & \ref{sec_Nariai-Bach} 
				\\[1mm]
				$[0,2]^\infty$  &   $ (-2,-2/L)_{\bar r_0}$  &  disc.& higher-order discrete Nariai--Bach solutions (nS)  & \ref{sec_Nariai-Bach}
				\\[1mm]
$[0,2]^\infty$  &   --  & $\tfrac{3}{8k}$  & Bachian generalization of the Nariai spacetime  (K, nS)  & \ref{sec_[0,2]nekb} 
\\[1mm]
				$[0,1]^\infty$  & --   & $\tfrac{3}{8k}$ &
				Pleba\'{n}ski--Hacyan spacetime
				 (K, nS)  & \ref{sec_[0,m2]infty}\\[1mm]
				$[0,0]^\infty$  & --   & $\tfrac{3}{8k}$ & Pleba\'{n}ski--Hacyan spacetime (K, nS)  & \ref{sec_[0,m2]infty}\\[1mm]
				$\left[\tfrac{J}{2},J+2\right]^\infty$ 
				 &   $(-2,(4,6] )_\infty $  & disc.& solutions with regular Bachian infinity (nS)   & \ref{sec_[N,2N+2]infty}\\[1mm]
                		$J\in \mathbb{N}$,  $J\ge2$
                &     & &    &  \\[1mm]
				\hline
			\end{tabular} \\[2mm]
			\caption{All  solutions to QG that can be written as the power series \eqref{rozvojomegINF}--\eqref{rozvojcalHINF},
				expanded in the Kundt coordinates as ${r\to\infty}$. For some solutions only discrete values of $\Lambda$ are allowed (indicated by ``disc."). The symbols ``(S)" and ``(nS)" indicate that the class contains or does \emph{not} contain the Schwarzschild--(A)dS black hole as a special case. Note that some of these solutions are written only in the Kundt form (indicated by``(K)") and cannot be transformed to the standard spherically symmetric coordinates.
				}
			\label{tbl:02}
		\end{footnotesize}
	\end{center}
\end{table}

\newpage

\begin{table}[h!]
	\begin{center}
		\begin{footnotesize} 
			\begin{tabular}{|l|l|l|c|l|}
				\hline
				Family &  $[n,p]$ or $[N,P]^\infty\!\!$ & Parameters  & Free param. & Interpretation
				\\
				\hline
				\hline
				$(s,t)$ & \multicolumn{4}{|c|}{ ${\bar r\rightarrow 0}$ }\\
				\hline
				$(2,2)_0$& $[1,0]$&
				$a_0,c_0,c_1,c_2,r_0 
				$ & $6\rightarrow 4$& Bachian singularity \\
				& & $\Lambda\in\mathbb{R}$ & & (nS) \\
				$(2,2)_{0,E}$& $[1,0]_{E} 
				$ &
				$a_0,c_0,r_0 
				 $ & $4\rightarrow 2$ & Bachian singularity \\
				 & ${ c_1=0=c_3}$ & $\Lambda\in\mathbb{R}$& & (nS) \\
				 $(1,-1)_0$&  $[-1,3]^\infty$&
				$A_0,C_0,C_1,C_3 
				$& $5\rightarrow 3$& Schwa--Bach--(A)dS black hole \\
				& & $\Lambda\in\mathbb{R}$ & & (S) \\
				$(0,0)_0$	& $[-1,2]^\infty$&
				$A_0,C_1,C_2 
				$& $4\rightarrow 2$& Bachian--(A)dS vacuum \\
				& & $\Lambda\in\mathbb{R}$ & & (nS) \\
				\hline
				$(w,t)$ & \multicolumn{4}{|c|}{  ${\bar r\rightarrow \bar r_0}$    }\\
				\hline
				$(1,1)_{\bar r_0}$&$[0,1]$ &
				$a_0,c_0,c_1,r_h 
				$ & $5\rightarrow 3$& Schwa--Bach--(A)dS black hole \\
				& & $\Lambda\in\mathbb{R}$ & &(S) \\
				$(\frac{3}{2},\frac{1}{2})_{\bar r_0,1/2}\!\!$ & $[0,1] 
				$&
				$a_0,c_0,r_0$& $4\rightarrow 2$& ``unusual'' horizon \\
				& ${a_1=0}$ & $\Lambda\in\mathbb{R}$ & & (nS)\\
				$(0,0)_{\bar r_0}$&$[0,0]$&
				$a_0,a_1,c_0,c_1,c_2,r_0
				\!$& $7\rightarrow 5$ & generic solution \\
				& & $\Lambda\in\mathbb{R}$ & & (S)\\
				$(1,0)_{\bar r_0,1/2}$&$[0,0]
				$&
				$a_0,c_0,c_1,c_2,r_0
				$& $6\rightarrow 4$& half-integer wormhole  \\
				& ${a_1=0}$&
				 $\Lambda\in\mathbb{R}$ & &(nS) \\
				$(1,0)_{\bar r_0,E}$&$[0,0] 
				$&
				$a_0,c_0,r_0 
				$& $4\rightarrow 2$& symmetric wormhole  \\
				&${a_1=0=c_1=c_3}$ & $\Lambda\in\mathbb{R}$ & &(nS) \\
				$(\frac{4}{3},0)_{\bar r_0,1/3}$&  $[0,0] 
				$&
				$a_0,c_0,c_1,r_0 
				$& $5\rightarrow 3$ &  not known  \\
				& ${a_1=0=a_2}$ &  $\Lambda\in\mathbb{R}$ & & (nS)\\
				$(2,2)_{\bar r_0}$ & $[0,2]$ & $c_1,r_h 
				$ & $3\rightarrow 1$ &extreme Schwarzschild--dS black hole \\
				& & $\Lambda\geq 0$ & &(S) \\
				$(2,2)_{\bar r_0}$ & $[0,2]$& $c_1,c_{L},r_h$ & $3\rightarrow 1$ & h-o discrete extreme Schwa--Bach--dS \\
				& & $\Lambda\geq 0$ disc. & &(S) \\
				$(2,2)_{\bar r_0}$ &$[0,2]$ & $a_0,c_1,r_h$ & $3 \rightarrow 1$ &  extreme Bachian--dS black hole  \\
				& & $\Lambda=3/(8k)$ & &(nS/S) \\
 				$(2,-\frac{2}{L})_{\bar r_0,1/L}$&  $ [0,2]^\infty
					$&
			$C_1,C_{L},r_0$ & $ 3\rightarrow 1$ & h-o discrete Nariai--Bach solutions  \\
				&
				$ 
				{A_1= \dots =A_{L-1}=0}$
				 & $\Lambda\geq 0$ disc.& &(nS)\\
				\hline
				$(s,t)$ & \multicolumn{4}{|c|}{  ${\bar r\rightarrow \infty}$    }\\
				\hline 	
				$(0,0)_\infty$ & $[-1,2]$&
				$a_0,c_1,r_0$ & $3\rightarrow 1$& Schwarzschild black hole  \\
				& & $\Lambda=0$ & & (S)\\
				$(-2,2)_\infty$ & $[-1,0]$ &
				$a_0,c_3,r_0
				$ & $4\rightarrow 2$& Schwarzschild--(A)dS black hole  \\
				& & $\Lambda\in\mathbb{R}$ & & (S)\\
				$(-2,2)_\infty$& $[-1,0]$&
				$a_0,c_3,c_{L+3},r_0$ & $4\rightarrow 2$& h-o discrete  Schwa--Bach--(A)dS  \\
				& & $\Lambda$ disc. & & (S)\\
				${\left(-2,	\left(4\tfrac{J-1}{J}\right)
					 \right)_\infty}$	& $\left[-\tfrac{J}{2},2-J \right]$& $a_0,c_{2J-1},r_0$
				& $3\rightarrow 1$ & solutions with  regular Bachian infinity  \\
				$\qquad t\in \left[ \tfrac{8}{3},4\right)$ & \qquad  $J\geq 3$ & $\Lambda$ disc.& &(nS)\\
				${\left(-2, \left(4\tfrac{J+1}{J} \right)\right)_\infty }$ & $\left[\tfrac{J}{2},J+2 \right]^\infty$& $A_0,C_1,C_{2J+1}$
				& $3\rightarrow 1$  & solutions with regular Bachian infinity  \\
				$\qquad t\in (4, 6]  $ & \qquad $J\geq 2$ &$\Lambda$ disc. & &(nS) \\
				\hline
			\end{tabular}  \\[2mm]
			\caption{
				All non-Kundt solutions, found and analyzed in this paper, sorted according to the physical regions in which the expansions are taken { in the standard spherically symmetric coordinates}.		
				Subscripts ``$\,_0$'', ``$\,_{{\bar r}_0}$'', and
				``$\,_\infty$'' denote solutions $(s,t)$ or $(w,t)$ (cf. \eqref{rcef2}, \eqref{rceh2}) near ${\bar r=0}$,  ${\bar r={\bar r_0}}$, and  ${\bar r\rightarrow\infty}$, respectively. Subscript ``$\,_E$'' indicates that only even powers are present in the expansion, while ``$\,_{1/2}$'',  and ``$\,_{1/3}$'', and ``$\,_{1/L}$'' indicate that fractional powers are present. The number of free parameters is given before and after removing two parameters by the gauge freedom \eqref{scalingfreedom} in the Kundt coordinates. In usual coordinates, only one parameter can be removed by rescaling \eqref{scaling-t}. The symbols ``(S)'' or ``(nS)'' indicate that a class of solutions contains or does \emph{not} contain the Schwarzschild--(A)dS black hole as a special case, respectively. ``h-o'' stands for the abbreviation of ``higher-order''.}
			\label{tab:3}
		\end{footnotesize}  	
	\end{center}
\end{table}

\

\newpage


\section*{Acknowledgements}

This work has been supported by the Czech Science Foundation Grants
No.~19-09659S (VP, AP) and No.~20-05421S (JP, R{\v S}), and the Research Plan RVO:
67985840 (VP, AP).


\appendix

\section{Derivation and simplification of the field equations }
\label{analysingFE}

To derive the QG  field equations \eqref{Eq1} and \eqref{Eq2} (with the condition \eqref{trace}) for the static spherically symmetric metric in the Kundt coordinates \eqref{BHmetric}, we employ Appendices A--C of \cite{PodolskySvarcPravdaPravdova:2020}.
After substituting (A13)--(A16), (B5), and  (B7)--(B10) of \cite{PodolskySvarcPravdaPravdova:2020} into the field equations \eqref{fieldeqsEWmod},
\BE
R_{ab}-\Lambda \, g_{ab}=4k\, B_{ab}\,,
\label{EWfield equations}
\EE
 we obtain
\begin{align}
 \Omega\Omega''-2{\Omega'}^2 & = \tfrac{1}{3}k\, {\H}'''' \,, \label{Neq_rr} \\
 \big(\Omega^2 {\H}\big)'' -2\Lambda \Omega^4 & = -\tfrac{2}{3}k \big(2\,{\H}{\H}''''+{\H}'{\H}'''
 -{\textstyle\frac{1}{2}}{{\H}''}^2 +2\big) \,, \label{Neq_ru} \\
 \big({\H}\Omega\Omega'\big)'+\Omega^2 -\Lambda \Omega^4 & = \tfrac{1}{3}k \,\big({\H}{\H}''''+{\H}'{\H}'''
 -{\textstyle\frac{1}{2}}{{\H}''}^2 +2 \big) \,, \label{Neq_xx}
\end{align}
which are the nontrivial components $rr$, $ru$, and $xx$, respectively. The $yy$ component is identical to the $xx$ one, and $uu$ is a multiple of  the $ru$ component.

Using (B11) of \cite{PodolskySvarcPravdaPravdova:2020}, the trace \eqref{traceEW} of the field equations  \eqref{EWfield equations}, ${{R}=4\Lambda }$, takes the form
\begin{equation}
\T\equiv{\H}\Omega''+{\H}'\Omega'+{\textstyle \frac{1}{6}} ({\H}''+2)\Omega =
{\textstyle \frac{2}{3}\Lambda \,\Omega^3 } \,,
\label{traceC}
\end{equation}
which indeed follows from (\ref{Neq_rr})--(\ref{Neq_xx}).

In what follows, our goal is to show that  the three nontrivial field equations (\ref{Neq_rr})--(\ref{Neq_xx})
for the two functions $\Omega(r)$ and ${\H}(r)$ \emph{can be reduced just to two equations}.

Introducing a symmetric tensor $J_{ab}$ as
\BE
	J_{ab}\equiv  R_{ab}-\pul  R\, g_{ab}+\Lambda \,  g_{ab} - 4k\,  B_{ab} \,,
	\label{defJab}
\EE
the vacuum QG  field equations  \eqref{fieldeqsEW}, assuming $R=$const.,  take the form
\BE
 J_{ab}=0 \,.
 \label{Jab=0}
\EE
For the metric \eqref{BHmetric}, the non-trivial components of \eqref{defJab} are
\begin{equation}
    J_{rr}\,, \qquad
	J_{uu}=-{\H}\, J_{ru}\,, \qquad
    J_{xx}=\mathcal{J}(r)\, g_{xx}= J_{yy}  \,, \label{EqDep}
\end{equation}
where the function $\mathcal{J}(r)$ is defined as
\begin{equation}
	\mathcal{J} \equiv \Omega^{-2}\Big[\big({\H}\Omega\Omega'\big)'+\Omega^2+\Lambda \Omega^4
	-3\,\T\,\Omega  -\tfrac{1}{3}k \,\big({\H}{\H}''''+{\H}'{\H}'''-{\textstyle\frac{1}{2}}{{\H}''}^2 +2 \big)\Big] \,, \label{calJ}
\end{equation}
and
\bea
J_{rr} \rovno  2\Omega^{-2} \Big[-\Omega\Omega''+2{\Omega'}^2+\tfrac{1}{3}k\,{\H}''''\Big] \,, \label{bJrr}\\
J_{ru} \rovno  \Omega^{-2} \Big[-\tfrac{1}{2}\big(\Omega^2 {\H}\big)''-\Lambda \Omega^4 +3\,\T\,\Omega
	-\tfrac{1}{3}k \big(2\,{\H}{\H}''''+{\H}'{\H}'''-{\textstyle\frac{1}{2}}{{\H}''}^2 +2\big)\Big] \label{bJru}\,.
\eea

Since the Bach tensor is conserved, ${\nabla^b B_{ab}=0}$, see \eqref{Bachproperties}, the contracted Bianchi identities ${\nabla^b R_{ab}=\frac{1}{2} R_{,a}}$  then yield
\BE
	\nabla^b J_{ab}\equiv 0 \,,
	\label{BianchiIdEW}
\EE
which is valid regardless of the form of the field equations.

For the metric \eqref{BHmetric}, this leads to identities
\begin{align}
	& \nabla^b J_{rb} = - \Omega^{-3}\Omega'\big(J_{ij}\,{g}^{ij}
	+{\H} J_{rr}\big)-\Omega^{-2}\big({\H} J_{rr,r}+ J_{ru,r}+\tfrac{3}{2}{\H}' J_{rr}\big)
	\hspace{-25mm}&\equiv 0 \,, \label{BI_r} \\
	& \nabla^b J_{ub} = -2\Omega^{-3}\Omega'\big(J_{uu}+{\H} J_{ru}\big)-\Omega^{-2}\big(J_{uu}+
	{\H} J_{ru}\big)_{,r} &\equiv 0 \,, \label{BI_u} \\
	& \nabla^b J_{ib}\, = \Omega^{-2} J_{ik||l}\,{g}^{kl} &\equiv 0 \,, \label{BI_i}
\end{align}
where the spatial covariant derivative $_{||}$ is calculated with respect to the spatial (2-sphere) part $g_{ij}$ of the Kundt seed metric \eqref{BHmetric}.

Using  \eqref{EqDep}, equations (\ref{BI_u}) and (\ref{BI_i}) are identically satisfied, while \eqref{BI_r} is the only nontrivial one. If the field equations ${J_{rr}\!=\!0}$ and ${J_{ru}\!=\!0}$ hold,
then from \eqref{BI_r} it necessarily follows that ${J_{ij}\,g^{ij} =  J_{xx} \,g^{xx}  + J_{yy}\,g^{yy} =  2\mathcal{J}(r)\equiv0}$ and thus ${J_{xx}=0=J_{yy}}$. Therefore, due to the Bianchi identities, there are only two equations that have to be satisfied, namely ${J_{rr}=0}$ yielding \eqref{Neq_rr}, and ${J_{ru}=0}$ implying \eqref{Neq_ru}, where ${\T= {\textstyle \frac{2}{3}\Lambda \,\Omega^3 } }$ given by \eqref{traceC} has been used. They completely determine all  vacuum QG  solutions of the form \eqref{BHmetric}. The remaining equations ${J_{xx}=0= J_{yy}}$ are then automatically satisfied since necessarily ${\mathcal{J}=0}$, i.e., using (\ref{calJ})
	\begin{equation}
	\big({\H}\Omega\Omega'\big)'+\Omega^2+\Lambda \Omega^4 -3\,\T\,\Omega
	= \tfrac{1}{3}k \,\big({\H}{\H}''''+{\H}'{\H}'''-{\textstyle\frac{1}{2}}{{\H}''}^2 +2 \big) \,. \label{J=0}
	\end{equation}
By substituting ${\T={\textstyle \frac{2}{3}\Lambda \,\Omega^3 }}$, see \eqref{traceC}, into \eqref{J=0} we immediately obtain equation \eqref{Neq_xx}.

Thus solving the QG field equations \eqref{fieldeqsEWmod} for the metric \eqref{BHmetric} is equivalent to solving  \eqref{Neq_rr} and
\be
\big(\Omega^2 {\H}\big)'' +2 \Lambda \Omega^4-6\T\Omega =
-\tfrac{2}{3}k \big(2\,{\H}{\H}''''+{\H}'{\H}'''-{\textstyle\frac{1}{2}}{{\H}''}^2 +2\big) \label{bJru0}\,.
\ee
Substituting for  ${{\H}''''}$ from \eqref{Neq_rr}, these two equations \eqref{Neq_rr} and \eqref{bJru0} can be simplified to the final set of the field equations \eqref{Eq1} and \eqref{Eq2} for the metric functions $\Omega(r)$ and ${\H}(r)$, namely
\begin{align}
\Omega\Omega''-2{\Omega'}^2 = &\ \tfrac{1}{3}k\,{\H}'''' \,, \label{Eq1C}\\
\Omega\Omega'{\H}'+3\Omega'^2{\H}+\Omega^2 -\Lambda \Omega^4
= &\ \tfrac{1}{3}k \big({\H}'{\H}'''-{\textstyle\frac{1}{2}}{{\H}''}^2 +2 \big)\,. \label{Eq2C}
\end{align}

Let us also note that, instead of the system \eqref{Eq1C} and \eqref{Eq2C}, one can alternatively solve equation \eqref{Eq1C} and any two equations  from the set (\ref{Neq_ru}), (\ref{Neq_xx}), (\ref{traceC}).

\newpage


\end{document}